\documentclass[twocolumn]{aastex62}

\usepackage{graphicx}

\usepackage{aas_macros}
\usepackage{amsmath}
\usepackage{cases}
\usepackage{apjfonts}
\usepackage{courier}
\usepackage[caption=false]{subfig}

\usepackage{booktabs}
\usepackage{threeparttable}
\usepackage{tabularx}
\usepackage{multirow}

\usepackage{color}

\DeclareMathAlphabet{\mathsc}{OT1}{cmr}{m}{sc}
\def\testbx{bx}%
\DeclareRobustCommand{\ion}[2]{%
\relax\ifmmode
\ifx\testbx\f@series
{\mathbf{#1\,\mathsc{#2}}}\else
{\mathrm{#1\,\mathsc{#2}}}\fi
\else\textup{#1\,{\mdseries\textsc{#2}}}%
\fi}

\begin{document}

%\preprint{APS/123-QED}

\title{A Self-Consistent Framework for Multi-Line Modeling in Line Intensity Mapping Experiments}

\author{Guochao Sun}
\email{gsun@astro.caltech.edu}
\affiliation{California Institute of Technology, 1200 E. California Blvd., Pasadena, CA 91125, USA}

\author{Brandon S. Hensley}
\affiliation{Spitzer Fellow, Department of Astrophysical Sciences, Princeton University, Princeton, NJ 08544, USA}
\affiliation{Jet Propulsion Laboratory, California Institute of Technology, 4800 Oak Grove Dr., Pasadena, CA 91109, USA}

\author{Tzu-Ching Chang}
\affiliation{California Institute of Technology, 1200 E. California Blvd., Pasadena, CA 91125, USA}
\affiliation{Jet Propulsion Laboratory, California Institute of Technology, 4800 Oak Grove Dr., Pasadena, CA 91109, USA}

\author{Olivier Dor\'e}
\affiliation{California Institute of Technology, 1200 E. California Blvd., Pasadena, CA 91125, USA}
\affiliation{Jet Propulsion Laboratory, California Institute of Technology, 4800 Oak Grove Dr., Pasadena, CA 91109, USA}

\author{Paolo Serra}
\affiliation{Jet Propulsion Laboratory, California Institute of Technology, 4800 Oak Grove Dr., Pasadena, CA 91109, USA}

\begin{abstract}
Line intensity mapping (LIM) is a promising approach to study star formation and the interstellar medium (ISM) in galaxies by measuring the aggregate line emission from the entire galaxy population. In this work, we develop a simple yet physically-motivated framework for modeling the line emission as would be observed in LIM experiments. It is done by building on analytic models of the cosmic infrared background that connect total infrared luminosity of galaxies to their host dark matter halos. We present models of the \ion{H}{i} 21\,cm, CO(1-0), [\ion{C}{ii}] 158\,$\mu$m, and [\ion{N}{ii}] 122 and 205\,$\mu$m lines consistent with current observational constraints. With four case studies of various combinations of these lines that probe different ISM phases, we demonstrate the potential for reliably extracting physical properties of the ISM, and the evolution of these properties with cosmic time, from auto- and cross-correlation analysis of these lines as measured by future LIM experiments.
\end{abstract}


% =====================   S1: Introduction   ===================== %

\section{Introduction}

Line intensity mapping (LIM) is an emerging observational technique developed to statistically measure the intensity field fluctuations of a given spectral line (see \citealt{Kovetz_2017} for a recent review). While traditional galaxy surveys are restricted by the detection limit of individual sources, LIM is sensitive to the emission from {\it all} galaxies, providing a complementary probe of faint objects. Due to its statistical nature, LIM is most effective at constraining how {\it average} physical properties, including the star formation rate, ISM conditions, luminosity function, spatial distribution, etc., of the source galaxy population evolve over cosmic time (\citealt{Serra_2016}, hereafter S16; \citealt{Kovetz_2017}; \citealt{Chang_2019BAAS}).

LIM was first pioneered with the redshifted \ion{H}{i} 21\,cm line signal. It serves as a probe of both the matter density distribution as traced by the atomic hydrogen gas in the interstellar medium (ISM), for example, the baryon acoustic oscillation (BAO) feature in galaxy power spectrum \citep{Chang_2010, Switzer_2013}, and the structure of neutral intergalactic medium (IGM) at high redshift, in particular during cosmic reionization \citep{MMR_1997, FZH_2004, FOB_2006, PL_2012}. Recently, the application of LIM to other emission lines has gained increasing attention, including CO rotational lines \citep{pullen2013, Breysse_2014, Mashian_2015, Li_2016}, far-infrared (FIR) fine-structure lines of \ion{C}{ii}, \ion{N}{ii}, \ion{O}{i} and others (\citealt{Gong_2012}; \citealt{Uzgil_2014}; \citealt{Silva_2015}; \citealt{Yue_2015}; S16), and bright optical/UV emission lines such as Ly$\alpha$ and H$\alpha$ \citep{Silva_2013, Pullen_2014, CF_2016,Gong_2017,Silva_2018}.

Substantial theoretical and experimental efforts have been devoted to the detection and interpretation of LIM signals of individual lines. However, a simple, physical model that allows multiple line signals, presumably originating from and thus probing different ISM phases, to be modeled in a self-consistent manner is still lacking. The goal of this work is to develop such a self-consistent framework for determining the integrated line intensities of galaxies observed in the intensity mapping regime. This framework is intended to bridge the gap between commonly-used approaches anchored on scaling relations empirically determined from observations (e.g., \citealt{VL_2010}; \citealt{pullen2013}; \citealt{Silva_2015}; \citealt{Li_2016}; S16) and sophisticated simulations of galaxy-scale hydrodynamics and radiative transfer \cite[e.g.,][]{Pallottini_2019, Popping_2019}. More specifically, it should be sophisticated enough to capture the relevant ISM physics and employ meaningful physical parameters, yet simple enough to interpret the auto/cross-correlation of the intensities of various lines observed in the intensity mapping regime in terms of coarse-grained galactic ISM properties. Some examples include the mass fraction of different ISM phases, as well as quantities like the photoelectric (PE) heating efficiency and the CO-to-$\rm H_2$ ratio, which are closely related to exact physical conditions of the ISM (e.g., temperature, density, radiation field) and therefore of particular interest to LIM surveys of the corresponding lines. Furthermore, this analytical framework should also allow mock signal maps to be readily constructed from given information about the position and physical properties of source populations, thereby enabling straightforward implementation in semi-analytic models of the LIM signals.  

We build such a formalism using the information from the cosmic infrared background (CIB). The CIB has, on account of sensitive FIR observations from experiments like {\it Planck} \citep{planckXXX} and {\it Herschel} \citep{Viero_2013CIB}, been the subject of detailed modeling efforts. In particular, analytic models connecting the infrared (IR) luminosity of galaxies to the mass and redshift of their host halos have been successful in reproducing the statistical properties of the CIB (e.g., \citealt{Shang_2012}; S16; \citealt{WD_2017}). In this work, we follow and extend the ideas presented in S16 by employing the CIB model as a starting point for models of both line and continuum emission from galaxies as a function of redshift and halo mass. Taken at face value, the IR luminosities assumed in these models imply a corresponding dust mass, gas mass, and metallicity, which in turn can inform predictions of emission from various interstellar lines, including \ion{H}{i}, [\ion{C}{ii}], [\ion{N}{ii}] and CO(1-0). We work through these consequences, with an eye toward testable predictions from upcoming intensity mapping experiments of these lines. 

This paper is organized as follows. In Section~\ref{sec:halos}, we present a simple analytic model that describes a variety of physical properties of dark matter halos hosting the line-emitting galaxies, such as their star formation rate, dust mass, metallicity and so forth. We then discuss in Section~\ref{sec:lines} how we model the emission of \ion{H}{i}, [\ion{C}{ii}], [\ion{N}{ii}] and CO lines as tracers of different phases of the ISM, based on our model of halo properties. In Section~\ref{sec:lim}, we review the theoretical framework of estimating the power spectrum signal of intensity mapping experiments, as well as the uncertainty associated with the measurements. We compare the predicted strengths of different lines to constraints from the literature in Section~\ref{sec:results}. We then present four case studies in Section~\ref{sec:case_studies} to demonstrate how physical conditions of multi-phase ISM may be probed by and extracted from with intensity mapping experiments. We outline prospects for further improving and extending our simple modeling framework, before briefly concluding in Section \ref{sec:disncon}. Throughout the paper, we assume a flat, $\Lambda$CDM cosmology consistent with the measurement by the \cite{Planck2016XIII}. 

% =====================   S2: Analytic Model of Radiation Backgrounds   ===================== %

\section{A Simple Analytic Model of Mean Halo Properties} \label{sec:halos}

An important criterion for choosing first targets for LIM surveys is the overall brightness of the spectral line, which is determined by many different factors, including abundance, excitation potential, critical density, destruction and/or scattering and so forth. In many cases, nevertheless, the line signal either directly traces the star-forming activity (e.g., [\ion{C}{ii}], [\ion{N}{ii}]) or indirectly probes the gas reservoir closely associated with star formation (e.g., \ion{H}{i}, CO). Therefore, it is critical to understand and model the star formation of galaxies well enough in order to properly estimate the production of lines in LIM. 

The majority of starlight from young stars at optical/UV wavelengths is absorbed and reprocessed into IR radiation by interstellar dust, naturally giving rise to the connection between IR observations of galaxies and their star formation rate. Because the fraction of spatially resolved galaxies decreases rapidly with increasing wavelength in the IR/sub-millimeter regime, the observed CIB mean intensity and fluctuations provide a useful probe of global star-forming activities. Combining the halo model formalism describing the clustering of galaxies at different angular scales \citep{CS_2002} and the observed angular anisotropy of the CIB, \citet{Shang_2012} developed a simple parametric form for the infrared luminosity of galaxies as a function of halo mass and redshift, which has been successfully applied to reconstruct the observed angular CIB auto- and cross-power spectra (\citealt{planckXXX}; S16; \citealt{WD_2017}). In this section, we extend the discussion in S16 and present a simple, CIB-based model for the mean properties of dark matter halos, such as their infrared luminosity, dust and gas mass, metallicity, etc., which are essential ingredients for the line emission models in this work.

\begin{table}[h!]
\centering
\caption{Fiducial Parameters of CIB Model}
\begin{tabular}{cccc}
\hline
\hline
\bf{Parameter} & \bf{Description} & \bf{Value} & \bf{Reference} \\
\hline
$L_0$ & $L_{\rm IR}$ normalization & $0.0135\,L_{\odot}/M_\odot$ & Eq.~\ref{eq:ml_relation} \\
$s$ & $z$ evolution of $L_{\rm IR}$ & 3.6 & Eq.~\ref{eq:cib_z_evol} \\
$T_0$ & $T_{\rm dust}$ at $z=0$ & $24.4$\,K & Eq.~\ref{eq:cib_Td} \\
$\alpha$ & $z$ evolution of $T_{\rm dust}$ & 0.36 & Eq.~\ref{eq:cib_Td} \\
$\beta$ & RJ-side index & 1.75 & Eq.~\ref{eq:cib_Theta} \\
$\gamma$ & Wien-side index & 1.7 & Eq.~\ref{eq:cib_Theta}, \ref{eq:cib_nu0} \\
$M_{\mathrm{eff}}$ & effective halo mass & $10^{12.6}\,M_{\odot}$ & Eq.~\ref{eq:cib_Sigma} \\
$\sigma_{L/M}$ & log scatter & 0.5 & Eq.~\ref{eq:cib_Sigma} \\
\hline
\hline
\label{tb:cib}
\end{tabular}
\end{table}

% ----------   LIR   ---------- %

\subsection{IR luminosity}

We work in the aforementioned framework of the halo model for CIB anisotropies introduced by \cite{Shang_2012}, which has been exploited in various contexts, including the modeling of high-redshift emission lines (e.g., \citealt{planckXXX}; S16; \citealt{WD_2017}; \citealt{Pullen_2018}). In this model, the specific luminosity emitted by a galaxy hosted by a halo of mass $M$ at redshift $z$ at the observed frequency $\nu$ is given by
\begin{equation}
\label{eq:ml_relation}
L_{{\rm IR},\left(1+z\right)\nu}\left(M, z\right) = L_{{\rm IR},0} \Phi\left(z\right) \Sigma\left(M\right) \Theta\left[\left(1+z\right)\nu\right]~,
\end{equation}
where $L_{{\rm IR},0}$ is a normalization constant (see Table~\ref{tb:cib} for a summary of fiducial parameter values taken for the CIB model), whereas $\Phi\left(z\right)$, $\Sigma\left(M\right)$, and $\Theta\left[\left(1+z\right)\nu\right]$ are functions to be specified. $\Phi(z)$ governs the evolution of the luminosity--mass relation with redshift, driven, e.g., by an increase in the star formation rate at fixed halo mass with increasing redshift. This is modeled as a power law
\begin{equation}
\Phi\left(z\right) = \left(1+z\right)^s~,
\label{eq:cib_z_evol}
\end{equation}
where \citet{WD_2017} found a best-fit value of $s = 3.6$. However, we note that the exact value of $s$ is not well-constrained by the integrated CIB intensity and less steep slopes have indeed been suggested by some other CIB analyses and galaxy evolution models (see discussion in S16).

$\Theta\left[\left(1+z\right)\nu\right]$ describes the frequency dependence of the dust emission as a function of redshift. Over most of the FIR frequency range, the dust emission in a galaxy is modeled as a modified blackbody of temperature $T_{\rm d}$ and spectral index $\beta$, 
\begin{equation}
\label{eq:mbb}
I_\nu \propto \nu^\beta\ B_\nu\left[T_{\rm d}(z)\right]~,
\end{equation}
where $B_\nu\left[T_{\rm d}(z)\right]$ is the Planck function at a dust temperature
\begin{equation}
T_{\rm d}(z) = T_0(1+z)^\alpha~,
\label{eq:cib_Td}
\end{equation}
where $T_0 = 24.4$\,K is the typical dust temperature in a star-forming galaxy at $z=0$, and the redshift dependence is taken to be $\alpha=0.36$ following \cite{planckXXX} and \cite{WD_2017}. The high frequency component is modeled as a power law to account for emission from small, stochastically-heated grains. The full SED is given by
\begin{equation}
  \Theta\left[\left(1+z\right)\nu\right] = A\left(z\right) \times
  \begin{cases}
  \nu^\beta\ B_\nu\left[ T_{\rm d}(z) \right] & \text{$\nu < \nu_0$} \\
  \nu^{-\gamma} & \text{$\nu \geq \nu_0$}
  \end{cases}
  ~,
  \label{eq:cib_Theta}
\end{equation}
where the frequency $\nu_0$ at any given redshift is determined by having
\begin{equation}
\frac{{\rm d}\ \ln \left\{\nu^\beta B_\nu \left[ T_{\rm d}(z) \right] \right\}}{{\rm d}\ \ln\nu} = -\gamma
\label{eq:cib_nu0}
\end{equation}
satisfied at $\nu = \nu_0$. We adopt $\beta = 1.75$ and  $\gamma = 1.7$ \citep{planckXXX, WD_2017}, which yield $\nu_0=3.3, 2.1, 1.6$ and 1.3\,THz or wavelength equivalents 92, 143, 185, and 222\,$\mu$m at redshifts $z=0, 1, 2$ and 3, respectively. The redshift-dependent normalization factor $A\left(z\right)$ is defined such that
\begin{equation}
    \int \Theta\left(\nu,z\right)\,d\nu = 1
\end{equation}
for all $z$.

$\Sigma(M)$ links the IR luminosity to the halo mass and is modeled as a lognormal relation:
\begin{equation}
\Sigma(M) = M \frac{1}{\sqrt{2\pi\sigma^2_{L/M}}}\mathrm{exp}\Big[-\frac{(\mathrm{log}_{10}M - \mathrm{log}_{10}M_{\mathrm{eff}})^2}{2\sigma_{L/M}^2}\Big]~,
\label{eq:cib_Sigma}
\end{equation}
where $M_{\rm eff}$ describes the most efficient halo mass at hosting star formation, and $\sigma_{L/M}$ accounts for the range of halo masses mostly contributing to the infrared luminosity. This functional form captures the fact that the star formation efficiency is suppressed for halo masses much lower or much higher than $M_{\rm eff}$ \citep{Mo_book, Furlanetto_2017, KVM_2018}, due to various feedback mechanisms such as input from supernova explosions and active galactic nuclei (AGNs). The total infrared luminosity (8--1000\,$\mu$m) is then
\begin{equation}
L_{\rm IR}(M, z) = \int^{\rm 37.5\,THz}_{\rm 300\,GHz} \mathrm d \nu L_{(1+z)\nu}(M, z)~.
\end{equation}

% ----------   SFH   ---------- %

\begin{figure}[h!]
\centering
\includegraphics[width=0.48\textwidth]{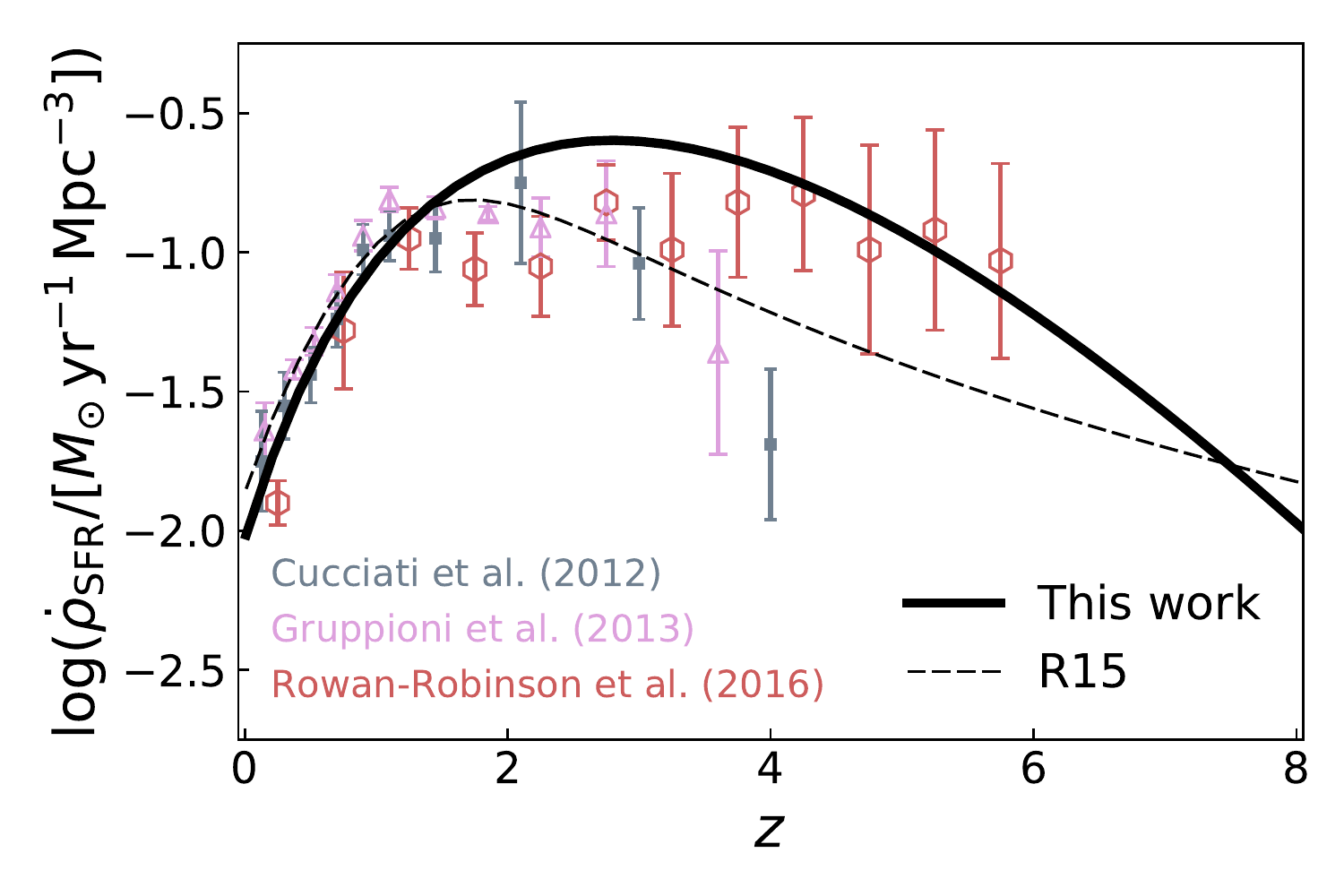}
\caption{Cosmic star formation history implied by our CIB model, compared with that inferred from UV \citep{Cucciati_2012} and IR \citep{Gruppioni_2013, RR_2016} data. Also shown for comparison is the maximum-likelihood model from \cite{Robertson_2015}, which is a fit to the SFRD estimates based on IR and (primarily) optical/UV data.}
\label{fig:sfrd}
\end{figure}

\subsection{Star Formation History}

From the total infrared luminosity, it is straightforward to derive the star formation rate as a function of halo mass and redshift thanks to the well-established correlation between them \citep{Kennicutt_1998, MD_2014}. In this work, we simply take
\begin{equation}
\label{eq:sfr_lir}
\dot{M}_{\star}(M, z) = \mathcal{K}_{\rm IR} L_{\rm IR}~,
\end{equation}
where $\mathcal{K}_{\rm IR} = 1.73 \times 10^{-10}\,M_{\odot}\,\mathrm{yr^{-1}}\,L^{-1}_{\odot}$, consistent with a stellar population with a Salpeter initial mass function (IMF) and solar metallicity. The star formation rate density (SFRD) can consequently be written as
\begin{equation}
\dot{\rho}_{\star}(M, z) = \int_{M_{\rm min}}^{M_{\rm max}} \frac{\mathrm d N}{\mathrm d M} \dot{M}_{\star}(M, z)~,
\end{equation}
where $\mathrm d N / \mathrm d M$ is the dark matter halo mass function defined for the virial mass $M_{\rm vir}$ \citep{Tinker_2008}. Figure~\ref{fig:sfrd} shows a comparison between cosmic SFRDs predicted by the adopted CIB model and those from the literature. The data points represent estimated SFRDs based on both dust-corrected UV observations \citep{Cucciati_2012} and infrared/sub-millimeter observations of obscured star formation \citep{Gruppioni_2013, RR_2016}. Also shown is the maximum-likelihood model of cosmic SFRD from \citet{Robertson_2015} based on extrapolating the galaxy IR and UV luminosity functions down to $10^{-3}L_\star$. The agreement between the CIB-derived SFRD and the optical/UV-derived SFRD may be improved with different modeling choices \cite[e.g.,][]{Maniyar_2018}. However, this comes at the expense of phenomenological parameterizations of the effective bias factor of dusty galaxies, and we therefore do not follow that approach here. 

% ----------   Dust Mass   ---------- %

\begin{figure}[h!]
\centering
\includegraphics[width=0.48\textwidth]{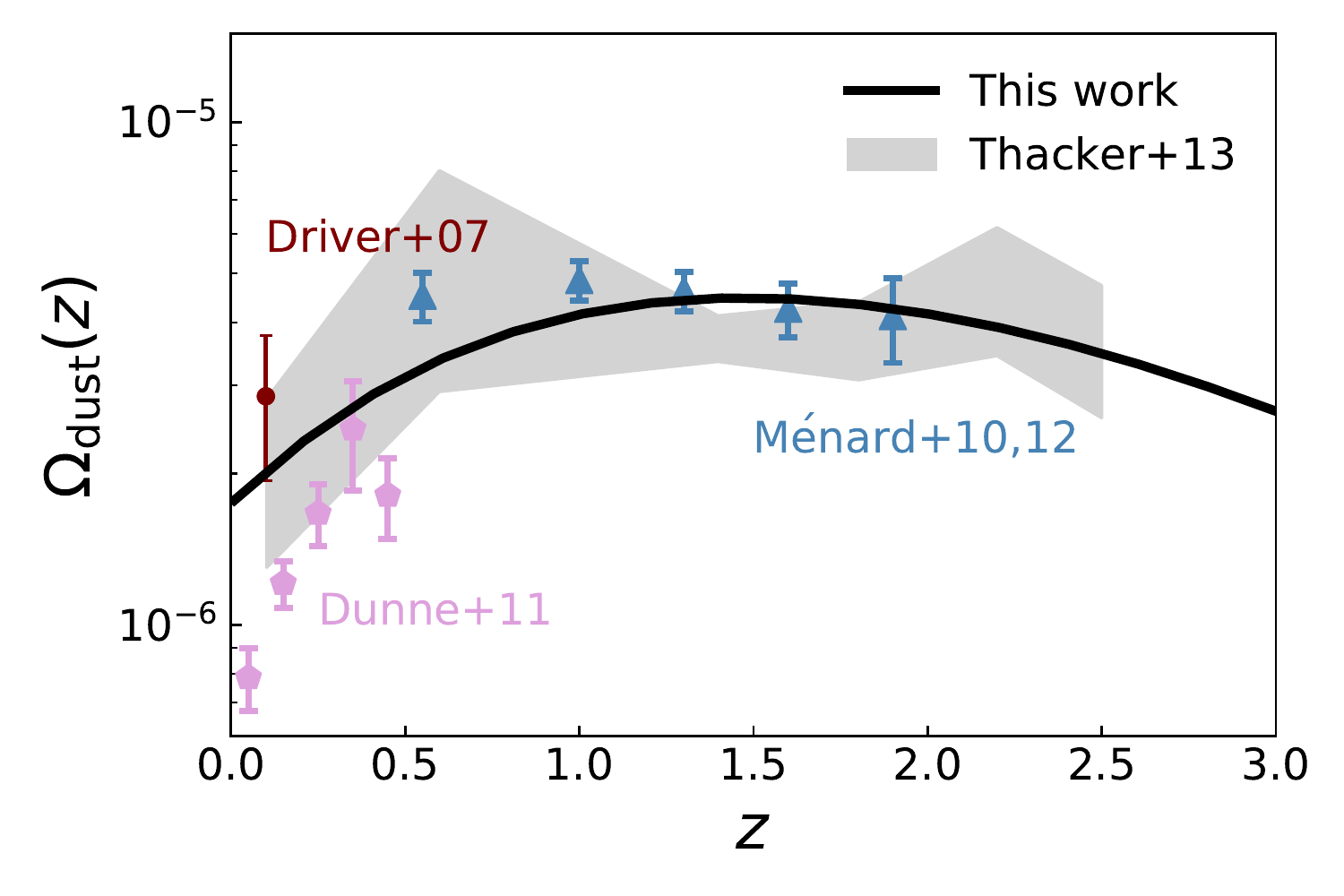}
\caption{Redshift evolution of the dust density parameter $\Omega_{\rm d}$, compared with various dust abundance constraints from the literature \citep{Driver_2007, Dunne_2011, Menard_2010, Menard_2012, Thacker2013}.}
\label{fig:omega_dust}
\end{figure}

\subsection{Dust Mass}
Since we have specified both the dust luminosity and the dust temperature from the CIB model, it is possible to estimate the implied dust mass. Assuming that the dust mass is dominated by larger grains whose emission can be described by a modified blackbody with a single dust temperature, the dust luminosity, mass, and temperature can be related via

\begin{equation}
L_{\rm IR}\left( M, z \right) = P_0 M_{\rm d}\left( M, z \right) \left[ \frac{T_{\rm d}(z)}{T_0} \right]^{4+\beta}\,~,
\end{equation}
where the normalization constant $P_0$ is the power emitted per mass of dust at temperature $T_0$.

To estimate $P_0$, we note that \citet{Planck_Int_XVII} found the Galactic \ion{H}{i}-correlated dust emission to be well-described by Equation~\ref{eq:mbb} with $T_{\rm d} \simeq 20$\,K and $\beta \simeq 1.6$. Further, they derived an 857\,GHz dust emissivity per H of $\epsilon_{857} = 4.3\times10^{-21}$\,MJy\,sr$^{-1}$\,cm$^2$\,H$^{-1}$. Thus,

\begin{align}
P_0 &= 4\pi\epsilon_{857}\frac{M_{\rm H}}{M_{\rm d}}\frac{1}{m_{\rm p}}\int \left(\frac{\nu}{\rm 857\,GHz}\right)^{1.6} \frac{B_\nu\left(20\,{\rm K}\right)}{B_{857}\left(20\,{\rm K}\right)}\,{\rm d}\nu \nonumber \\
& \simeq 110\,L_\odot / M_\odot~,
\end{align}
where we have assumed a gas-to-dust mass ratio of 100 \citep{Draine_2007}. By using this formalism to estimate the dust mass, we are implicitly assuming that physical properties (e.g., composition) of dust grains do not evolve systematically with redshift or metallicity, only their abundance per H atom. 

Because $\Theta$ is normalized to unity (see Equation~\ref{eq:cib_Theta}), the redshift dependence of $L_{\rm IR}$ is determined entirely by $\Phi(z)$, and thus
\begin{equation}
M_{\rm d} \propto \Sigma(M) (1+z)^{s-\alpha(4+\beta)}\,.
\end{equation}

The implied cosmic density of dust, $\Omega_{\rm d}$, as a function of redshift is
\begin{eqnarray}
\Omega_{\rm d}\left(z\right) = \frac{1}{\rho_{\rm crit,0}}\int \mathrm d M \frac{\mathrm d N}{\mathrm d M} M_{\rm d}(M,z)~,
\end{eqnarray}
where $\rho_{\rm crit,0}$ denotes the critical density of the universe at the present time. In Figure~\ref{fig:omega_dust}, we plot the redshift evolution of the dust density parameter $\Omega_{\rm d}$, which is compared with a compilation of previous dust abundance measurements by \citet{Thacker2013}, including constraints from integrating low-$z$ dust mass functions \citep{Dunne_2011}, extinction measurements from the Sloan Digital Sky Survey\footnote{The combined data set from \citet{Thacker2013} is adopted here, which assumes that the halo dust content does not evolve significantly with redshift.} \citep{Menard_2010, Menard_2012} and 2dF \citep{Driver_2007}, and cosmic far-infrared background anisotropy \citep{Thacker2013}.

% ----------   Hydrogen Mass   ---------- %

\subsection{Hydrogen Mass}

\begin{figure}[h!]
\centering
\includegraphics[width=0.45\textwidth]{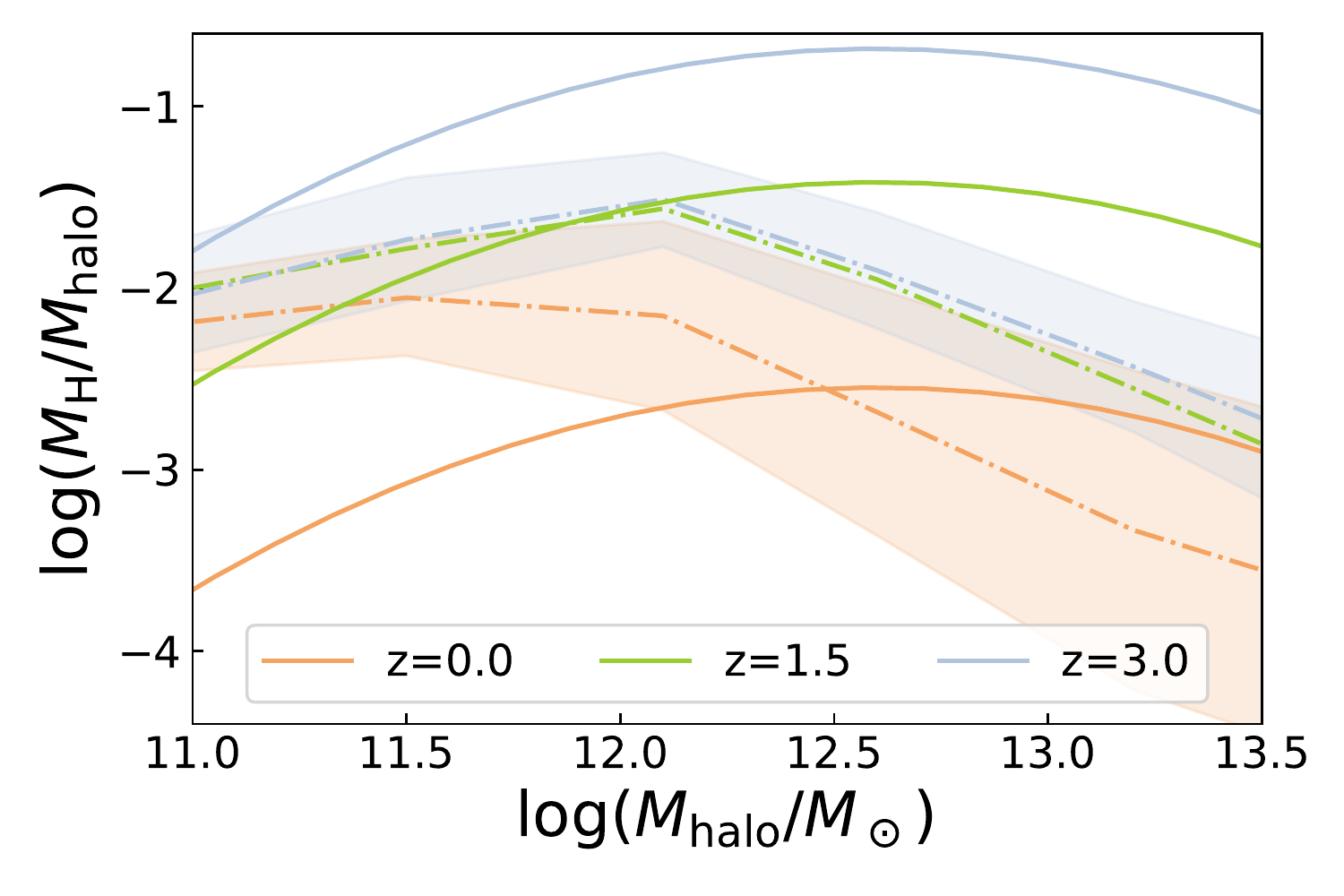}
\caption{Hydrogen--halo mass relation predicted by our $K \Sigma(M) \Phi(z)$ parameterization at different redshifts (solid curves), compared with the semi-empirical estimates from \cite{Popping_2015} shown by the dash-dotted curves and shaded bands (at $z=0$ and 3 only, 95\% confidence intervals). }
\label{fig:hydrogen_mass}
\end{figure}

Insofar as gas and dust are well mixed, $\Phi\Sigma$ encodes the total hydrogen mass in the halo. However, the correspondence is not direct since the dust luminosity depends on not only the amount of dust present but also the dust temperature, which is assumed to evolve with redshift (see Equation~\ref{eq:cib_Theta}). We therefore introduce the modification

\begin{equation}
M_{\rm H}(M,z) = K(z)\Sigma(M)\Phi(z)~,
\label{eq:hydrogen_mass}
\end{equation}
where $K(z) = \zeta(1+z)^\xi$ is a normalization factor that sets the total amplitude of $M_{\rm H}$. The amplitude and redshift dependence of $K$ are determined by approximately matching the hydrogen--halo mass relation over $0<z<3$ predicted by \cite{Popping_2015} as shown in Figure~\ref{fig:hydrogen_mass}, while at the same time yielding a gas metallicity of approximately $Z_\odot$ at $z=0$ (discussed in next section). For our fiducial model, we take $\zeta = 0.005$, $\xi = -1$ for $0<z<1$ and $\zeta = 0.0025$, $\xi = 0$ otherwise. These values are chosen such that the overall redshift dependence of $M_{\rm H}$ roughly agrees with the product of inferred growth rate of halo mass, which scales as $(1+z)^{1.5}$ at $z \la 1$ and $(1+z)^{2.5}$ at higher redshifts \citep{McBride_2009}, and the average star formation efficiency, which may carry an extra factor of $(1+z)^{1-1.5}$ depending on the exact physical mechanisms coupling the stellar feedback (e.g., supernova explosions) to galaxies \citep{SF_2016, Furlanetto_2017}. Indeed, the gas-to-stellar mass ratio of $M>10^{11}\,M_\odot$ halos of interest has been found to be only weakly dependent on redshift \citep{Popping_2015}. The mass dependence, on the other hand, is motivated since the same physical mechanisms preventing star formation at both ends of halo masses also play a role in regulating the hydrogen mass in a galaxy.

The total mass of hydrogen in our model can be written as
\begin{equation}
M_{\rm H} = M_{\ion{H}{i}} + M_{\rm H_2} + M_{\ion{H}{ii}}~.
\end{equation}
If we express the fractions of molecular and ionized hydrogen as $f_{\rm H_2}$ and $f_{\ion{H}{ii}}$ respectively, then the masses of hydrogen in three different phases become
\begin{align}
M_{\rm H_2}(M, z) & = f_{\rm H_2} M_{\rm H}(M, z)~, \\
M_{\ion{H}{ii}}(M, z) & = f_{\ion{H}{ii}} M_{\rm H}(M, z)~, \\
M_{\ion{H}{i}}(M, z) & = (1 - f_{\rm H_2} - f_{\ion{H}{ii}}) M_{\rm H}(M, z)~.
\end{align}
As a fiducial value, we set $f_{\rm H_2}=0.2$, typical for most galaxies up to $z\sim1$ and the most massive ones up to $z\sim2$ (see, e.g., \citealt{Popping_2012}). Likewise, we adopt $f_{\ion{H}{ii}}=0.1$, based on the estimated masses of different ISM phases from \citet{Tielens_2005}. We note that a factor of 1.36 accounting for the helium abundance is needed to connect the total hydrogen mass to the total gas mass, i.e., $M_{\rm gas} = 1.36 M_{\rm H}$ \citep{Draine_2007}. 

\begin{figure}[h!]
\centering
\includegraphics[width=0.45\textwidth]{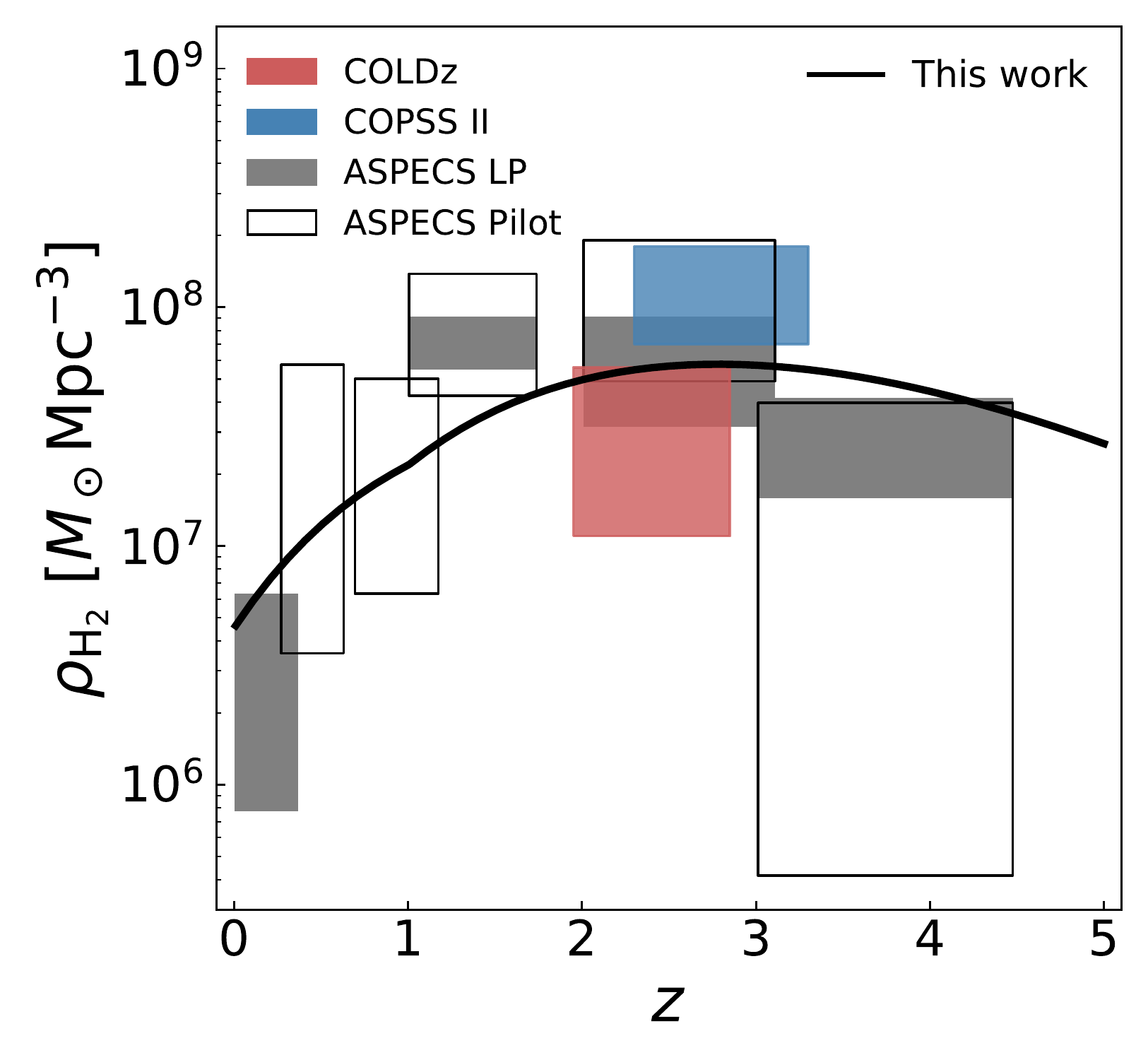}
\caption{Cosmic evolution of the molecular gas density $\rho_{\rm H_2}$ predicted by our reference model with $f_{\rm H_2} = 0.2$, compared with observational constraints from COLDz \citep{Riechers_2019}, COPSS~II \citep{Keating_2016}, ASPECS Pilot \citep{Decarli_2016} and ASPECS large program \citep{Decarli_2019}.}
\label{fig:rhoH2}
\end{figure}

Using Eq.~\ref{eq:hydrogen_mass} and the fiducial molecular gas fraction $f_{\rm H_2} = 0.2$, we can also obtain the cosmic evolution of the molecular gas density $\rho_{\rm H_2}$, whose comparison against the cosmic SFRD (especially the peak of star formation at $z\sim2$) provides vital information about the fueling and regulation of star formation by cold gas. Constraints on $\rho_{\rm H_2}$ have so far been placed primarily by observations of the CO rotational transitions. Figure~\ref{fig:rhoH2} shows how $\rho_{\rm H_2}$ as a function of redshift, computed with our fiducial choice of $f_{\rm H_2} = 0.2$, compares with constraints derived from various CO LIM experiment and deep galaxy surveys, including COLDz \citep{Riechers_2019}, COPSS~II \citep{Keating_2016}, ASPECS Pilot \citep{Decarli_2016} and ASPECS large program \citep{Decarli_2019}. Planned LIM experiments such as COMAP \citep{Li_2016} and TIME \citep{Crites_2014SPIE} and next-generation Very Large Array (ngVLA) concepts \citep{Walter_2019BAAS} are expected to greatly reduce the substantial uncertainties present in current limits.

% ----------   Metallicity   ---------- %

\subsection{Metallicity}

If the dust-to-metals ratio is assumed to be constant, the metallicity $Z$ of interstellar gas in our model can be expressed as a function of the dust mass as
\begin{eqnarray}
\frac{Z}{Z_{\odot}}(z) \sim 100 \frac{M_{\rm d}(M, z)}{M_{\rm H}(M, z)}~.
\end{eqnarray}
Recent hydrodynamic galaxy formation simulations have indeed found little variation in the dust-to-metals ratio with redshift or metallicity above 0.5$Z_\odot$ \citep{Li_2019}.

\begin{figure}[h!]
\centering
\includegraphics[width=0.45\textwidth]{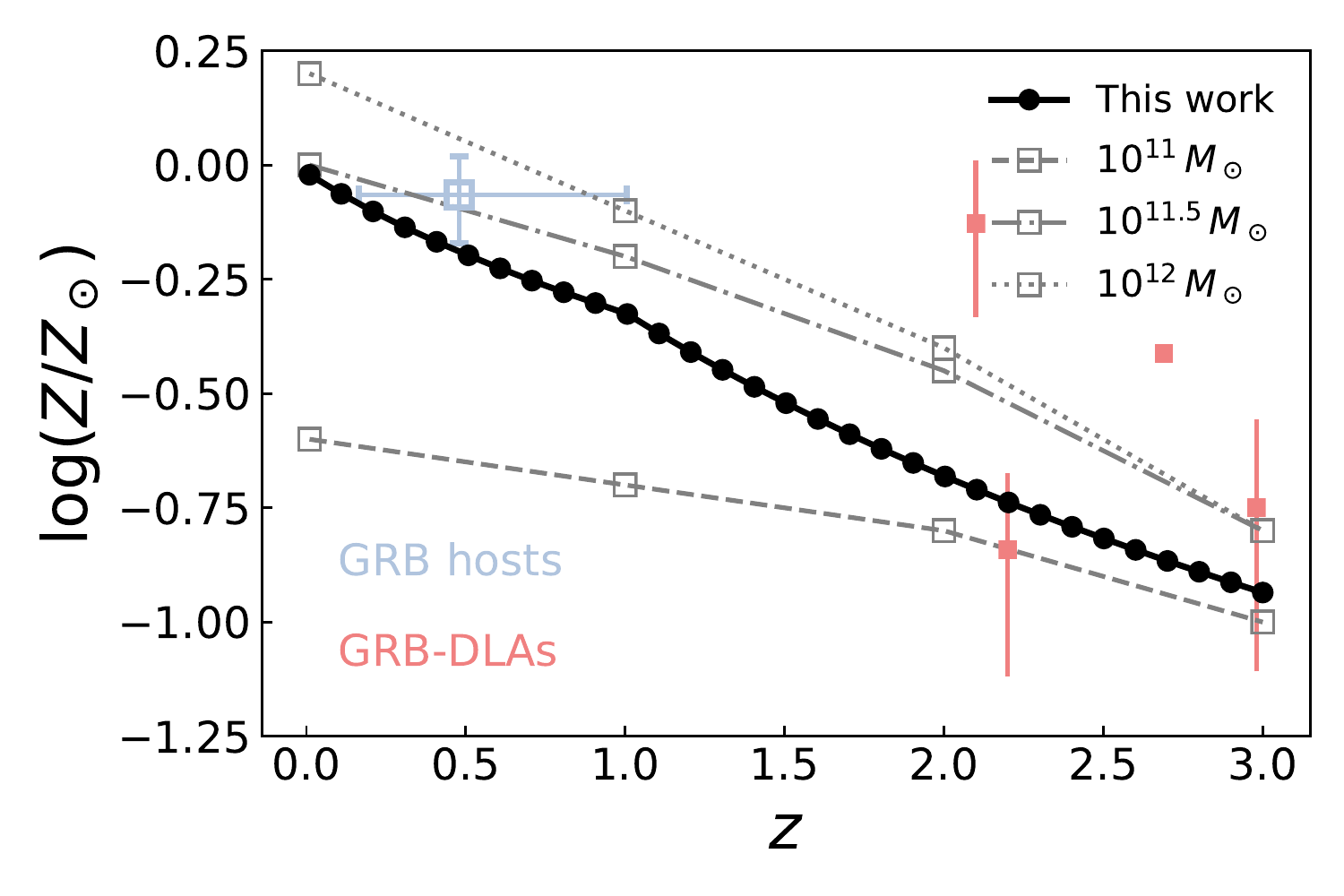}
\caption{Redshift evolution of the metallicity $Z$ derived from our model, compared with semi-analytic estimates of gas-phase $Z$ from \citet{Fu_2013}, evaluated at different halo masses ranging from $10^{11}$ to $10^{12}\,M_\odot$. Also shown are inferred metallicities of the warm ISM of $z<1$ GRB host galaxies and the neutral ISM of GRB-DLAs from \citet{Savaglio_2006NJPh}. }
\label{fig:metallicity}
\end{figure}

Our simple model of the gas-phase metallicity gives no halo mass dependence, which is likely an oversimplification given that effects like galactic winds regulating the metallicity of galaxies may evolve with halo mass in a non-trivial way. Nevertheless, as shown in Figure~\ref{fig:metallicity}, our predicted redshift evolution of metallicity is broadly consistent with that estimated semi-analytically by \citet{Fu_2013}, using the Millennium-II Simulation \citep{BK_2009MSII} combined with an $\rm H_2$ prescription specified by the gas surface density, metallicity and a constant clumping factor \citep{KMT_2009, MK_2010}. We note that $Z$ only evolves moderately for $M > 10^{11.5}\,M_\odot$, a halo mass range that our CIB model is calibrated and most sensitive to. Therefore, in the context of the CIB model a mass-independent gas metallicity is likely a fair approximation. Figure~\ref{fig:metallicity} also shows the cosmic metallicity evolution inferred from gamma-ray burst (GRB) observations for comparison. By analogy to damped Ly$\alpha$ (DLA) systems of quasars, \citet{Savaglio_2006NJPh} uses strong absorption lines due to the intervening neutral gas to estimate the metallicity evolution of GRB-DLA systems and compare it with the average metallicity derived for a sample of GRB hosts at $z<1$.

\section{Models of Emission Lines} \label{sec:lines}

Based on the mass and redshift dependencies of a wide range of halo properties derived in Section~\ref{sec:halos}, we construct a model of the emission lines that trace star formation and the ISM in galaxies. In this section, we present our line emission models of the \ion{H}{i} 21\,cm line, [\ion{C}{ii}] 158\,$\mu$m line, the 122 and 205\,$\mu$m [\ion{N}{ii}] lines, and the CO(1-0) 2.6\,mm line. Each of these lines probes a somewhat different phase of the ISM, ranging from the coldest molecular gas to the warm ionized medium (WIM). As such, their joint analysis can reveal rich information about the multi-phase ISM, as will be illustrated in the following sections. 

\begin{figure}[h!]
\centering
\includegraphics[width=0.48\textwidth]{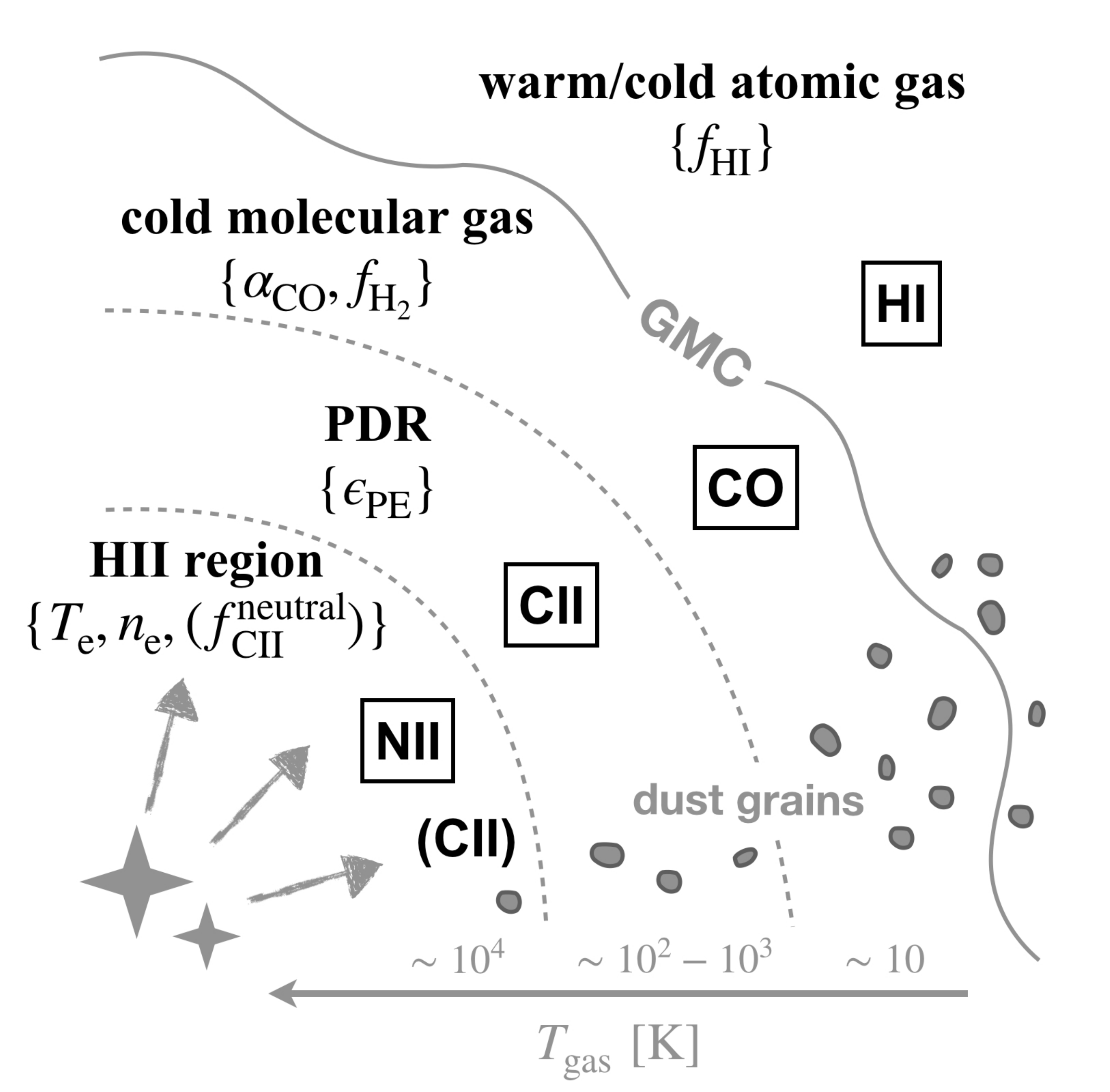}
\caption{Cartoon illustration of how physical parameters describing different ISM phases in our simplistic model are mapped to the emission lines considered. }
\label{fig:sketch}
\end{figure}

Figure~\ref{fig:sketch} illustrates how our modeling framework connects the emission from each of these lines to the phases of the ISM. Young stars formed in dense regions of a giant molecular cloud (GMC) are surrounded by \ion{H}{ii} regions ionized by UV radiation, whose physical conditions may be probed by FIR [\ion{N}{ii}] and [\ion{C}{ii}] lines. Photodissociation regions (PDRs) occupy the interface of \ion{H}{ii} regions and cold molecular gas traced by CO lines and produce the majority of [\ion{C}{ii}] emission, which is the main cooling mechanism balancing the photoelectric heating by dust grains. Together molecular gas clouds compose roughly half of the total ISM mass, whereas warm/cold atomic gas contributing most of the \ion{H}{i} mass is responsible for the remaining half. 

\begin{table}[!t]
\centering
\caption{Physical Parameters of the Reference ISM Model. \label{tb:line_params}}
\begin{threeparttable}
\begin{tabularx}{0.45\textwidth}{cccc}
\hline
\hline
\bf{Signal} & \bf{Parameter} & \bf{Symbol} & \bf{Value} \\
\hline
-- & Molecular gas fraction\tnote{$\star$} & $f_{\rm H_2}$ & 0.2 \\
-- & Ionized gas fraction\tnote{$\star$} & $f_{\ion{H}{ii}}$ & 0.1 \\
\,CO & $L$--$M$ conversion\tnote{$\star$} & $\alpha_{\rm CO}$ & $\frac{4.4\,M_\odot}{\mathrm{K\,km\,s^{-1}\,pc^2}}$ \\
 & Excitation temperature & $T_{\rm exc}$ & 10\,K \\
 & $\rm H_2$ number density & $n_{\rm H_2}$ & $2\times10^3\,\mathrm{cm^{-3}}$ \\
\,[\ion{C}{ii}] & PE efficiency\tnote{$\star$} & $\epsilon_{\rm PE}$ & $5\times10^{-3}$ \\
\,[\ion{N}{ii}] & Gas temperature\tnote{$\star$} & $T_{\rm gas, \ion{H}{ii}}$ & $10^4$\,K \\
 & Electron number density\tnote{$\star$} & $n_{\rm e, \ion{H}{ii}}$ & $10^2\,\mathrm{cm^{-3}}$ \\
\hline
\hline
\end{tabularx}
\begin{tablenotes}
\item [$\star$] {\footnotesize Varied as free parameters in the case studies presented in Section~\ref{sec:case_studies}.}
\end{tablenotes}
\end{threeparttable}
\end{table}

% ----------   HI   ---------- %

\subsection{\ion{H}{i} 21cm Line} \label{sec:lines-hi}

The hyperfine structure \ion{H}{i} 21\,cm line serves as a direct probe of the atomic hydrogen content of galaxies, so its abundance and clustering properties can be straightforwardly modeled with the \ion{H}{i}--halo mass relation derived. The \ion{H}{i} mass is related to the mean brightness temperature, the relevant observable for \ion{H}{i} maps, via \cite[e.g.,][]{Bull_2015, Wolz_2017}
\begin{equation}
\bar{T}_{\ion{H}{i}} = \mathcal{C}_{\ion{H}{i}} \bar{\rho}_{\ion{H}{i}}(z) = \frac{3 h c^3 A_{21}}{32\pi k_{\rm B} m_{\rm p} \nu^2_{21}} \frac{(1+z)^2}{H(z)} \bar{\rho}_{\ion{H}{i}}(z)~,
\label{eq:T_HI}
\end{equation}
where $\mathcal{C}_{\ion{H}{i}}$ is the conversion factor from the mean \ion{H}{i} density to the mean brightness temperature and $A_{21} = 2.88\times10^{-15}\,\mathrm{s^{-1}}$ is the Einstein coefficient corresponding to the 21\,cm line. The mean \ion{H}{i} mass density is expressed as \citep{Padmanabhan_2017}
\begin{equation}
\bar{\rho}_{\ion{H}{i}}(z) = \int \mathrm d M \frac{\mathrm d N}{\mathrm d M} M_{\ion{H}{i}}(M, z)~.
\end{equation}

% ----------   CII   ---------- %

\subsection{{\rm [\ion{C}{ii}]} 158\,$\mu$m Line} \label{sec:lines-cii}

The 158\,$\mu$m [\ion{C}{ii}] line is one of the most important metal cooling lines in the interstellar medium and can alone account for $\sim0.1$\% of the total FIR emission of a galaxy \citep{Stacey1991,Malhotra1997}. Empirically, the emission in the [\ion{C}{ii}] line correlates with both FIR dust emission \citep{Crawford1985,Wright1991} and star formation \citep{Stacey1991,DL_2014}.

The strong correlation between the [\ion{C}{ii}] and IR luminosity can be understood with a model in which the cooling of interstellar gas is dominated by [\ion{C}{ii}] emission and the heating is dominated by photoelectric emission from dust grains. If the dust converts a fraction $\epsilon_{\rm PE} \ll 1$ of UV and optical radiation absorbed into photoelectric heating and the remainder into infrared emission, then the total heating rate is proportional to $\epsilon L_{\rm IR}$. We can therefore approximate

\begin{equation}
L_{[\ion{C}{ii}]} = \left(1-f_{{\rm H}_2}\right)\epsilon_{\rm PE} L_{\rm IR}
~,
\label{eq:PE_eff_def}
\end{equation}
where the factor $\left(1-f_{{\rm H}_2}\right)$ accounts for the fact that dust is present and will radiate in molecular clouds where there is little atomic C. $\epsilon_{\rm PE}$ is taken to be a free parameter in the model with a fiducial value of $5\times10^{-3}$, which yields an $L_{[\ion{C}{ii}]}$/$L_{\rm IR}$ ratio consistent with that estimated from observations of the LMC \citep[e.g.,][]{Rubin_2009} and nearby galaxies \citep[e.g.,][]{DL_2014}. We note that the observed proportionality between SFR and $L_{[\ion{C}{ii}]}$ is reproduced here since SFR is correlated with $L_{\rm IR}$ (Equation~\ref{eq:sfr_lir}).

A number of simplifications are inherent in this prescription. For instance, other cooling lines (e.g., [\ion{O}{i}]) can be important relative to [\ion{C}{ii}] \citep{Tielens1985,YoungOwl2002}. Second, the photoelectric efficiency of dust grains is a function of the grain charge. As gas density and radiation intensity increase, $\epsilon_{\rm PE}$ is expected to decrease \citep{BT_1994}, and so we might expect systematic changes in the $L_{\rm IR}$--$L_{[\ion{C}{ii}]}$ relation with galaxy properties just from this effect. Finally, unlike the dust emission, the [\ion{C}{ii}] line can saturate at high gas temperatures and radiation intensities, breaking the linear correlation \citep{Munoz2016,Rybak_2019}. These effects are most pronounced in gas of extreme density and temperature and may account for the breakdown of the $L_{\rm IR}$--$L_{[\ion{C}{ii}]}$ correlation in luminous and ultraluminous galaxies. We do not incorporate these effects into our model at this time, but we discuss potential implementation in Section~\ref{sec:disncon}.

% ----------   NII   ---------- %

\subsection{{\rm [\ion{N}{ii}]} 122 and 205$\mu$m Lines} \label{sec:lines-nii}

The emission from singly ionized nitrogen, which has an ionization potential of 14.53\,eV, traces \ion{H}{ii} regions (see Figure~\ref{fig:sketch}). When the density is lower than the critical density, collisional de-excitation can be neglected and the luminosity of the [\ion{N}{ii}] 122 and 205\,$\mu$m lines can be approximated by the balance between the rates of collisional excitation and radiative de-excitation. For an ionized gas cloud of volume $V$,
\begin{equation}
L_{[\ion{N}{ii}]} \simeq n_{\rm e, \ion{H}{ii}} n_{\rm N^+} q_{\nu} h \nu_{[\ion{N}{ii}]} V~, 
\end{equation}
where $q_{\nu}$ denotes the collisional excitation coefficient, with $q_{122} = 2.57\times10^{-8}\,\mathrm{cm^3\,s^{-1}}$ and $q_{205} = 6.79\times10^{-8}\,\mathrm{cm^3\,s^{-1}}$ \citep[e.g.,][]{HC_2016}. Meanwhile, the ionization equilibrium of \ion{H}{ii} regions gives
\begin{equation}
Q_0 = n_{\rm e, \ion{H}{ii}} n_{\rm H^+} \alpha_{\rm B} \left( T_{\rm gas, \ion{H}{ii}} \right) V~, 
\end{equation}
where $Q_0$ is the rate of hydrogen photoionization sourced by UV photons from O and B stars and $\alpha_{\rm B} = 2.6\times10^{-13} \left( T_{\rm gas, \ion{H}{ii}} / 10^4\,\mathrm{K} \right)^{-0.76}\,\mathrm{cm^3\,s^{-1}}$ is the case B recombination coefficient, a reasonable assumption for typical \ion{H}{ii} regions where the mean free path of ionizing photons is small. For Population~II stars with a Salpeter IMF, each stellar baryon produces $N_{\rm ion} \simeq 4000$ ionizing photons on average \citep{LF_2013}, in which case $Q_0$ can be related to the star formation rate by
\begin{equation}
Q_0(M, z) =  \frac{N_{\rm ion} \dot{M}_{\star}}{m_{\rm p}/(1-Y)} \simeq 1.14 \times 10^{53} \left[ \frac{\dot{M}_{\star}(M, z)}{M_{\odot}/\mathrm{yr}} \right]\,\mathrm{s^{-1}}~, 
\label{eq:Q_0}
\end{equation}
where we take the helium mass fraction to be $Y=0.25$. The ionization rate can then be related to the luminosity of [\ion{N}{ii}] lines by
\begin{equation}
L_{[\ion{N}{ii}]} \simeq \frac{q_\nu h \nu_{[\ion{N}{ii}]}}{\alpha_{\rm B}\left( T_{\rm gas, \ion{H}{ii}} \right)} \frac{n_{\rm N^+}}{n_{\rm H^+}} \frac{N_{\rm ion} \dot{M}_{\star}}{m_{\rm p}/(1-Y)}~
\label{eq:lum_lowd}
\end{equation}
which gives
\begin{equation}
L^{\rm tot}_{[\ion{N}{ii}]} = 9 \times 10^6 L_{\odot} \left( \frac{T_{\rm gas, \ion{H}{ii}}}{10^4\,\mathrm{K}}\right)^{0.76} \times  \frac{\dot{M}_\star}{M_{\odot}/\mathrm{yr}} \times \frac{Z}{Z_{\odot}}~,
\end{equation}
where $n_{\rm N^+} / n_{\rm H^+}$, under the assumption that the second ionization of nitrogen ($\rm N^+ \rightarrow N^{++}$) with a potential of 29.6\,eV is negligible, can be approximated by the $\rm N/H$ ratio $\mathrm{N/H} = \mathrm{(N/H)_{\odot}} \times \left[ Z(z) / Z_{\odot} \right] \simeq 7.4\times10^{-5} \left[ Z(z) / Z_{\odot} \right]$ \citep{Asplund_2009}.

\begin{figure}[h!]
\centering
\includegraphics[width=0.45\textwidth]{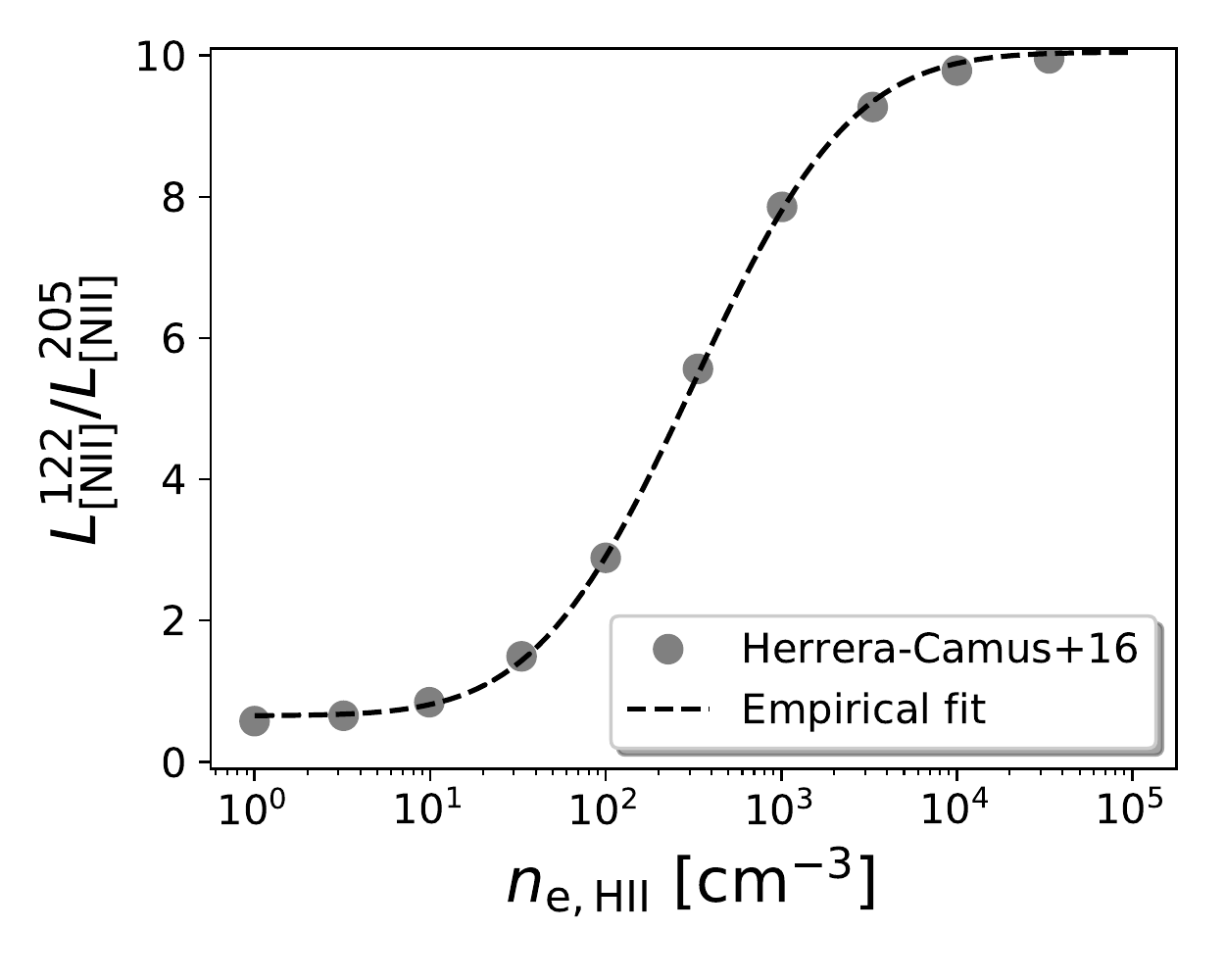}
\caption{Line ratio of [\ion{N}{ii}] 122$\mu$m and 205$\mu$m lines as a function of \ion{H}{ii} region electron number density. The dashed line shows our model parameterization given by Equation~\ref{eq:niiratio}.}
\label{fig:niiratio}
\end{figure}

In order to model the impact of electron number density $n_{\rm e,\ion{H}{ii}}$ on the strength of [\ion{N}{ii}] line emissions, we exploit a simple parameterization of the [\ion{N}{ii}] 122\,$\mu$m/205\,$\mu$m line ratio as a function of $n_{\rm e,\ion{H}{ii}}$
\begin{equation}
R^\prime_{[\ion{N}{ii}]} = R_{[\ion{N}{ii}]} + A_R \times \left\{ 1 + \mathrm{erf} \left[ \frac{\log(n_{\rm e,\ion{H}{ii}}/n^0_{\rm e,\ion{H}{ii}})}{\sigma_R} \right] \right\}~,
\label{eq:niiratio}
\end{equation}
where $R_{[\ion{N}{ii}]} = L^{122}_{[\ion{N}{ii}]}/L^{205}_{[\ion{N}{ii}]} \sim 0.65$ is the line ratio in the low-density limit discussed above. We further take the normalization factor $A_R$ to be 4.7, the characteristic density $n^0_{\rm e,\ion{H}{ii}}$ to be $10^{2.5}\,\mathrm{cm^{-3}}$ and the transition width $\sigma_R$ to be 1, in order to obtain a good fit to the results from \cite{HC_2016} over $1\,\mathrm{cm^{-3}} \la n_{\rm e, \ion{H}{ii}} \la 10^5\,\mathrm{cm^{-3}}$, as illustrated in Figure~\ref{fig:niiratio}. Meanwhile, there is also a non-trivial evolution of the total [\ion{N}{ii}] luminosity with the electron number density (or effectively $R^\prime_{[\ion{N}{ii}]}$) owing to the increasingly important collisional de-excitation at higher densities, whose effect can be approximated by
\begin{equation}
{L^\prime}^{\rm tot}_{[\ion{N}{ii}]} \simeq L^{\rm tot}_{[\ion{N}{ii}]} \left( T_{\rm gas, \ion{H}{ii}} \right) \times 10^{0.12\left[R_{[\ion{N}{ii}]} - R^\prime_{[\ion{N}{ii}]}(n_{\rm e, \ion{H}{ii}})\right]}~.
\end{equation}
The resulting [\ion{N}{ii}] line luminosities depend on both the temperature and the density of \ion{H}{ii} regions:
\begin{equation}
{L^\prime}^{\rm 205}_{[\ion{N}{ii}]} = \frac{1}{1+R^\prime_{[\ion{N}{ii}]}(n_{\rm e, \ion{H}{ii}})} \times {L^\prime}^{\rm tot}_{[\ion{N}{ii}]} \left( T_{\rm gas, \ion{H}{ii}}, n_{\rm e, \ion{H}{ii}} \right)~,
\end{equation}
and
\begin{equation}
{L^\prime}^{\rm 122}_{[\ion{N}{ii}]} = \frac{R^\prime_{[\ion{N}{ii}]}(n_{\rm e, \ion{H}{ii}})}{1+R^\prime_{[\ion{N}{ii}]}(n_{\rm e, \ion{H}{ii}})} \times {L^\prime}^{\rm tot}_{[\ion{N}{ii}]} \left( T_{\rm gas, \ion{H}{ii}}, n_{\rm e, \ion{H}{ii}} \right)~.
\end{equation}

In our model of [\ion{N}{ii}] emission, we set the gas temperature to be $T_{\rm gas, \ion{H}{ii}} \simeq 10^4$\,K, which is a characteristic temperature of $\ion{H}{ii}$ regions where ionized nitrogen is expected to be found \cite[e.g.,][]{Goldsmith_2015, HC_2016}. Meanwhile, it is important to note that, alternatively to the empirical prescription presented, the dependence of [\ion{N}{ii}] line ratio on $n_{\rm e,\ion{H}{ii}}$ may also be derived ab initio from the transition rates of collisionally coupled states of [\ion{N}{ii}] (see, e.g., \citealt{Goldsmith_2015}). 

% ----------   CO   ---------- %

\subsection{CO(1-0) Line}

The CO(1-0) rotational transition ($\lambda = 2.6$\,mm) is a powerful tracer of the molecular gas content of both individual molecular clouds and of galaxies \citep[e.g.,][]{Solomon1987,Dame2001,Ivison2011,Saintonge2011}. In molecular clouds, the CO(1-0) line is generally optically thick, and so the line luminosity $L_{\rm CO}$ is independent of the CO abundance. For a virialized molecular cloud, it can be shown that $L_{\rm CO}$ is proportional to the cloud mass \citep[e.g.,][]{Draine_2011,Bolatto+Wolfire+Leroy_2013}, with the constant of proportionality designated $\alpha_{\rm CO}$. Even in this idealized case of a homogeneous cloud, $\alpha_{\rm CO}$ depends on the precise conditions within the cloud. \citet{Draine_2011} derives the dependence of $\alpha_{\rm CO}$ on the excitation temperature $T_{\rm exc}$ and molecular gas density $n_{{\rm H}_2}$ as

\begin{equation}
    \alpha_{\rm CO} = 4.2\left(\frac{n_{{\rm H}_2}}{10^3\,{\rm cm}^{-3}}\right)^{1/2}\left(e^{5.5\,{\rm K}/T_{\rm exc}} - 1\right)\,\frac{M_\odot}{{\rm K}\,{\rm km}\,{\rm s}^{-1}\,{\rm pc}^2}~,
\end{equation}
where we have adopted a factor of 1.36 to convert from hydrogen mass to total gas mass, which accounts for the abundance of He \citep{Bolatto+Wolfire+Leroy_2013}. We note that for a density $n_{\rm H_2} = 2\times10^3\,\mathrm{cm^{-3}}$, typical of GMCs, $T_{\rm exc} = 10$\,K implies a CO-to-$\rm H_2$ conversion factor of $\alpha_{\rm CO} \approx 4.4\,M_\odot (\mathrm{K\,km\,s^{-1}\,pc^2})^{-1}$, consistent with the value inferred from observations \citep{Bolatto+Wolfire+Leroy_2013}. 

A population of virialized molecular clouds will likewise have a linear relationship between the total molecular gas mass and the integrated CO(1-0) line luminosity provided that the covering factor is low enough just that the CO emission from one cloud is unlikely to be absorbed by another cloud \citep{Dickman1986,Bolatto+Wolfire+Leroy_2013}.

Under these assumptions, we can write the CO luminosity directly in terms of the molecular gas mass $M_{{\rm H}_2} \equiv f_{{\rm H}_2}M_{\rm H}$:

\begin{equation}
    L_{\rm CO}\left(M,z\right) = \alpha_{\rm CO}^{-1} f_{{\rm H}_2}M_{\rm H}\left(M,z\right)
    ~~~.
\end{equation}
We treat $\alpha_{\rm CO}$ as a parameter to be fit. While there are indications that $\alpha_{\rm CO}$ may vary systematically with other galaxy properties, e.g., metallicity \citep{Genzel2012,Bolatto+Wolfire+Leroy_2013,Sandstrom2013}, we do not consider such variations here. 

\section{Intensity Mapping Framework} \label{sec:lim}

\subsection{Modeling the Fluctuation Signals}

In this section, we introduce a simple, generic halo occupation distribution (HOD) model, which is used to compute the power spectra that describe the spatial fluctuations of various signals emitted from discrete galaxies. Incorporating the correlation of subhalo structure (e.g., satellite galaxies) via such an HOD model is non-trivial, since both observational and theoretical studies have shown that massive dark matter halos tend to host more than one galaxy at low redshifts \cite[e.g.,][]{Gao_2011, McCracken_2015}, with a peak in subhalo abundance for a given halo mass at $z\sim2-3$ as found by \citet{Wetzel_2009}. The original HOD model describes the occupation of halos by central and satellite galaxies \citep{Kravtsov_2004}. Here, we generalize it to describe the fluctuations in line signals associated with the clustering of both central and satellite galaxies by weighting the galaxy number counts by a measure of the signal strength $S_{\nu}$ at observed frequency $\nu$ for a given halo mass and redshift, which means slightly differently for different signals (see later text). In particular, we define the number-count-weighted signal strengths of central and satellite galaxies, 
\begin{equation}
f^{\mathrm{cen}}_{\nu} (M, z) = N_{\mathrm{cen}} S_{\nu} (M, z)~, 
\end{equation}
and
\begin{equation}
f^{\mathrm{sat}}_{\nu} (M, z) = \int_{M_{\mathrm{min}}}^{M} \mathrm d m \frac{\mathrm d n}{\mathrm d m} (m, z | M) S_{\nu} (m, z)~, 
\end{equation}
where $N_\mathrm{cen}$ is the number of central galaxies in a halo, which is equal to 1 for $M>M_\mathrm{min} = 10^{10}\,M_\odot$ and 0 otherwise \citep{WD_2017}, and $\mathrm d n / \mathrm d m$ represents the subhalo mass function, for which we adopt the fitting function in \citet{TW_2010}. We consequently define the mean radiation strength as
\begin{equation}
\bar{j}_{\nu}(z) = \int_{M_{\mathrm{min}}}^{M_{\mathrm{max}}} \mathrm d M \frac{\mathrm d N}{\mathrm d M} \left[ f^{\mathrm{cen}}_{\nu} (M, z) + f^{\mathrm{sat}}_{\nu} (M, z) \right]~.
\end{equation}
We note that in our expression, for different signals, $\bar{j}_{\nu}$ represents slightly different physical quantities and thus carries different units by convention. Specifically, $\bar{j}_{\nu}$ denotes the mean \textit{volume emissivity}, \textit{intensity}, and \textit{brightness temperature} for the CIB,\footnote{This shows how the HOD formalism is originally defined in the CIB anisotropy model, provided here for completeness and better illustrating our generalization.} [\ion{C}{ii}]/[\ion{N}{ii}]/CO lines and \ion{H}{i} 21cm line, respectively. For the signals under consideration, we have
\begin{equation}
S_{\nu} (M, z) = \frac{L_{(1+z)\nu} \left( M, z \right)}{4\pi}           \hspace{0.8em} \left( \rm CIB \right)~, 
\end{equation}
\begin{equation}
S_{\nu} (M, z) = \frac{L_{\rm line} (M, z)}{4\pi D^2_{L}} y(z) D^2_A           \hspace{0.8em} \left(\rm [\ion{C}{ii}],\ [\ion{N}{ii}]\ and\ CO \right)~, 
\end{equation}
\begin{equation}
S_{\nu} (M, z) = \mathcal{C}_{\ion{H}{i}} M_{\mathrm{HI}}(M, z)         \hspace{0.8em} \left( \rm \ion{H}{i} \right)~, 
\end{equation}
where the units of signal strengths are $\rm erg\ s^{-1}\ Hz^{-1} sr^{-1}$, $\rm cm\ erg\ s^{-1}\ Hz^{-1} sr^{-1}$ and $\rm mK\ cm^3$, respectively. The mapping from frequency to line-of-sight distance is given by $y(z) = \mathrm d \chi / \mathrm d \nu = c(1+z) / \left[ \nu H(z) \right]$, where $\chi$ denotes the comoving radial distance.

Generally, the power spectrum of a pair of signals at frequencies $\nu$ and $\nu'$ (auto-correlation if $\nu=\nu'$ and cross-correlation otherwise) can be expressed as the sum of one-halo, two-halo, and shot-noise components, namely
\begin{equation}
P_{\nu \nu'}(k, z) = P_{\mathrm{1h, \nu\nu'}} (k, z) + P_{\mathrm{2h, \nu\nu'}} (k, z) + P_{\mathrm{SN, \nu\nu'}} (k, z)~.
\end{equation}
The one-halo term characterizes the contribution to the fluctuations from emitters residing in the same halo. Assuming that the occupation statistics of central and satellite galaxies are independent and that the latter is Poissonian, we have
\begin{align}
P_{\mathrm{1h, \nu\nu'}} (k, z) = & \int_{M_{\mathrm{min}}}^{M_{\mathrm{max}}} \mathrm d M \frac{\mathrm d N}{\mathrm d M}\ \times \\ \nonumber
& \Big[ f^{\mathrm{cen}}_{\nu} (M, z) f^{\mathrm{sat}}_{\nu'} (M, z) u(k | M, z)\ + \\ \nonumber
& \ f^{\mathrm{cen}}_{\nu'} (M, z) f^{\mathrm{sat}}_{\nu} (M, z) u(k | M, z)\ + \\ \nonumber
& \ f^{\mathrm{sat}}_{\nu} (M, z) f^{\mathrm{sat}}_{\nu'} (M, z) u^2(k | M, z) \Big]~,
\end{align}
where $u(k|M,z)$ is the normalized Fourier transform of the halo density profile \citep{NFW_1997,Bhattacharya_2013}. The two-halo component describes the contribution from emitters residing in different halos, 
\begin{equation}
P_{\mathrm{2h, \nu\nu'}} (k, z) = D_{\nu}(k, z) D_{\nu'}(k, z) P_{\delta \delta} (k, z)~, 
\end{equation}
where $P_{\delta \delta} (k, z)$ is the dark matter power spectrum and
\begin{align}
D_{\nu}(k, z) = & \int_{M_{\mathrm{min}}}^{M_{\mathrm{max}}} \mathrm d M \frac{\mathrm d N}{\mathrm d M} b(M, z) u(k | M, z) \\ \nonumber
& \times \left[ f^{\mathrm{cen}}_{\nu} (M, z) + f^{\mathrm{sat}}_{\nu} (M, z) \right]~, 
\end{align}
with $b(M,z)$ being the halo bias factor \citep{Tinker_2008}. Finally, the shot-noise component describes the self-correlation of emitters due to their discrete nature, 
\begin{equation}
P_{\mathrm{SN, \nu\nu'}} (z) = \int_{M_{\mathrm{min}}}^{M_{\mathrm{max}}} \mathrm d M \frac{\mathrm d N}{\mathrm d M} f^{\mathrm{cen}}_{\nu} (M, z) f^{\mathrm{cen}}_{\nu'} (M, z)~, 
\end{equation}
which can be considered as the $k\rightarrow0$ limit of the one-halo term in the absence of satellite galaxies \cite[see, e.g.,][]{Wolz_2017}. 

\begin{figure}[h!]
\centering
\includegraphics[width=0.45\textwidth]{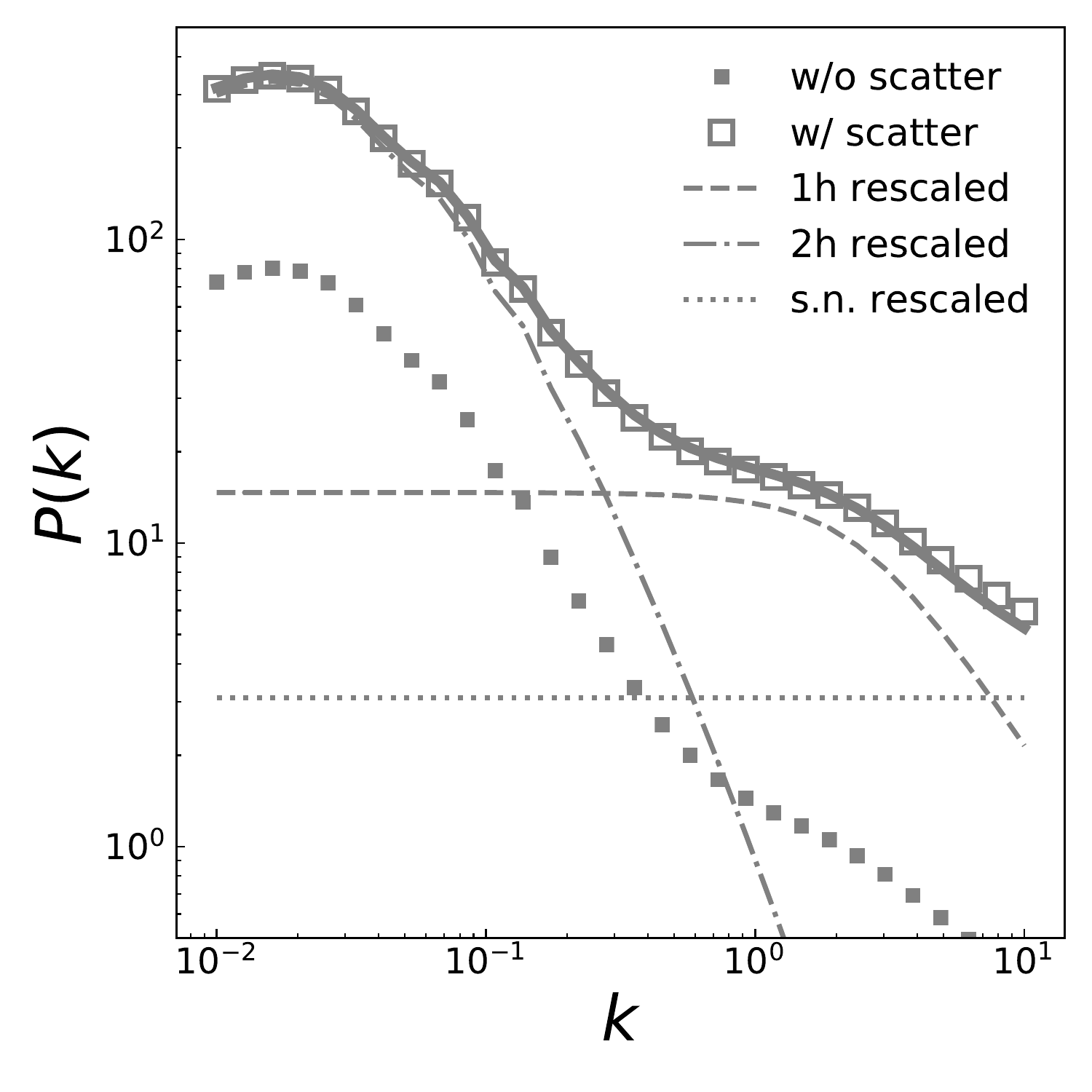}
\caption{Effect of a $\sigma_\nu = 0.5$\,dex scatter on the power spectrum. The two-halo (dash-dotted curve) term is rescaled by the correction factor defined by Eq.~\ref{eq:corr_1st}, whereas the one-halo (dashed curve) and shot-noise (dotted curve) terms are rescaled by the correction factor defined by Eq.~\ref{eq:corr_2nd}. The total power spectrum (solid curve) rescaled from the one without scatter (filled squares) matches well with that derived from averaging over 1000 random realizations (open squares).}
\label{fig:scatter_demo}
\end{figure}

Finally, following \cite{Sun_2018}, in order to take into account of the stochasticity of individual galaxies, we introduce a simple parameterization of a lognormal distribution of line brightness below for a given halo mass and redshift. The probability density can be expressed as
\begin{equation}
P\left( x | \mu_\nu, \sigma_\nu \right) = \frac{1}{\sqrt{2\pi}\sigma_\nu} \exp \left[ - \frac{(x-\mu_\nu)^2}{2\sigma_\nu^2} \right]~,
\end{equation}
where $\mu_\nu = \log[S_\nu(M,z)]$ is the aforementioned mean line strength and $\sigma_\nu = 0.3$\,dex is our fiducial level of scatter reflecting the typical galaxy-to-galaxy variation in line production. It is straightforward to show that the power spectrum averaged over the lognormal distribution is essentially a rescaling of the power spectrum without scatter, specified by the additive correction factors in the following relations:
\begin{equation}
\langle \mu_\nu \rangle = \mu_\nu + \frac{\sigma_\nu^2}{2} \ln 10~,
\label{eq:corr_1st}
\end{equation}
which applies to the two-halo term of power spectrum scaling as the square of the first luminosity moments, and
\begin{equation}
\langle 2\mu_\nu \rangle = 2\mu_\nu + 2 \sigma_\nu^2 \ln 10~,
\label{eq:corr_2nd}
\end{equation}
which applies to the one-halo and shot-noise terms of power spectrum scaling as the second luminosity moment. Figure~\ref{fig:scatter_demo} shows how the power spectrum is affected by the above correction factors in the presence of a non-trivial scatter $\sigma_\nu$. For comparison, the open squares indicate the average of power spectra directly drawn from 1000 random realizations of the lognormally distributed $S_\nu(M,z)$ relation, which agrees well with the one analytically derived using $\langle \mu_\nu \rangle$ and $\langle 2\mu_\nu \rangle$ as shown by the solid curve (a sum of rescaled one-halo, two-halo, and shot-noise components). In our power spectrum analysis, we include these correction factors to obtain constraints on the lognormal scatter together with physical properties of the ISM. We further assume, for simplicity, that similar physical processes (e.g., regulations of galaxy evolution by star formation, outflows and interactions, variations of stellar population and ISM properties) give rise to the stochasticity for a given halo mass and redshift, and therefore line luminosities considered in this work all share the same lognormal scatter $\sigma$.

\subsection{Sensitivity Analyses}

\begin{table*}[!t]
\centering
\caption{Reference Instrumental Parameters for Case Studies}
\begin{tabular}{ccccccccccc}
\hline
\hline
\bf{Parameter} & $t_{\rm obs}$ & $D_{\rm ap}$ & $\Omega_{\rm survey}$ & $N_{\rm feeds}$ & \{$\nu_{\rm min}$, $\nu_{\rm max}$\} & $\Delta z$ & $\delta_\nu$ & $V_{\rm vox}$ & NEFD ($T_{\rm sys}$) & $\rm SNR_{tot}$ \\
\bf{Units} & [hr] & [m] & [deg$^2$] &  & [GHz] &  & [GHz] & [Mpc$^3$] & [$\rm mJy\,s^{1/2}$] ([K]) &  \\
\hline
 & \multicolumn{10}{c}{\textit{Case I: $P_{\ion{H}{i}}$ and $P_{\mathrm{CO}}$ at $z=2$}} \\
\cmidrule(l{2em}r{2em}){2-11}
\ion{H}{i} & 2000 & 1000 & 1 & 100 & \{0.44, 0.52\} & $\pm 0.25$ & 0.003 & 22.8 & (50) & 7.6 \\
CO & 500 & 10 & 1 & 100 & \{35, 42\} & $\pm 0.25$ & 0.05 & 7.11 & (40) & 7.7 \\
 & \multicolumn{10}{c}{\textit{Case II: $P_{\ion{C}{ii}}$, $P_{\ion{H}{i}}$, $P_{\mathrm{CO}}$, $P_{\ion{C}{ii},\ion{H}{i}}$ $P_{\ion{H}{i},\mathrm{CO}}$, and $P_{\ion{C}{ii},\mathrm{CO}}$ at $z=2$}} \\
\cmidrule(l{2em}r{2em}){2-11}
\ion{H}{i} & 2000 & 1000 & 1 & 100 & \{0.44, 0.52\} & $\pm 0.25$ & 0.003 & 22.8 & (50) & 7.6, \ion{H}{i}$\times$[\ion{C}{ii}]: 5.0 \\
{[\ion{C}{ii}]} & 1000 & 12 & 1 & 100 & \{585, 691\} & $\pm 0.25$ & 2 & 0.04 & 50 & 8.0, \ion{H}{i}$\times$[\ion{C}{ii}]: 6.6 \\
CO & 500 & 10 & 1 & 100 & \{35, 42\} & $\pm 0.25$ & 0.05 & 7.11 & (40) & 7.7, \ion{H}{i}$\times$CO: 5.1 \\
 & \multicolumn{10}{c}{\textit{Case III: $P^{\ion{N}{ii}}_{122}$, $P^{\ion{N}{ii}}_{205}$ and $P^{\ion{N}{ii}}_{122 \times 205}$ at $z=2$}} \\
\cmidrule(l{2em}r{2em}){2-11}
{[\ion{N}{ii}]}\,122 & 2000 & 10 & 1 & 400 & \{757, 894\} & $\pm 0.25$ & 1 & 0.02 & 10 & 5.2 \\
{[\ion{N}{ii}]}\,205 & 2000 & 10 & 1 & 400 & \{450, 531\} & $\pm 0.25$  & 1 & 0.07 & 10 & 4.7, cross: 6.7 \\
 & \multicolumn{10}{c}{\textit{Case IV: $P_{\ion{C}{ii}}$ and $P^{\ion{N}{ii}}_{205}$ at $z=2$}} \\
 \cmidrule(l{2em}r{2em}){2-11}
{[\ion{C}{ii}]} & 1000 & 12 & 1 & 100 & \{585, 691\} & $\pm 0.25$ & 2 & 0.04 & 50 & 8.0 \\
{[\ion{N}{ii}]}\,205 & 2000 & 10 & 1 & 400 & \{450, 531\} & $\pm 0.25$ & 1 & 0.07 & 10 & 4.7 \\
\hline
\hline
\end{tabular}
\label{tb:inst_params}
\end{table*}

In this section, we describe the formalism to forecast the sensitivity to the power spectrum signal, assuming a given experimental setup. For a three-dimensional survey of volume $V_\mathrm{s} = L_x L_y L_z$, the observed 3D power spectrum $\mathcal{P}(K)$ for a given mode $K$ in the Fourier space of the observing frame is related to the true, spherically-averaged power spectrum $\Delta^2(k) = k^3 P(k) / 2\pi^2$ by
\begin{equation}
\mathcal{P}(K) = V_\mathrm{s} \int_{-\infty}^{\infty} \mathrm d \ln k \Delta^2(k) W(k, K)~, 
\end{equation}
where $W(k, K)$ is a convolution kernel commonly referred to as the ``window function,'' which is determined by the survey geometry. Here we only consider the simple situation that the survey volume is large enough such that $W(k, K)$ can be well-approximated by a function sharply peaking at $k \sim K$, which yields $\mathcal{P}(K) \approx P(k)$. Following S16, we write the uncertainty of the power spectrum $P(k)$ as the sum of a sample variance (i.e., cosmic variance) term and a thermal noise term. In particular, for the auto power spectrum $P_{\nu\nu}(k)$, we have
\begin{equation}
\delta P_{\nu\nu}(k) = \frac{ P_{\nu\nu}(k) + P_{\nu\nu}^\mathrm{noise}(k) }{G(k)\sqrt{N_{\rm modes}(k)}}~,
\end{equation}
where $G(k)$ denotes a smoothing factor due to finite spatial and spectral resolutions, which attenuates the power spectrum at large $k$ values beyond resolvable scales and is defined as \cite[][]{Li_2016}
\begin{equation}
G(k) = e^{-k^2 \sigma^2_{\perp}} \int_0^1 e^{-k^2 (\sigma^2_{\parallel}-\sigma^2_{\perp})\mu^2} \mathrm d \mu~,
\end{equation}
where $\mu = \cos \theta$ is the cosine of the angle a given $k$ vector makes with respect to the line of sight. For any given frequency channel width $\delta_\nu$, the spatial and spectral resolutions in physical units are given by $\sigma_{\parallel}(z) = k^{-1}_{\parallel, \mathrm{max}}(z) = y(z)\delta_\nu$ and $\sigma_{\perp}(z) = k^{-1}_{\perp, \mathrm{max}}(z) = \chi(z) \sqrt{\Omega_{\rm beam}}$, respectively. For the cross power spectrum $P_{\nu\nu'}(k)$, we have
\begin{equation}
\delta P_{\nu\nu'}(k) = \frac{\left[ P^2_{\nu\nu'}(k) + \delta P_{\nu}(k) \delta P_{\nu'}(k) \right]^{1/2}}{G(k)\sqrt{2N_{\rm modes}(k)}}~, 
\end{equation}
where
\begin{equation}
\delta P_{\nu}(k) = P_{\nu\nu}(k) + P^\mathrm{noise}_{\nu\nu}(k)~. 
\end{equation}
The (averaged) power spectrum of thermal noise is scale-independent and can be expressed as
\begin{equation}
P^\mathrm{noise}_{\nu\nu} = \sigma_{\rm noise}^2 V_{\rm vox}~.
\end{equation}
Using the radiometer equation, we can compute the on-sky sensitivity from the noise equivalent flux density (NEFD) or system temperature $T_{\rm sys}$, the beam size
\begin{equation}
\Omega_{\rm beam} = \left( \frac{\theta_{\rm FWHM}}{2.355} \right)^2 = \left( \frac{1.15 \lambda_{\rm obs} / D_{\rm ap}}{2.355} \right)^2~, 
\end{equation}
and the observing time per voxel
\begin{equation}
t_{\rm obs} = \left( N_{\rm feeds} \Omega_{\rm beam}/\Omega_{\rm survey} \right) t_{\rm survey}
\end{equation}
as
\begin{equation}
\sigma_{\rm noise} = \frac{\mathrm{NEFD}}{\Omega_{\rm beam} \sqrt{t_{\rm obs}}} = \frac{T_{\rm sys}}{\sqrt{\delta_\nu t_{\rm obs}}}~, 
\end{equation}
where $D_{\rm ap}$ and $N_{\rm feeds}$ represent the instrument's effective aperture size and number of feeds (i.e., the number of spatial channels or spectrometers simultaneously on sky), respectively; the radio astronomy convention is adopted in the second equality. The voxel size can be derived from the spectral and angular resolutions as
\begin{equation}
V_{\rm vox} = \sigma^2_{\perp} \sigma_{\parallel} = \chi(z)^2 \Omega_{\rm beam} y(z) \delta_\nu~. 
\end{equation}
As long as the survey has proper spectral and angular resolutions to sample the $k$ space in a roughly isotropic manner, the number of (independent) modes $N_{\rm modes}$ can be calculated as (e.g., \citealt{FL_2007}; \citealt{Li_2016}; S16)
\begin{equation}
N_{\rm modes}(k) = \frac{1}{2} \times 4\pi k^2 \Delta k \frac{V_\mathrm{s}}{(2\pi)^3} = \ln(10) k^3 \Delta \log k \frac{V_\mathrm{s}}{4\pi^2}~, 
\end{equation}
where the factor of $1/2$ comes from the fact that the power spectrum is the Fourier transform of a real-valued function and thus only half of the Fourier plane contains independent information. The total signal-to-noise ratio (S/N) of a measured power spectrum is then defined to be \citep{Gong_2012, Li_2016}
\begin{equation}
\mathrm{SNR_{tot}} = \sqrt{\sum_{k\  \mathrm{bins}}\left[ \frac{P(k)}{\delta P(k)}\right]^2}~.
\label{eq:tot_snr}
\end{equation}

The values of relevant instrumental parameters, adopted to guarantee significant detections of the LIM signals with comparable total S/N in our analysis, are summarized in Table~\ref{tb:inst_params} for each of the four case studies to be discussed in Section~\ref{sec:case_studies}. We note that certain requirements presented exceed the scope of planned surveys and better resemble future mission concepts, for instance, a 10\,m class, FIR telescope in space like the \textit{Origins Space Telescope (OST)} for measuring [\ion{C}{ii}] and [\ion{N}{ii}] at intermediate redshifts, as well as the large number of feeds that will be enabled by the successful deployment of broadband, on-chip spectrometers like SuperSpec \citep{HD_2014SPIE} and DESHIMA \citep{Endo_2019NatAs}. Meanwhile, the detector noise levels assumed for some signals (e.g., [\ion{N}{ii}]) are substantially more optimistic than what may be achieved from the ground, and therefore require observations in space, in which case an NEFD of order of $\rm 10\,mJy\,s^{1/2}$, corresponding to a noise equivalent power (NEP) of a few times $\rm 10^{-19}\,W\,Hz^{-1/2}$, is achievable \citep{Bradford_2008SPIE, Bradford_2018SPIE}. Even though we make no effort to carefully build these case studies on existing or planned experiments, auto-/cross-correlation opportunities based on real experiments in similar contexts will be described. 

\begin{figure}
\centering 
\includegraphics[width=0.45\textwidth]{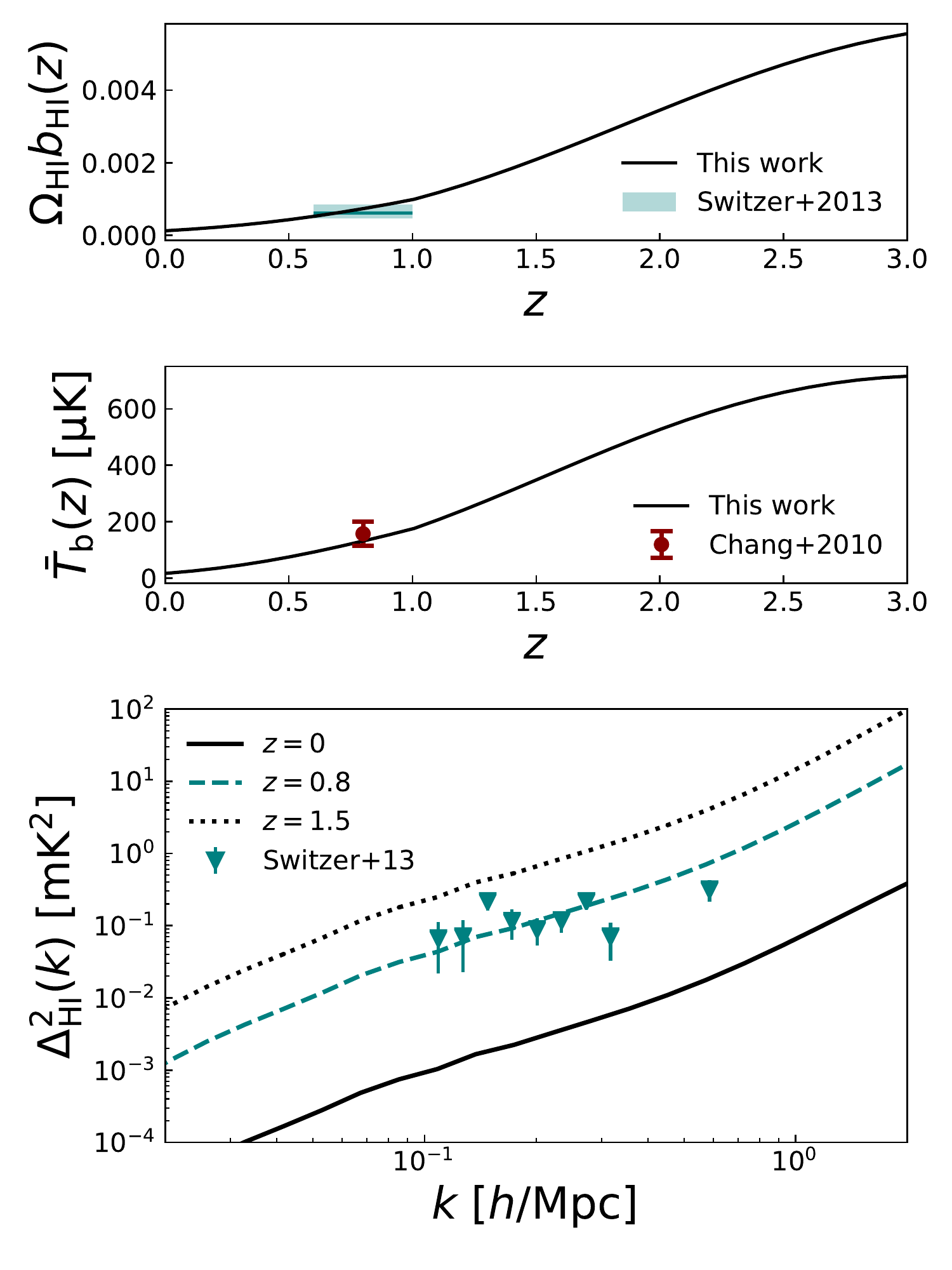}
\caption{\textit{Top}: observational constraints from the literature \citep{Switzer_2013} on the product $\Omega_{\ion{H}{i}} b_{\ion{H}{i}}$ of \ion{H}{i} density parameter and bias factor at $z\sim0.8$, compared with our model prediction. \textit{Middle}: redshift evolution of \ion{H}{i} brightness temperature, compared with the constraint from \cite{Chang_2010} at $z\sim0.8$. \textit{Bottom}: \ion{H}{i} power spectrum at different redshifts predicted by our HOD model. For comparison, deep-field results from \citet{Switzer_2013} are shown by the teal triangles, which shall be interpreted as upper limits when residual foreground is present. All the data from observations are shown with their 68\% confidence level. }
\label{fig:HI_21cm}
\end{figure}

\begin{figure}
\centering 
\includegraphics[width=0.45\textwidth]{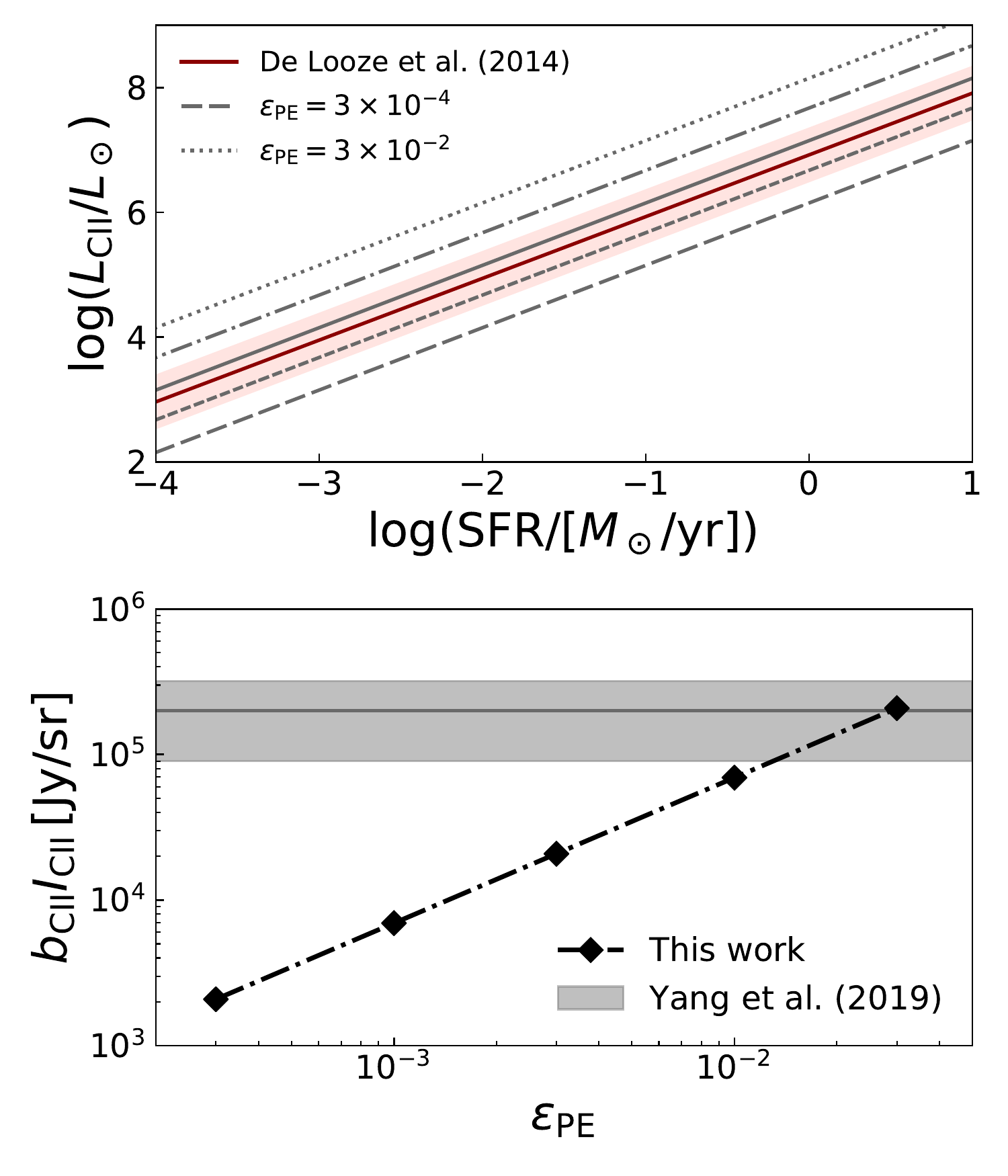}
\caption{\textit{Top}: $L_{\ion{C}{ii}}$--SFR relation from our model evaluated at different values of the photoelectric heating efficiency (from bottom to top, $\epsilon_{\rm PE} = 3\times10^{-4}, 1\times10^{-3}, 3\times10^{-3}, 1\times10^{-2}$ and $3\times10^{-2}$), compared with the best-fit relation with a $0.4\,$dex scatter to the entire galaxy sample from \citet{DL_2014}. \textit{Bottom}: products of the mean [\ion{C}{ii}] intensity and the bias factor $b_{[\ion{C}{ii}]} I_{[\ion{C}{ii}]}$ predicted by our model at $z\sim2.6$ for the five different values of $\epsilon_{\rm PE}$. The latest observational constraint on $b_{[\ion{C}{ii}]} I_{[\ion{C}{ii}]}$ (95\% confidence level) inferred from the cross-correlation between Planck maps and galaxy surveys \citep{YPS_2019}. }
\label{fig:CII_COMP}
\end{figure}

\begin{figure}[h!]
\centering
\vspace{0.5cm}
\includegraphics[width=0.45\textwidth]{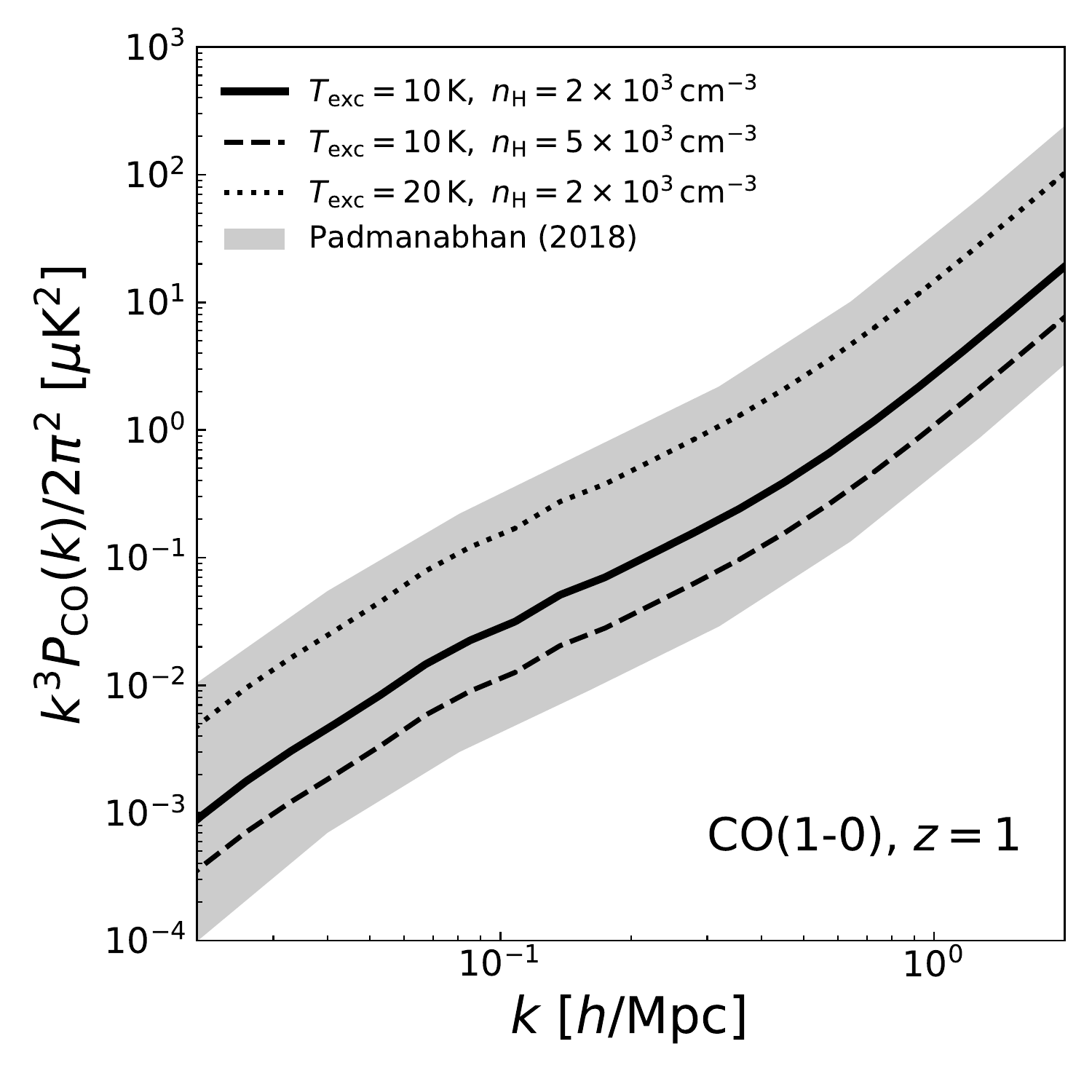}
\caption{Power spectra of CO(1-0) emission at $z=1$ for different values of the molecular gas density $n_{\rm H_2}$ and the excitation temperature $T_{\rm exc}$ as predicted by our HOD model. Constraints (68\% confidence level) from a compilation of observations by \cite{Padmanabhan_2018} are also shown by the shaded region for comparison.}
\label{fig:cops}
\end{figure}

\begin{figure}[h!]
\centering
\includegraphics[width=0.45\textwidth]{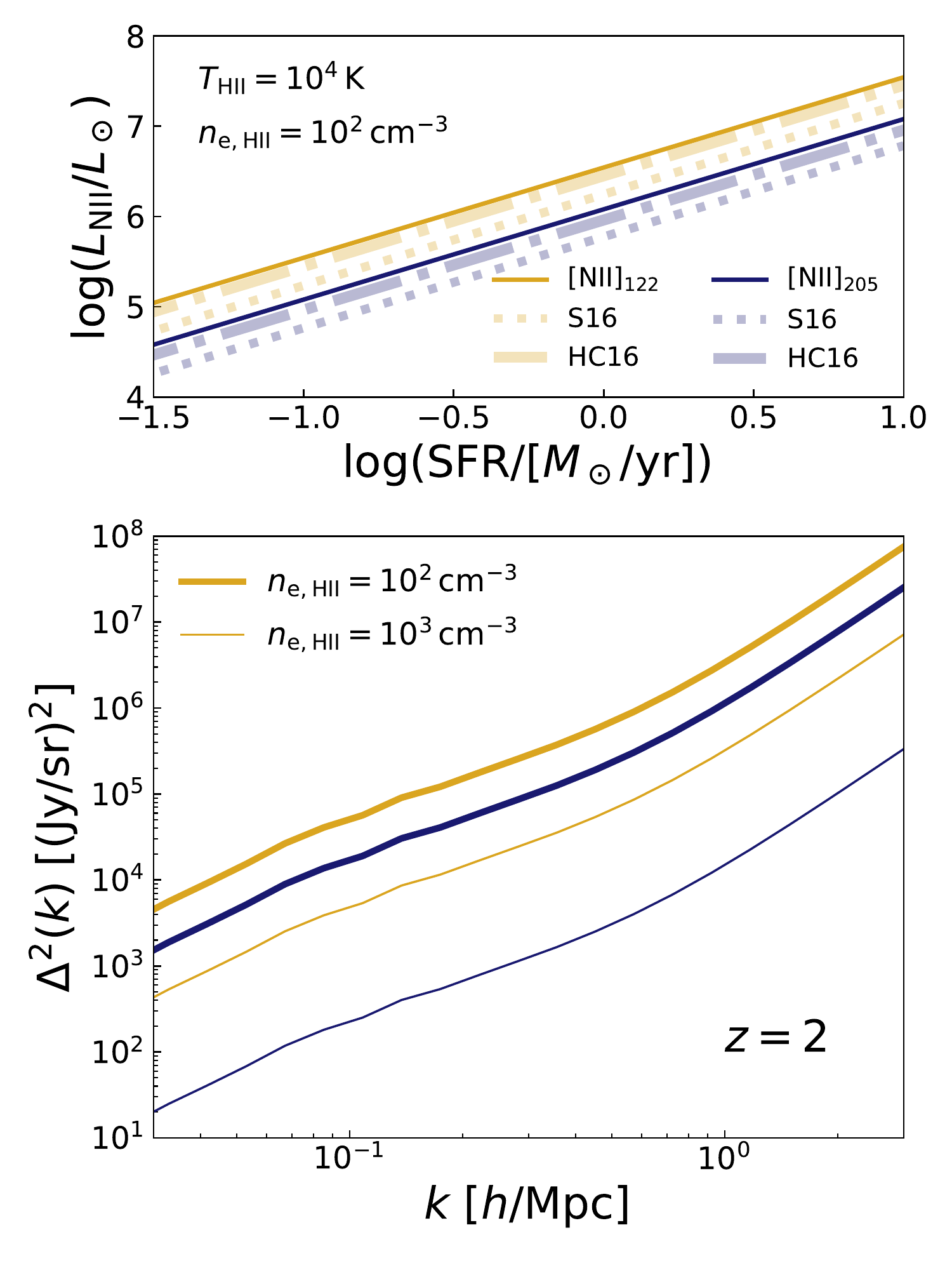}
\caption{\textit{Top}: correlations between [\ion{N}{II}] 122$\mu$m and 205$\mu$m line luminosities and the star formation rate, compared with those taken from S16 and \citet{HC_2016}. Fiducial values of \ion{H}{ii} region temperature and electron density from the reference ISM model are assumed. \textit{Bottom}: [\ion{N}{ii}] power spectra at $z=2$ predicted by our HOD model. Two sets of curves with different thicknesses are shown to illustrate the density effect on the ratio of [\ion{N}{II}] lines. }
\label{fig:niips}
\end{figure}

\section{Comparison to Existing Observational Constraints} \label{sec:results}

From LIM observations of the large-scale distribution of neutral hydrogen, constraints have been placed on the \ion{H}{i} density parameter, defined as the ratio of the \ion{H}{i} density to the critical density of the universe at $z=0$, namely, $\Omega_{\ion{H}{i}} = \rho_{\ion{H}{i}} / \rho_{\rm c, 0}$ or equivalently the mean \ion{H}{i} brightness temperature $\bar{T}_{\ion{H}{i}}$ as defined in Equation~\ref{eq:T_HI}. The top two panels of Figure~\ref{fig:HI_21cm} show the product of \ion{H}{i} density parameter and bias factor, degenerate when constrained by the large-scale clustering of \ion{H}{i}, and the mean 21cm brightness temperature predicted by our model, respectively, which are found to be in good agreement with observed values at $z\sim0.8$ \citep{Chang_2010, Switzer_2013}. The corresponding \ion{H}{i} power spectra $\Delta^2_{\ion{H}{i}}$ derived from our HOD model at $z=0, 0.8$ and 1 are shown in the bottom panel of Figure~\ref{fig:HI_21cm}, together with the deep-field results (detections only) from \citet{Switzer_2013}. While the detections shall be interpreted as upper limits since residual, correlated foregrounds are very likely present, predictions by our reference ISM model are still broadly consistent with \ion{H}{i} observations available to date. 

The top panel of Figure~\ref{fig:CII_COMP} shows the $L_{[\ion{C}{ii}]}$--SFR relations derived from our model assuming different photoelectric heating efficiency (from bottom to top, $\epsilon_{\rm PE} = 3\times10^{-4}, 1\times10^{-3}, 3\times10^{-3}, 1\times10^{-2}$ and $3\times10^{-2}$) and how they compare with the best-fit relation to a large sample of galaxies of various populations (starburst galaxies, dwarfs, ULIRGs, AGNs, high-$z$ galaxies, etc.) taken from \citet{DL_2014}. Recently, \citet{Pullen_2018} and \citet{YPS_2019} report a tentative detection of excess emission in the 545\,GHz Planck map that can be attributed to redshifted [\ion{C}{ii}] line emission. From angular cross-power spectra of high-frequency Planck maps with BOSS quasars and CMASS galaxies, a joint constraint on the product of mean [\ion{C}{ii}] intensity and bias factor $b_{[\ion{C}{ii}]} I_{[\ion{C}{ii}]} = 2.0^{+1.2}_{-1.1}\times10^5\,\mathrm{Jy\,sr^{-1}}$ is inferred at 95\% confidence level. In the bottom panel of Figure~\ref{fig:CII_COMP}, we compare our model predictions at the five different $\epsilon_{\rm PE}$ values against the measurement from \citet{YPS_2019}. We note that a relatively high $\epsilon_{\rm PE}$ is required to match the measured level of $b_{[\ion{C}{ii}]} I_{[\ion{C}{ii}]}$, which may lead to tension with the observed $L_{[\ion{C}{ii}]}$--SFR relation. Such a discrepancy is also observed by \citet{Pullen_2018} and \citet{YPS_2019} when comparing against phenomenological models \cite[e.g.,][]{Gong_2012, Silva_2015} based on local observations. While it is possible that the $L_{\rm IR}$--$L_{[\ion{C}{ii}]}$ relation is different at these redshifts, in which case a deviation from the proportionality $L_{[\ion{C}{ii}]} \propto \mathrm{SFR}$ may be implied (see, e.g., the data-driven model of [\ion{C}{ii}] emission presented by \citealt{Padmanabhan_2019}), the observed excess may also be produced by non-[\ion{C}{ii}] factors such as interloper lines or redshift evolution of CIB parameters. Future, high-resolution [\ion{C}{ii}] LIM surveys will help clarify this discrepancy. 

Measuring CO power spectrum from dedicated LIM experiments, such as COPSS~II \citep{Keating_2016}, COMAP~\citep{Li_2016} and Y.~T.~Lee Array \citep{Ho_2009}, or galaxy surveys \citep[e.g.,][]{Uzgil_2019} is an emerging field. In Figure~\ref{fig:cops}, we show our model predictions of the CO(1-0) power spectrum at $z=1$, evaluated for three pairs of excitation temperature $T_{\rm exc}$ and molecular gas density $n_{\rm H_2}$ to illustrate how sensitive CO power spectrum is to these gas properties. For comparison, the best estimate from an empirical model fit to a compilation of existing observations, including constraints on CO luminosity function and power spectrum obtained at redshifts $0<z<3$, taken from \citet{Padmanabhan_2018} is shown by the shaded band. The prediction of our reference ISM model is in good agreement with the observational constraints. 

As there has not been any LIM measurement of [\ion{N}{ii}] lines because of their faintness, in Figure~\ref{fig:niips} we only compare our reference $L_{[\ion{N}{ii}]}$-$L_{\rm IR}$ model against results from the literature and then present the [\ion{N}{ii}] power spectra it predicts. The top panel of Figure~\ref{fig:niips} shows the relations between [\ion{N}{ii}] line luminosities and the star formation rate predicted by our reference ISM model assuming $T_{\ion{H}{ii}} = 10^4\,$K and $n_{\rm e, \ion{H}{ii}} = 10^2\,\mathrm{cm^{-3}}$. Estimates from previous work are shown for comparison, including scaling relations (\citealt{Spinoglio_2012}; S16)\footnote{The scaling relation for [\ion{N}{ii}] 205$\mu$m is not provided by \citet{Spinoglio_2012}, for which we assume a line ratio of $L^{122}_{[\ion{N}{ii}]}/L^{205}_{[\ion{N}{ii}]} = 3$ following S16 (corresponding to $n_{\rm e, \ion{H}{ii}}\sim100\,\mathrm{cm^{-3}}$, as can be seen from Figure~\ref{fig:niiratio}).} based on a sample of local galaxies observed with the \textit{ISO}-LWS spectrometer \citep{Clegg_1996} and compiled by \citet{Brauher_2008}, and relations derived by \citet{HC_2016} based on an observationally-motivated prescription assuming a uniform $n_{\rm e, \ion{H}{ii}} = 10^2\,\mathrm{cm^{-3}}$. Given the relatively large dispersion that exists in the existing data \cite[see, e.g.,][]{Spinoglio_2012}, our simple model is deemed satisfactory despite the fact that it may slightly overestimate the local [\ion{N}{ii}] luminosities. The bottom panel of Figure~\ref{fig:niips} shows the power spectra of [\ion{N}{ii}] 122\,$\mu$m and 205\,$\mu$m lines, evaluated at $z=2$ for two different values of the \ion{H}{ii} region electron number density to illustrate the density effect on the [\ion{N}{ii}] line ratio.

% ============================     S6: CASE STUDIES     ============================ %

\section{Inferring ISM Properties from Auto/Cross-Correlations} \label{sec:case_studies}

Both auto-correlation and cross-correlation analyses serve as a powerful tool to study the ISM physics when LIM data sets of multiple lines are available. The latter, however, has the advantage of avoiding contamination from uncorrelated foregrounds (line and continuum), which are usually a few orders of magnitude brighter than and/or spectrally blended with the signal of interest, therefore presenting a great challenge to reliably measuring the line intensity fluctuations (\citealt{Lidz_2009}; \citealt{pullen2013}; \citealt{Silva_2015}; S16; \citealt{BVL_2019}; see also \citealt{Switzer_2019}). In the rest of this section, we present several case studies in order to demonstrate how the population-averaged physical properties of different ISM phases, such as their gas temperature and density, might be reliably extracted by auto/cross-correlating the intensity fields of different tracers. 

We adopt a Bayesian analysis framework and fit parameters of ISM properties with the affine-invariant Markov Chain Monte Carlo (MCMC) code \texttt{emcee} \citep{FM_2013}. The likelihood function for fitting the mock power spectra can be expressed as
\begin{equation}
l \left( \hat{x} \big| \hat{\theta} \right) = \prod^{N_{s}}_{i=0} \prod^{N_{k}}_{j=0} p_{ij}(k)~,
\end{equation}
where $N_k$ is the number of $k$ bins in which auto or cross power spectra are measured and $N_s$ represents the number of auto/cross-correlation surveys being included. The probability of the data vector $\hat{x}$ is described by a normal distribution
\begin{equation}
p_{ij}= \frac{1}{\sqrt{2\pi} \sigma_{ij}(k)} \exp \left\{ - \frac{\left[ P(k) - P(k | \hat{\theta}) \right]^2}{2 \sigma^2_{ij}(k)} \right\}~,
\end{equation}
where $\sigma_{ij}$ represents the gaussian error associated with the measurement. Broad, uninformative priors on the model parameters $\hat{\theta}$ are used, whose values are to be stated below for each individual case study. Furthermore, for all the following case studies, we adopt the same range and binning scheme for $k$ which yield 15 bins evenly-spaced in $\log k$ over $-1.5 < \log [k/(h/\mathrm{Mpc})] < 1$. We stress that while all four case studies presented below are evaluated at $z\sim2$ for a redshift interval of $\Delta z = \pm 0.25$, the same exercise could be repeated at different redshifts in order to study the redshift evolution of different ISM properties, which is one of the most important applications of the modeling framework presented. 

It is also important to point out that our model implicitly enforces a linear relation between line luminosity and halo mass, which is likely an oversimplification given the complicated physics involved in line production. The only physical parameter that modifies the shape of line luminosity function (and therefore the shape of power spectrum) is the log scatter $\sigma_\nu$. Consequently, whether the constraining power comes from the clustering or shot-noise regime of the power spectrum only makes a moderate difference in our analysis, as will be shown in Case~I. The scale dependence of constraining power in a power spectrum analysis without such simplification can be found in recent studies \citep[e.g.,][]{YF_2019}. 

\begin{figure*}[h!]
\centering 
\subfloat{
  \includegraphics[width=0.95\columnwidth]{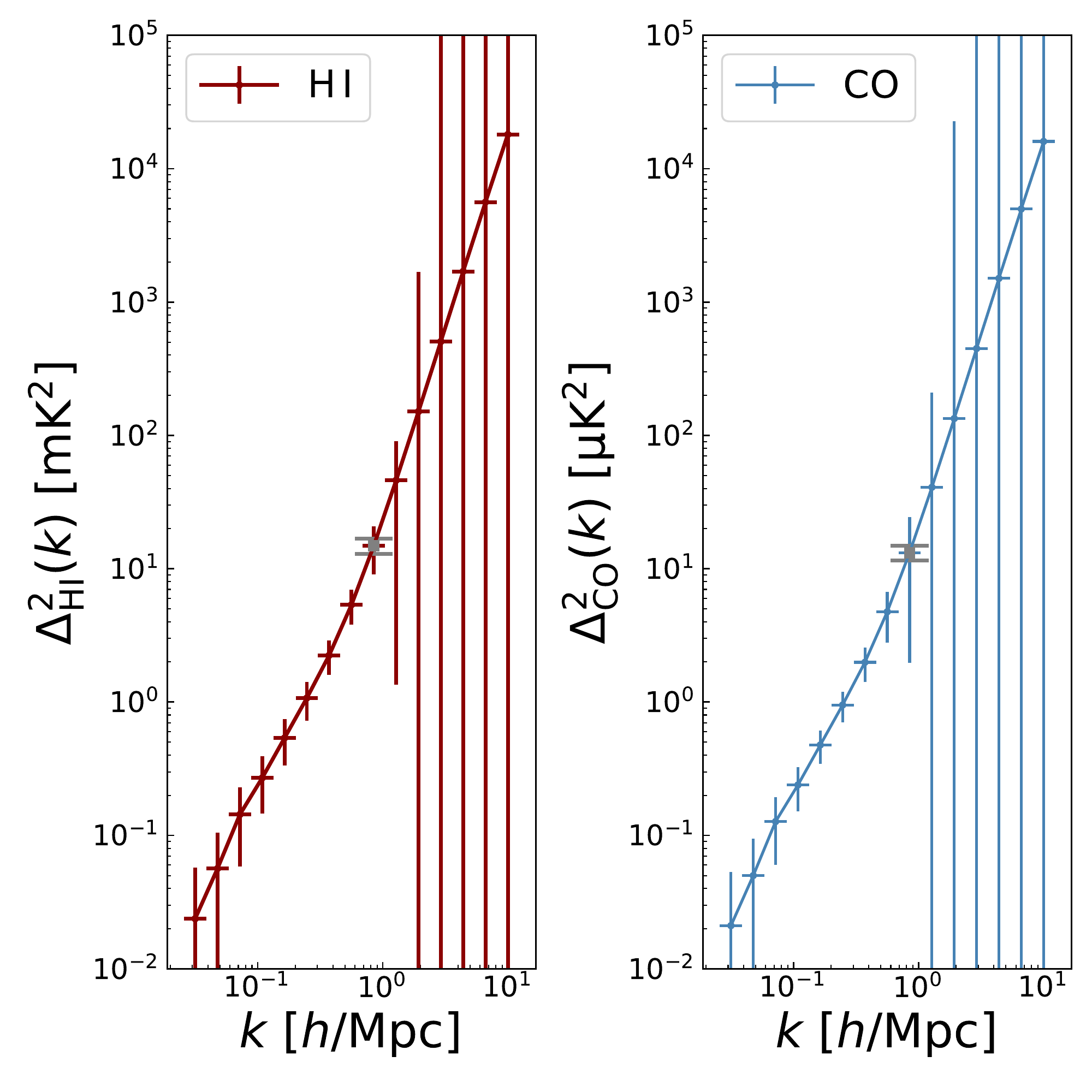}
}\qquad
\subfloat{
  \includegraphics[width=0.95\columnwidth]{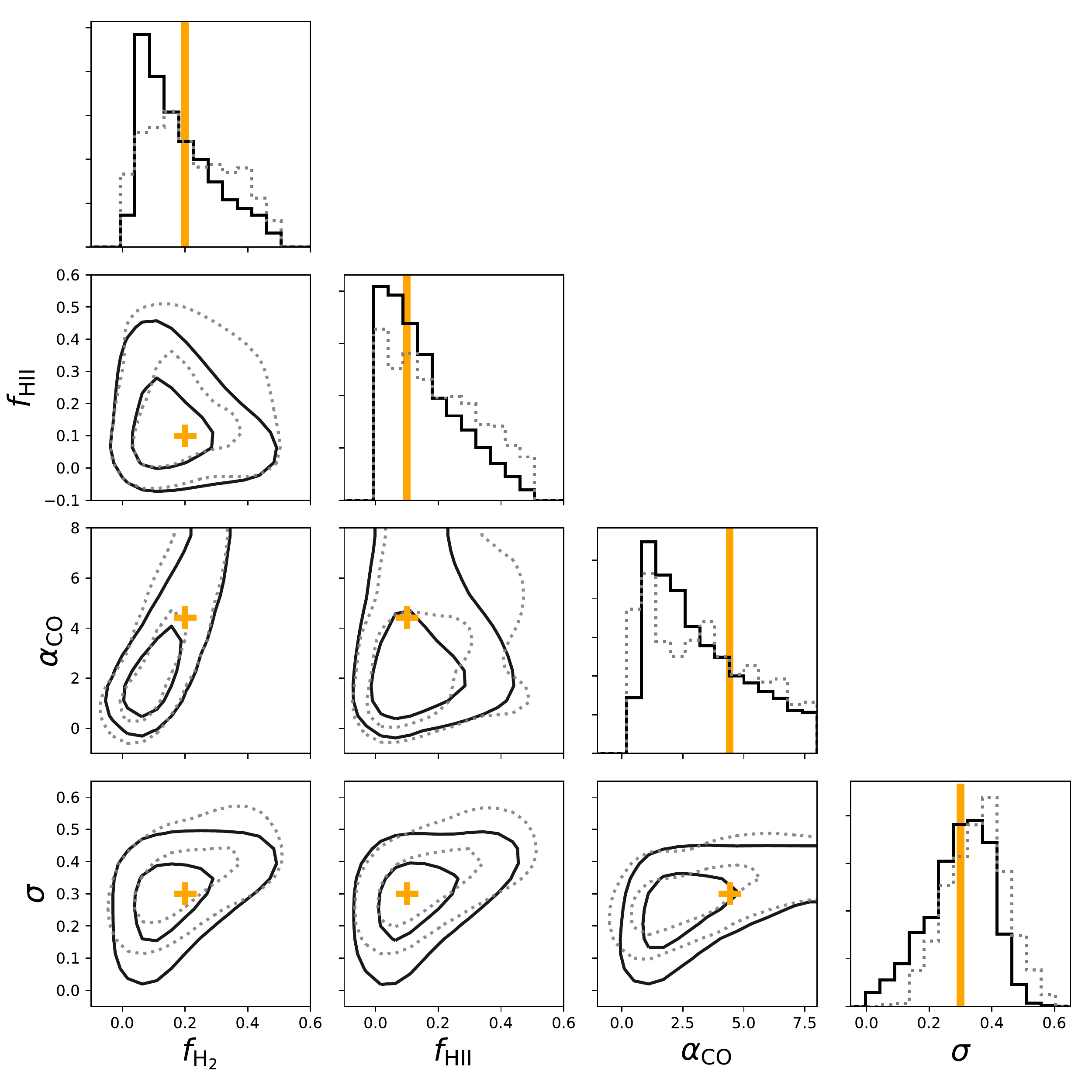}
}
\caption{\textit{Left}: mock data sets of the observed \ion{H}{i} and CO auto power spectra at $z\sim2$. The error bars are calculated via mode counting assuming the Case~I experimental setups specified in Table~\ref{tb:inst_params}. The gray error bar at $k \approx 1\,h/\mathrm{Mpc}$ indicates the shot-noise-only measurement with the same overall S/N. \textit{Right}: joint posterior distributions of the molecular gas fraction $f_{\rm H_2}$, the ionized gas fraction $f_{\ion{H}{ii}}$, the molecular gas density $n_{\rm H_2}$ and the scatter $\sigma$, shown for 68\% and 95\% confidence levels. The solid and dotted contours represent the constraints from measurements of the full power spectrum and only the shot noise, respectively. The true values in our reference ISM model used to generate mock observations are indicated by the orange plus signs. Diagonal panels show the marginalized distributions of each individual parameter. }
\label{fig:case1}
\end{figure*}

\begin{figure*}[h!]
\centering 
\subfloat{
  \includegraphics[width=1.1\columnwidth]{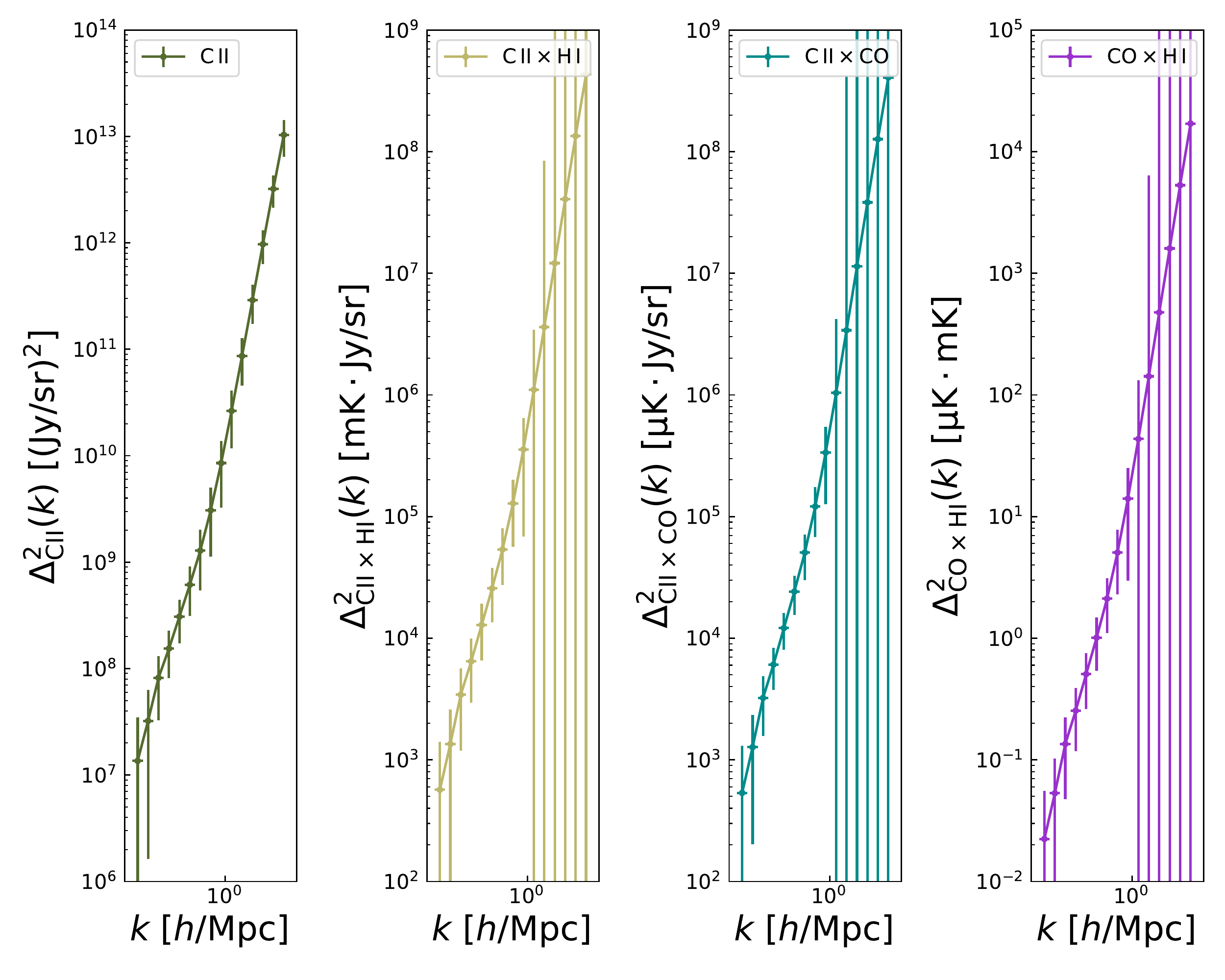}
}\qquad
\subfloat{
  \includegraphics[width=0.85\columnwidth]{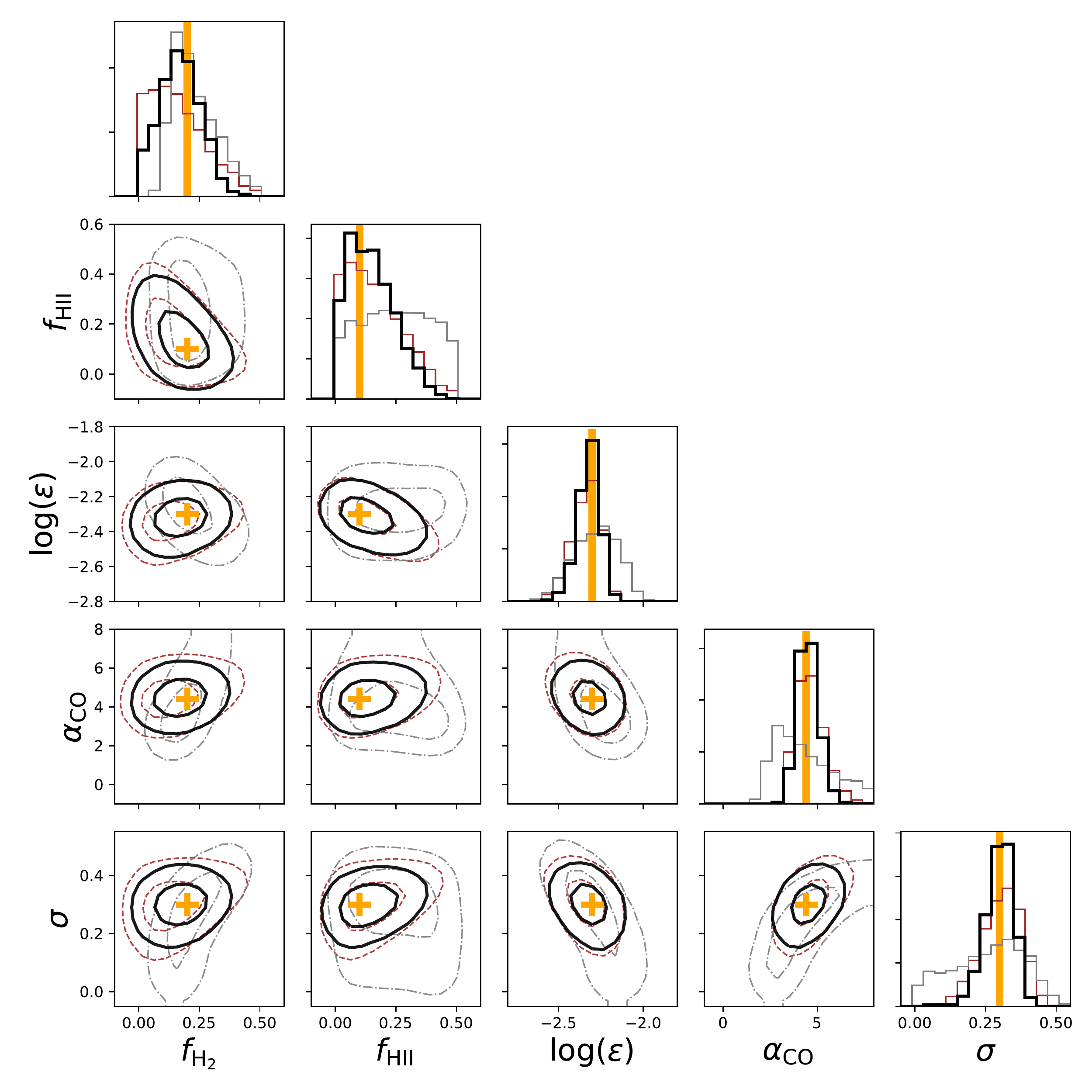}
}
\caption{\textit{Left}: mock data sets of the observed [\ion{C}{ii}] auto power spectrum and $\rm [\ion{C}{ii}] \times \ion{H}{i}$, $\rm [\ion{C}{ii}] \times CO$ and $\rm CO \times \ion{H}{i}$ cross power spectra at $z\sim2$. The error bars are calculated via mode counting assuming the Case~II experimental setups specified in Table~\ref{tb:inst_params}. \textit{Right}: joint posterior distributions of the molecular gas fraction $f_{\rm H_2}$, the ionized gas fraction $f_{\ion{H}{ii}}$, the photoelectric heating efficiency $\epsilon_{\rm PE}$, the molecular gas density $n_{\rm H_2}$ and the scatter $\sigma$, shown for 68\% and 95\% confidence levels as constrained by the auto-correlation (dashed contours), cross-correlation (dashed-dotted contours) and auto-and-cross combined data (solid contours). The true values in our reference ISM model used to generate mock observations are indicated by the orange plus signs. Diagonal panels show the marginalized distributions of each individual parameter. }
\label{fig:case2}
\end{figure*}

\subsection{Case I: Multi-Phase Diagnosis with \ion{H}{i} and CO}

As the first example, we investigate how the multi-phase ISM may be probed by a combination of \ion{H}{i} and CO LIM observations, which trace atomic and molecular hydrogen, respectively. Because the total gas mass is constrained implicitly by the CIB, the \ion{H}{i} measurement constrains both the atomic and molecular gas fraction. The CO measurement can then in principle break the degeneracy between the total amount of molecular gas and $\alpha_{\rm CO}$.

We consider two independent, mock measurements of \ion{H}{i} and CO auto power spectra, generated at $z\sim2$ assuming the reference ISM model described in Table~\ref{tb:line_params} and experimental setups specified in Table~\ref{tb:inst_params}, which yield a total S/N of approximately 8 for each signal\footnote{Summed over all $k$ bins; see Equation~(\ref{eq:tot_snr})}. Forthcoming single-dish/interferometric suveys, including FAST \citep{BigotSazy_2016ASPC}, CHIME \citep{CHIME_2014SPIE} and SKA \citep{SKA_2013} for \ion{H}{i} and COMAP \citep{Li_2016, Chung_2019} and mmIME (Keating et al. in preparation) for CO, will carry out these auto-correlation measurements directly, even though in both cases the signal is expected to be heavily contaminated by line/continuum foregrounds. Broad, flat priors over $0 < f_{\rm H_2} < 0.5$, $0 < f_{\rm \ion{H}{ii}} < 0.5$, $10^1 < n_{\rm H_2} / \mathrm{cm^{-3}} < 10^5$ and $0< \sigma < 1$ are assumed for the MCMC analysis. The MCMC sampling is constructed with 60 walkers, 500 burn-in steps---well above the estimated autocorrelation time ($\sim 50$ steps) \texttt{emcee} returned, and another 500 steps for sampling. 

The left panel of Figure~\ref{fig:case1} shows mock observed power spectra of \ion{H}{i} 21cm and CO auto-correlation signals at $z\sim2$, where the error bars are calculated from the assumed instrument parameters. The joint and marginalized posterior distributions of free parameters constrained by the mock auto power spectra under the MCMC framework are shown in the right panel of Figure~\ref{fig:case1}. Note that we have converted the posterior of $n_{\rm H_2}$ into the more commonly seen $\alpha_{\rm CO}$ factor using the assumed $T_{\rm exc} = 10\,$K. 

From the comparison between posterior distributions and true values (orange plus signs), as well as the fact that none of them are prior dominated, constraining power on all four parameters is observed. Even $f_{\rm H_2}$ and $\alpha_{\rm CO}$, though still strongly correlated, are individually constrained in this analysis. However, more precise estimation of the molecular gas content of galaxies from CO power spectrum measurements is conditional on how well $\alpha_{\rm CO}$ can be reliably determined, even if additional information about the atomic hydrogen content from \ion{H}{i} LIM is available. In practice, the exact value of $\alpha_{\rm CO}$ could vary in a non-trivial way with physical conditions of molecular gas in galaxies, especially the gas temperature distribution and metallicity. As a result, how LIM might be exploited to better determine its value is an interesting topic to be explored (see Section~\ref{sec:disncon} for further discussion). 

In addition to the default scenario using the full power spectrum in all $k$ bins, we consider an alternative scenario, where only the shot-noise power can be measured, while holding the overall S/N fixed. This resembles deep, targeted observations by, e.g., ALMA, from which information about large-scale intensity fluctuations is not available. As indicated by the gray error bars in the left panel of Figure~\ref{fig:case1}, we assume two $\rm S/N\sim8$ measurements of \ion{H}{i} and CO power spectrum at $k \approx 1\,h/\mathrm{Mpc}$ where shot noise is dominant. Due to the implicitly assumed linearity between line luminosity and halo mass, similar constraining power on the parameter space is observed, except that the measured log scatter $\sigma_\nu$ becomes biased, which can be easily understood given that in our model it is the only parameter sensitive to the shape of the power spectrum. 

\subsection{Case II: Multi-Phase Diagnosis with \ion{H}{i}, [\ion{C}{ii}] and CO}

Given the observed degeneracy between $f_{\rm H_2}$ and $\alpha_{\rm CO}$ in the previous case study, which introduces ambiguity to the interpretation of CO LIM results in terms of a molecular gas census, we investigate in this case how the inclusion of [\ion{C}{ii}] data, an indirect tracer of the molecular hydrogen fraction as indicated by Equation~(\ref{eq:PE_eff_def}), may help alleviate such a degeneracy. Additionally, we investigate how the constraining power on the parameter space may differ between using the three separate auto power spectra and using the $3(3-1)/2=3$ cross-correlation measurements available, which has the advantage of being immune to contamination from uncorrelated foregrounds as suggested in S16. 

Mock data sets of LIM observations are again created assuming the reference ISM model parameters and instrument parameters listed in Table~\ref{tb:line_params} and Table~\ref{tb:inst_params}, respectively. We note that when accounting for the effect of finite beam size in the cross-correlation sensitivity analysis, we conservatively evaluate for the coarser beam throughout our calculations. The overall S/N of cross-correlation data ($\rm SNR_{tot} \sim 5$ for each cross signal) is consequently lower than that of auto-correlation data. At intermediate redshifts, experiments like EXCLAIM \citep{EXCLAIM_2017} and TIM \citep{Aguirre_2018AAS} will measure [\ion{C}{ii}] in tomography, which, when spatially overlapped, may be combined with the \ion{H}{i} and CO surveys mentioned in the previous case to obtain their mutual cross-correlations. Broad, flat priors over $0 < f_{\rm H_2} < 0.5$, $0 < f_{\rm \ion{H}{ii}} < 0.5$, $10^{-4} < \epsilon_{\rm PE} < 10^{-1}$, $10^1 < n_{\rm H_2} / \mathrm{cm^{-3}} < 10^5$ and $0< \sigma < 1$ are assumed for the MCMC analysis. The MCMC sampling is constructed with 50 walkers, 500 burn-in steps---sufficiently larger than the estimated autocorrelation time ($\sim 60$ steps)---and another 500 steps for sampling. 

In the left panel of Figure~\ref{fig:case2}, we show the mock power spectrum data sets in addition to what has been shown in Figure~\ref{fig:case1}, including auto-correlation of [\ion{C}{ii}] and mutual cross-correlations of the three lines considered, all evaluated at $z\sim2$. The corresponding constraining power on the parameter space of our mock auto-correlation (brown dashed contours) and cross-correlation (gray dashed-dotted contours) and auto/cross-correlation-combined (black solid contours) data sets is presented in the right panel of Figure~\ref{fig:case2} as joint and marginalized posterior distributions. 

From the posterior constrained by auto-correlations, which becomes less biased from the true value after including [\ion{C}{ii}] data, it is clear that the degeneracy between $f_{\rm H_2}$ and $\alpha_{\rm CO}$ has been substantially reduced, although considerable uncertainty is still associated with $f_{\rm H_2}$. Other parameters, including $\alpha_{\rm CO}$, $\epsilon_{\rm PE}$ and $\sigma$, are well constrained by the auto-correlations from their marginalized posteriors, except for $f_{\ion{H}{ii}}$ which is not directly traced by any of the lines. The constraining power from cross-correlations, on the other hand, is not as good---particularly for $f_{\ion{H}{ii}}$ of which the constraint is prior dominated---yet still significant in general. While formally when $N \geq 3$, perfectly correlated lines are present, the mean line intensities shall be constrained equally well by their mutual cross-correlations; the poorer performance can be largely explained by the lower overall S/N of cross-correlation data. For completeness, we show also the total constraining power combining both auto-correlation and cross-correlation data sets, even though it is only slightly improved compared with the auto-only case. Both kinds of measurement are subject to realistic but different limitations --- while analysis based on auto correlations tends to be more foreground-contaminated in general, it has the advantage of not requiring the experiments to be spatially overlapped, as long as the cosmic variance of individual surveys can be properly accounted for. 

\begin{figure*}
\centering 
\subfloat{
  \includegraphics[width=0.95\columnwidth]{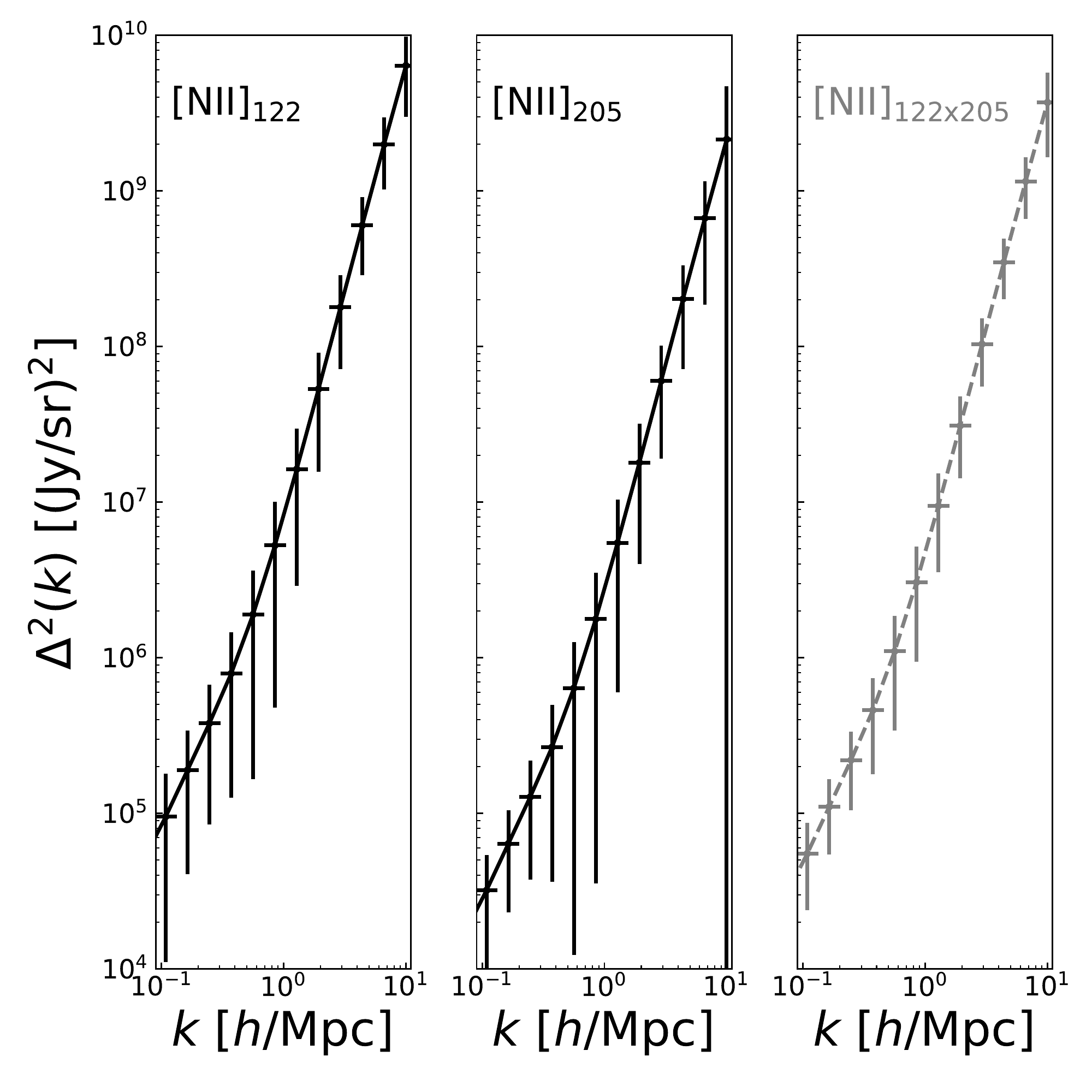}
}\qquad
\subfloat{
  \includegraphics[width=0.95\columnwidth]{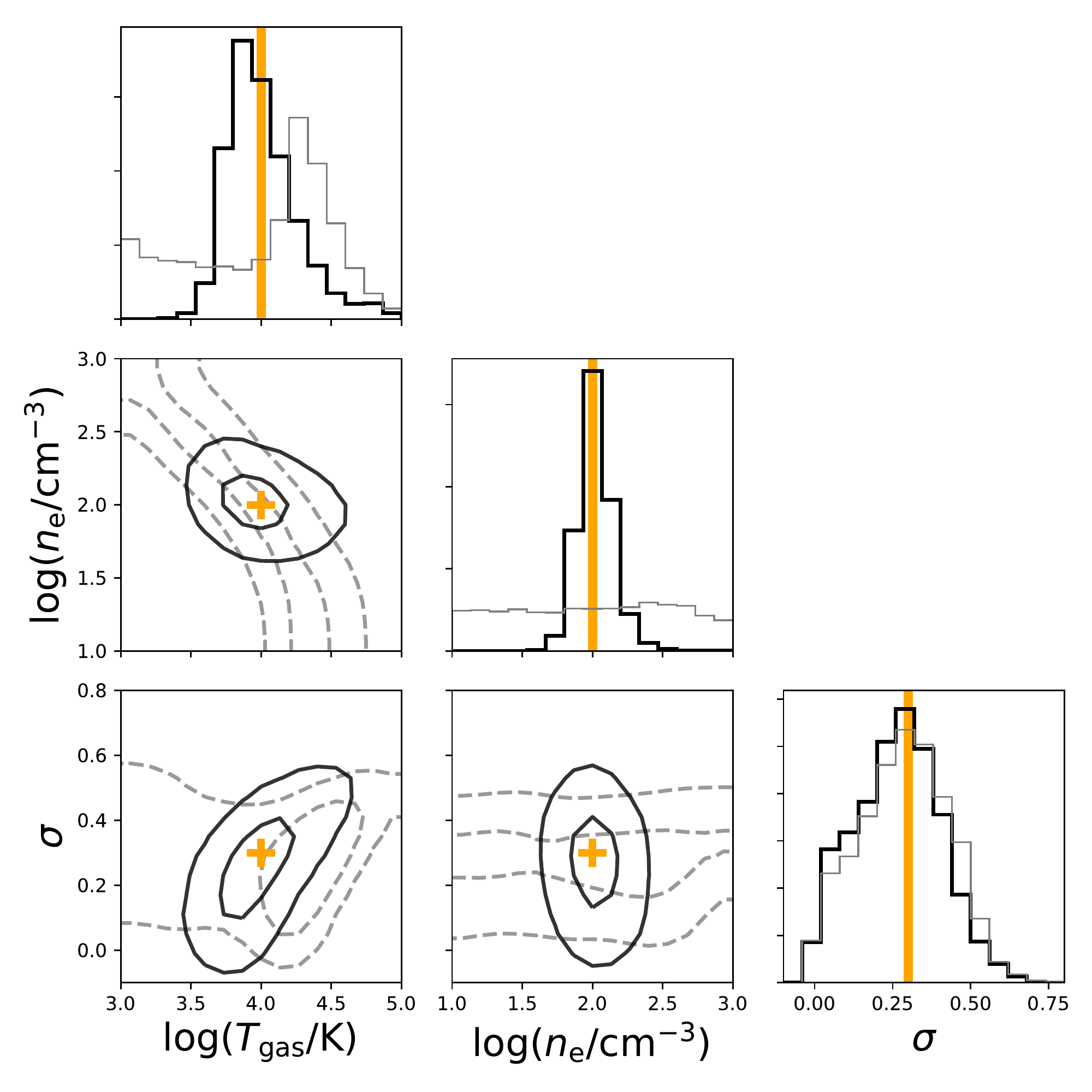}
}
\caption{\textit{Left}: mock data sets of the observed auto (solid) and cross (dashed) power spectra of [\ion{N}{ii}] 122\,$\mu$m and 205\,$\mu$m lines at $z\sim2$. The error bars are calculated via mode counting assuming the Case~III experimental setups specified in Table~\ref{tb:inst_params}. \textit{Right}: posterior distributions of the gas density $n_{\rm e, \ion{H}{ii}}$ and temperature $T_{\rm gas, \ion{H}{ii}}$ in \ion{H}{ii} regions, constrained by the cross power spectrum $P_{122 \times 205}$ (gray dashed contours) and the auto power spectra $P_{122}$ and $P_{205}$ (black solid contours), respectively. The inner and outer contours represent the 68\% and 95\% confidence intervals. The true values in our reference ISM model used to generate mock observations are indicated by the orange plus signs. Diagonal panels show the marginalized distributions of each individual parameter. }
\label{fig:corner_nii}
\end{figure*}

\subsection{Case III: Probing \ion{H}{ii} Regions with [\ion{N}{ii}] Lines.}

Another straightforward application of our line model is to use the two [\ion{N}{ii}] lines to constrain the state of ionized ISM, especially its electron number density $n_{\rm e, \ion{H}{ii}}$ directly probed by the [\ion{N}{ii}] fine-structure line ratio (see, e.g., \citealt{Goldsmith_2015} and \citealt{DiazSantos_2017} for applications of the $[\ion{N}{ii}]_{205}/[\ion{N}{ii}]_{122}$ ratio as a diagnostic of $n_{\rm e, \ion{H}{ii}}$ to the Galactic plane and local galaxies). Here, we consider two types of measurements, namely, the cross power spectrum of the two [\ion{N}{ii}] lines and their respective auto power spectra. 

Mock data sets of LIM observations are created assuming the reference ISM model ($n_{\rm e, \ion{H}{ii}} = 100\,\mathrm{cm^{-3}}$, $T_{\rm gas, \ion{H}{ii}} = 10^4$\,K and $\sigma = 0.3$\,dex), together with experimental setups specified in Table~\ref{tb:inst_params}. We note that while [\ion{N}{ii}] lines tend to be spectrally covered by [\ion{C}{ii}]-targeted experiments like EXCLAIM and TIM, at intermediate redshifts the required specifications in this (and the next) case study for a detection may only be achievable for next-generation space missions like \textit{OST} owing to the faintness of [\ion{N}{ii}] emission. Broad, flat priors over $1 < n_{\rm e, \ion{H}{ii}} / \mathrm{cm^{-3}} < 10^5$, $10^3 < T_{\rm gas, \ion{H}{ii}} / \mathrm{K} < 10^5$ and $0< \sigma < 1$ are assumed for the MCMC analysis. The MCMC sampling in either case is done with 100 walkers, 1000 burn-in steps---well above the estimated autocorrelation time ($\sim 100$ steps)---and another 1000 steps for sampling.

Figure~\ref{fig:corner_nii} shows the posterior distributions of the electron number density $n_{\rm e, \ion{H}{ii}}$, the gas temperature $T_{\rm gas, \ion{H}{ii}}$ and the lognormal scatter in line intensity $\sigma$, as constrained by the two types of observations, respectively. With the assumed model and survey parameters, the auto and cross power spectra $P^{[\ion{N}{ii}]}_{122}$, $P^{[\ion{N}{ii}]}_{205}$ and $P^{[\ion{N}{ii}]}_{122\times205}$ are measured at a total S/N of $\mathrm{SNR_{tot}} \sim 5.2$, 4.7 and 6.7, respectively. Both methods are able to determine the density and temperature without significant bias. Nevertheless, the auto power spectra of both [\ion{N}{ii}] lines together are much more effective than the just the cross power at breaking the degeneracy between the density and temperature as probed by the line ratio (see Figure~\ref{fig:niiratio}). We note that the difference in the constraining power between these two contrasting cases serves as an example of the importance of determining the amplitude of each individual tracer, through measurements of either the individual auto power spectra, or all mutual cross power spectra when $N \geq 3$ lines are detected, as suggested in S16 and demonstrated in the previous case study. 

\subsection{Case IV: Dissecting [\ion{C}{ii}] Origin with [\ion{C}{ii}]--[\ion{N}{ii}] 205$\mu$m Line Ratio}

\begin{figure*}
\centering 
\subfloat{
  \includegraphics[width=0.95\columnwidth]{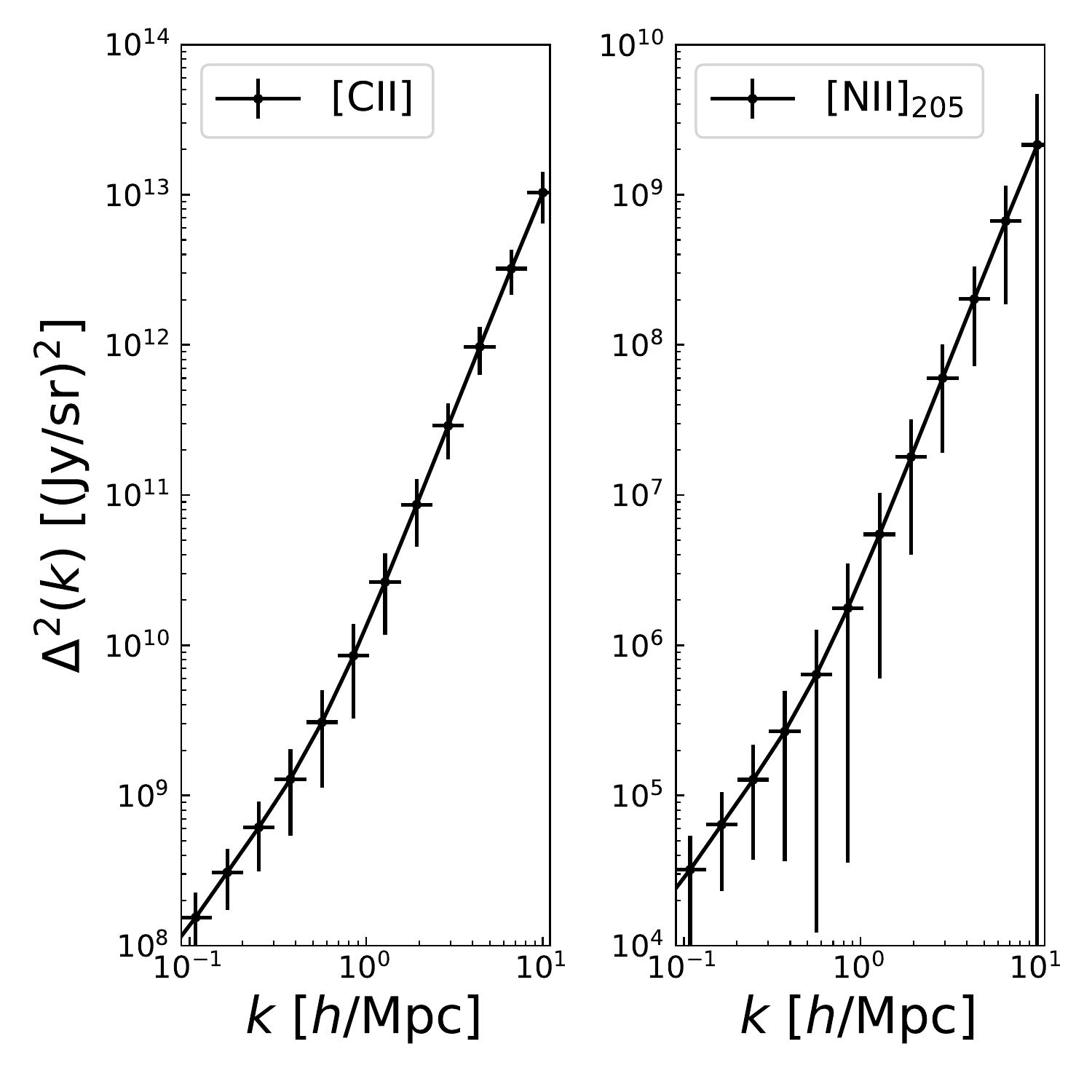}
}\qquad
\subfloat{
  \includegraphics[width=0.95\columnwidth]{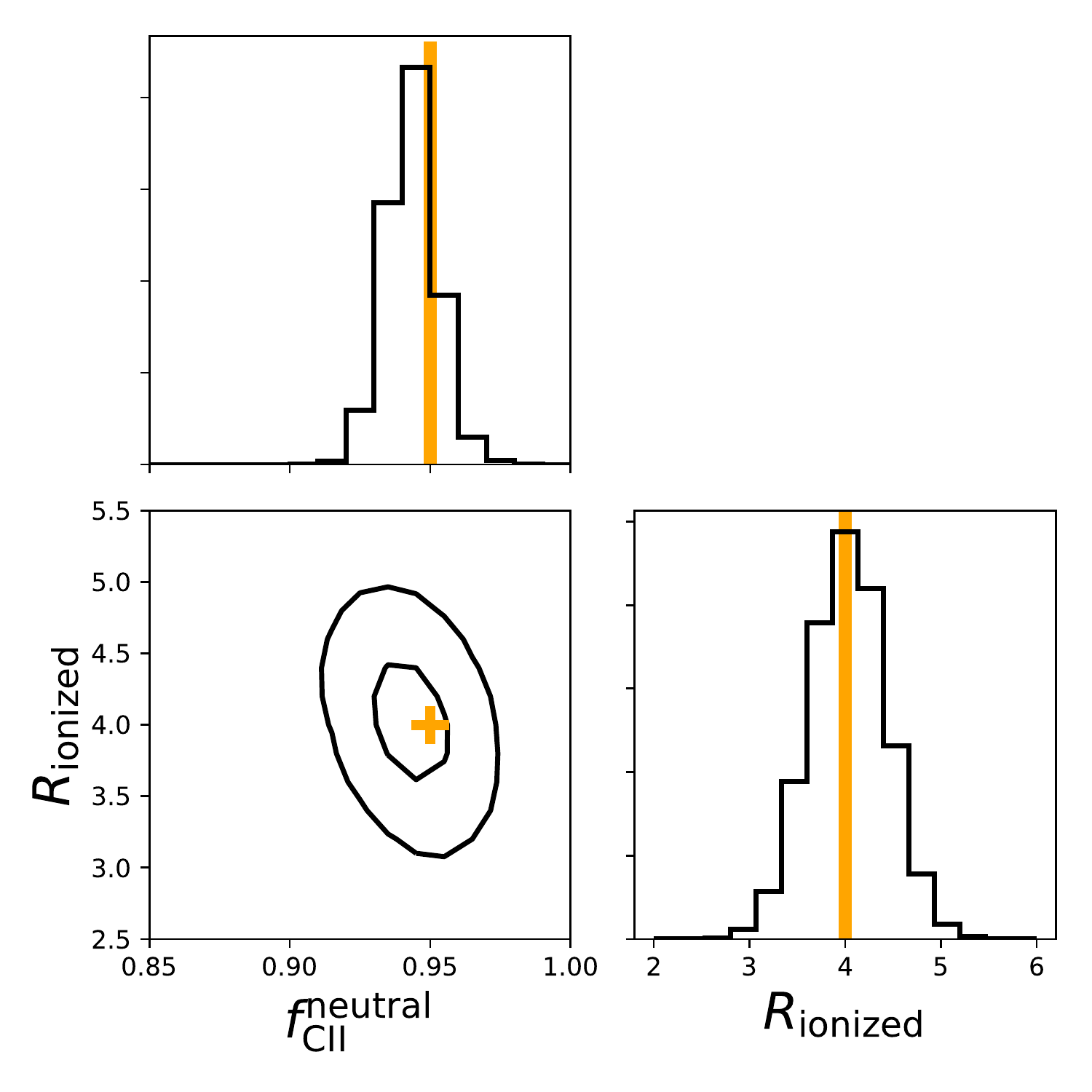}
}
\caption{\textit{Left}: mock data sets of the [\ion{C}{ii}] and [\ion{N}{ii}] 205\,$\mu$m auto power spectra at $z\sim2$. The error bars are calculated via mode counting assuming the Case~IV experimental setups specified in Table~\ref{tb:inst_params}. \textit{Right}: joint posterior distribution of the neutral-phase contribution, $f^{\rm neutral}_{[\ion{C}{ii}]}$, to the total [\ion{C}{ii}] line emission and the [\ion{C}{ii}]/[\ion{N}{ii}] line ratio, $R_{\rm ionized}$, inferred from collision rates, shown for 68\% and 95\% confidence levels. The (equivalent) true values $f^{\rm neutral}_{[\ion{C}{ii}]} \approx 0.95$ and $R_{\rm ionized} \approx 4$ in our reference ISM model used to generate mock observations are indicated by the orange plus sign. Diagonal panels show the marginalized distributions of each individual parameter. }
\label{fig:case3}
\end{figure*}

While in Section~\ref{sec:lines-cii} we have assumed that the [\ion{C}{ii}] line is solely attributed to the atomic gas in the PDRs so as to keep the line model simple, a small yet non-trivial fraction of the observed [\ion{C}{ii}] emission may actually originate in ionized gas phases as suggested by several recent studies \citep{Hughes_2015, Croxall_2017, Cormier_2019}. Therefore, in this final example we consider a slight extension of the [\ion{C}{ii}] model presented: we rewrite the total [\ion{C}{ii}] emission observed with LIM as $L^{\rm tot}_{[\ion{C}{ii}]} = L_{[\ion{C}{ii}]} / f_{[\ion{C}{ii}]}^{\rm neutral}$, where $L_{[\ion{C}{ii}]} = (1-f_{\rm H_2}) \epsilon_{\rm PE} L_{\rm IR}$ is the contribution from the neutral ISM (PDRs) defined in Section~\ref{sec:lines-cii} and $f_{[\ion{C}{ii}]}^{\rm neutral}$ is an extra parameter introduced here to describe the fraction of $[\ion{C}{ii}]$ emission contributed by the neutral ISM. Following \citet{Croxall_2017}, we use the ratio of [\ion{C}{ii}]/[\ion{N}{ii}] 205\,$\mu$m lines, whose critical densities for electron collisions are very similar ($n^{\rm crit}_{[\ion{C}{ii}]} \sim 45\,\mathrm{cm^{-3}}$ and $n^{\rm crit}_{[\ion{N}{ii}], 205} \sim 32\,\mathrm{cm^{-3}}$), as a diagnostic of $f_{[\ion{C}{ii}]}^{\rm neutral} = 1 - R_{\rm ionized} L_{[\ion{N}{ii}],205} / L_{[\ion{C}{ii}]}$, where $R_{\rm ionized} \approx 4$ denotes the ionized gas [\ion{C}{ii}]/[\ion{N}{ii}] ratio implied by their respective collision rates with electrons \citep{BP_1992, Tayal_2008, Tayal_2011}, assuming Galactic gas-phase abundances. 

Figure~\ref{fig:case3} demonstrates the constraining power on the neutral-phase contribution $f_{[\ion{C}{ii}]}^{\rm neutral}$ to the observed [\ion{C}{ii}] emission, estimated from the line ratio $L_{[\ion{N}{ii}],205} / L_{[\ion{C}{ii}]}$ inferred from mock LIM observations at $z\sim2$ (left panel). For the MCMC analysis, we assume a gaussian prior $\mathcal{N}(4, 0.4)$ for $R_{\rm ionized}$, whereas a broad, flat prior is used for $f_{[\ion{C}{ii}]}^{\rm neutral}$. The sampling is done with 100 walkers, 500 burn-in steps---sufficiently large compared with the estimated autocorrelation time ($\sim 30$ steps)---and another 500 steps for sampling. 

Our reference ISM model assumes a high $f_{[\ion{C}{ii}]}^{\rm neutral} \sim 0.9$ (orange plus sign), consistent with the finding that [\ion{C}{ii}] emission arises mostly from the neutral ISM. For the survey specifications given in Table~\ref{tb:inst_params}, we find that a population-averaged, neutral-phase contribution $f_{[\ion{C}{ii}]}^{\rm neutral}$ can be robustly determined by simultaneously observing [\ion{C}{ii}] and [\ion{N}{ii}] 205\,$\mu$m lines with LIM. We note that, in principle, the simple diagnostic described may be subject to density effects on the [\ion{C}{ii}]/[\ion{N}{ii}] line ratio and $R_{\rm ionized}$, although in both cases the dependence on $n_{\rm e}$ is found to be weak \citep{Croxall_2017}. 

% ============================     S7: DISCUSSION & CONCLUSION     ============================ %

\section{Discussion and Conclusion} \label{sec:disncon}

We have presented a simple analytical framework to self-consistently model the production of emission lines in the multi-phase ISM, based on the mean dark matter halo properties derived from a model fit to the observed CIB anisotropy. The redshift evolution of cosmic star formation, dust mass, gas (total and molecular) mass, gas-phase metallicity, and the strengths of \ion{H}{i}, [\ion{C}{ii}], [\ion{N}{ii}] and CO lines predicted by our model have been compared with observations, showing that our model, despite its simplicity, can describe the production of lines in the ISM in a physically-motivated way. We have illustrated how this modeling framework can be used to reconstruct the average properties of different ISM phases, such as the mass fractions and densities of neutral and ionized gas, the photoelectric heating efficiency in the PDRs, and so forth, over a wide range of redshifts from multi-tracer LIM observations. 

Our analysis underscores the importance of cross-correlation analyses. While equivalent information may be obtained from the auto-correlation of respective tracers, cross-correlation analysis in the same cosmological volume is minimally susceptible to foreground contamination. With the large number of upcoming LIM experiments targeting lines produced in different ISM phases, e.g., CCAT-Prime \citep{Stacey_2018SPIE}, CHIME \citep{CHIME_2014SPIE}, COMAP \citep{Li_2016}, CONCERTO \citep{Lagache_2018IAUS}, HIRAX \citep{Newburgh_2016SPIE}, SKA \citep{Santos_2015}, SPHEREx \citep{Dore_2014}, Tianlai \citep{Xu_2015}, TIM \citep{Aguirre_2018AAS} and TIME \citep{Crites_2014SPIE}, our understanding of the ISM evolution and physical processes dominating line emission over cosmic time is expected to be greatly deepened by the coarse-grained view built up from LIM surveys with multiple tracers. 

The simplicity and modularity of the model presented here lends itself to straightforward improvements and extensions to incorporate more sophisticated treatments of both galaxy evolution and ISM physics motivated by observational and theoretical studies. For instance, to more reliably apply this framework to galaxies at higher redshifts, including the reionization era, it would be valuable to introduce additional calibrations and constraints from data sets at other wavelengths. Currently the star formation history is anchored only to FIR emission constrained by the CIB anisotropy, which is sensitive mostly to galaxies at redshift $z\la3$ \cite[e.g.,][]{Viero_2013}. However, much of the information about the galaxy--halo connection, feedback-regulated galaxy and ISM evolution, and so forth, is encoded in data at shorter wavelengths, e.g., the galaxy UV luminosity function (UVLF). We therefore expect the exact mass and redshift dependence of mean halo properties (see Section~\ref{sec:halos}) to be better constrained out to the epoch of reionization by combining IR and UV data, which will be explored in future work.

Given the necessarily coarse-grained picture of galactic ISM properties painted by LIM, we have employed physically-motivated but ultimately simple prescriptions for the line emission physics. Particularly as new observations yield more model constraints, the line physics can be refined. Notably, our prescription for the [\ion{C}{ii}] emission does not account for the deficit relative to $L_{\rm IR}$ observed in luminous and ultraluminous galaxies \citep[e.g.,][]{Malhotra1997}. It should be explored whether this can be recovered within the modeling framework by introducing the effect of dust charging on $\epsilon_{\rm PE}$ \citep{BT_1994} and the saturation of [\ion{C}{ii}] at high gas temperatures \citep{Munoz2016}. Further, given that this effect appears to be a strong function of galaxy luminosity, this may have important testable implications for the predicted [\ion{C}{ii}] power spectra. 

Another important caveat to our simple prescription lies in the interpretation of LIM signals in terms of globally-averaged ISM properties. In reality, the ensemble of gas clouds within a galaxy, while all contributing to the same line emission, may have a wide distribution of physical properties (e.g., $\rm H_2$ gas temperature). Likewise, these distributions may vary significantly among different galaxy populations. As a result, interpreting LIM data in terms of a single ``mean'' property is an oversimplification. A more robust extraction of ISM physics from LIM data sets could be achieved by modeling these distributions directly, perhaps incorporating prior information about ISM conditions that are known to vary systematically in different galaxy populations, as well as how the line production is coupled to these distributions through radiative processes. Such modeling is beyond the scope of this paper and will be the subject of future investigation.

Finally, in this work we selected a small subset of available lines to illustrate the power of LIM to probe the multi-phase ISM. However, the model is readily extensible to other lines. For instance, the [\ion{O}{i}] 63\,$\mu$m and [\ion{O}{iii}] 88\,$\mu$m lines are also important cooling lines, and so the sum of emission from these lines and [\ion{C}{ii}] may yield a more robust correlation with $L_{\rm IR}$ \citep{DL_2014}, with their relative importance as a function of redshift and galaxy properties providing constraints on the physical state of the emitting gas. Simultaneous measurements of multiple CO rotational lines are both a powerful probe of the physics of molecular gas and a means of validation since the CO lines should be spatially correlated. In addition to CO, H$_2$ rotational lines can be used to constrain the molecular gas content of galaxies, meanwhile shedding light on the gas temperature distribution \citep{PS_2014, Togi2016}. Optical and UV lines of hydrogen such as Ly$\alpha$, H$\alpha$ and H$\beta$ are also being actively pursued by LIM experiments and should be incorporated, particularly given their potential to probe metal-poor environments in the very high redshift universe.

% ============================     Aknowledgment     ============================ %
\acknowledgments
We would like to thank the anonymous referee for comments that helped improve this paper. We would like to thank Hao-Yi (Heidi) Wu for helpful discussion on the CIB model, as well as Garrett (Karto) Keating and Ryan Keenan for compiling and sharing the constraints on cosmic molecular gas content. We are also grateful to Jamie Bock, Matt Bradford, Patrick Breysse, Paul Goldsmith, Adam Lidz, Lunjun Liu, Lluis Mas-Ribas and Anthony Pullen for constructive discussion and comments on this work. Part of the research described in this paper was carried out at the Jet Propulsion Laboratory, California Institute of Technology, under a contract with the National Aeronautics and Space Administration. 

\textit{Software:} \texttt{corner} \cite[][]{FM_2016}, \texttt{emcee} \cite[][]{FM_2013}, \texttt{hmf} \cite[][]{MPR_2013}, \texttt{matplotlib} \cite[][]{Hunter_2007}, \texttt{numpy} \cite[][]{Walt_2011} and \texttt{scipy} \cite[][]{Jones_2001}. 

\bibliography{imap}

\begin{thebibliography}{}
\expandafter\ifx\csname natexlab\endcsname\relax\def\natexlab#1{#1}\fi

\bibitem[{{Aguirre} \& {STARFIRE Collaboration}(2018)}]{Aguirre_2018AAS}
{Aguirre}, J., \& {STARFIRE Collaboration}. 2018, in American Astronomical
  Society Meeting Abstracts, Vol. 231, American Astronomical Society Meeting
  Abstracts \#231, 328.04

\bibitem[{{Asplund} {et~al.}(2009){Asplund}, {Grevesse}, {Sauval}, \&
  {Scott}}]{Asplund_2009}
{Asplund}, M., {Grevesse}, N., {Sauval}, A.~J., \& {Scott}, P. 2009, \araa, 47,
  481

\bibitem[{{Bakes} \& {Tielens}(1994)}]{BT_1994}
{Bakes}, E.~L.~O., \& {Tielens}, A.~G.~G.~M. 1994, \apj, 427, 822

\bibitem[{{Bandura} {et~al.}(2014){Bandura}, {Addison}, {Amiri}, {Bond},
  {Campbell-Wilson}, {Connor}, {Cliche}, {Davis}, {Deng}, {Denman}, {Dobbs},
  {Fandino}, {Gibbs}, {Gilbert}, {Halpern}, {Hanna}, {Hincks}, {Hinshaw},
  {H{\"o}fer}, {Klages}, {Landecker}, {Masui}, {Mena Parra}, {Newburgh}, {Pen},
  {Peterson}, {Recnik}, {Shaw}, {Sigurdson}, {Sitwell}, {Smecher}, {Smegal},
  {Vanderlinde}, \& {Wiebe}}]{CHIME_2014SPIE}
{Bandura}, K., {Addison}, G.~E., {Amiri}, M., {et~al.} 2014, in \procspie, Vol.
  9145, Ground-based and Airborne Telescopes V, 914522

\bibitem[{{Beane} {et~al.}(2019){Beane}, {Villaescusa-Navarro}, \&
  {Lidz}}]{BVL_2019}
{Beane}, A., {Villaescusa-Navarro}, F., \& {Lidz}, A. 2019, \apj, 874, 133

\bibitem[{{Bhattacharya} {et~al.}(2013){Bhattacharya}, {Habib}, {Heitmann}, \&
  {Vikhlinin}}]{Bhattacharya_2013}
{Bhattacharya}, S., {Habib}, S., {Heitmann}, K., \& {Vikhlinin}, A. 2013, \apj,
  766, 32

\bibitem[{{Bigot-Sazy} {et~al.}(2016){Bigot-Sazy}, {Ma}, {Battye}, {Browne},
  {Chen}, {Dickinson}, {Harper}, {Maffei}, {Olivari}, \&
  {Wilkinsondagger}}]{BigotSazy_2016ASPC}
{Bigot-Sazy}, M.~A., {Ma}, Y.~Z., {Battye}, R.~A., {et~al.} 2016, in
  Astronomical Society of the Pacific Conference Series, Vol. 502, Frontiers in
  Radio Astronomy and FAST Early Sciences Symposium 2015, ed. L.~{Qain} \&
  D.~{Li}, 41

\bibitem[{{Blum} \& {Pradhan}(1992)}]{BP_1992}
{Blum}, R.~D., \& {Pradhan}, A.~K. 1992, \apjs, 80, 425

\bibitem[{{Bolatto} {et~al.}(2013){Bolatto}, {Wolfire}, \&
  {Leroy}}]{Bolatto+Wolfire+Leroy_2013}
{Bolatto}, A.~D., {Wolfire}, M., \& {Leroy}, A.~K. 2013, \araa, 51, 207

\bibitem[{{Boylan-Kolchin} {et~al.}(2009){Boylan-Kolchin}, {Springel}, {White},
  {Jenkins}, \& {Lemson}}]{BK_2009MSII}
{Boylan-Kolchin}, M., {Springel}, V., {White}, S.~D.~M., {Jenkins}, A., \&
  {Lemson}, G. 2009, \mnras, 398, 1150

\bibitem[{{Bradford} {et~al.}(2008){Bradford}, {Kenyon}, {Holmes}, {Bock}, \&
  {Koch}}]{Bradford_2008SPIE}
{Bradford}, C.~M., {Kenyon}, M., {Holmes}, W., {Bock}, J., \& {Koch}, T. 2008,
  in Society of Photo-Optical Instrumentation Engineers (SPIE) Conference
  Series, Vol. 7020, \procspie, 70201O

\bibitem[{{Bradford} {et~al.}(2018){Bradford}, {Cameron}, {Moore}, {Amatucci},
  {Bradley}, {Corsetti}, {Leisawitz}, {Moseley}, {Staguhn}, {Tuttle}, {Brown},
  {Pope}, {Armus}, {Meixner}, \& {Pontoppidan}}]{Bradford_2018SPIE}
{Bradford}, C.~M., {Cameron}, B., {Moore}, B., {et~al.} 2018, in Society of
  Photo-Optical Instrumentation Engineers (SPIE) Conference Series, Vol. 10698,
  \procspie, 1069818

\bibitem[{{Brauher} {et~al.}(2008){Brauher}, {Dale}, \& {Helou}}]{Brauher_2008}
{Brauher}, J.~R., {Dale}, D.~A., \& {Helou}, G. 2008, \apjs, 178, 280

\bibitem[{{Breysse} {et~al.}(2014){Breysse}, {Kovetz}, \&
  {Kamionkowski}}]{Breysse_2014}
{Breysse}, P.~C., {Kovetz}, E.~D., \& {Kamionkowski}, M. 2014, \mnras, 443,
  3506

\bibitem[{{Bull} {et~al.}(2015){Bull}, {Ferreira}, {Patel}, \&
  {Santos}}]{Bull_2015}
{Bull}, P., {Ferreira}, P.~G., {Patel}, P., \& {Santos}, M.~G. 2015, \apj, 803,
  21

\bibitem[{{Chang} {et~al.}(2010){Chang}, {Pen}, {Bandura}, \&
  {Peterson}}]{Chang_2010}
{Chang}, T.-C., {Pen}, U.-L., {Bandura}, K., \& {Peterson}, J.~B. 2010, \nat,
  466, 463

\bibitem[{{Chang} {et~al.}(2019){Chang}, {Beane}, {Dore}, {Lidz}, {Mas-Ribas},
  {Sun}, {Alvarez}, {Thakur}, {Berger}, {Bethermin}, {Bock}, {Bradford},
  {Breysse}, {Burgarella}, {Charmandaris}, {Cheng}, {Cleary}, {Cooray},
  {Crites}, {Ewall-Wice}, {Fan}, {Finkelstein}, {Furlanetto}, {Hewitt},
  {Hunacek}, {Korngut}, {Kovetz}, {Hallinan}, {Heneka}, {Lagache}, {Lawrence},
  {Lazio}, {Liu}, {Marrone}, {Parsons}, {Readhead}, {Rhodes}, {Riechers},
  {Seiffert}, {Stacey}, {Visbal}, {Wu}, {Zemcov}, \& {Zheng}}]{Chang_2019BAAS}
{Chang}, T.-C., {Beane}, A., {Dore}, O., {et~al.} 2019, in \baas, Vol.~51,
  Bulletin of the American Astronomical Society, 282

\bibitem[{{Chung} {et~al.}(2019){Chung}, {Viero}, {Church}, {Wechsler},
  {Alvarez}, {Bond}, {Breysse}, {Cleary}, {Eriksen}, {Foss}, {Gundersen},
  {Harper}, {Ihle}, {Keating}, {Murray}, {Padmanabhan}, {Stein}, {Wehus}, \&
  {COMAP Collaboration}}]{Chung_2019}
{Chung}, D.~T., {Viero}, M.~P., {Church}, S.~E., {et~al.} 2019, \apj, 872, 186

\bibitem[{{Clegg} {et~al.}(1996){Clegg}, {Ade}, {Armand}, {Baluteau}, {Barlow},
  {Buckley}, {Berges}, {Burgdorf}, {Caux}, {Ceccarelli}, {Cerulli}, {Church},
  {Cotin}, {Cox}, {Cruvellier}, {Culhane}, {Davis}, {di Giorgio}, {Diplock},
  {Drummond}, {Emery}, {Ewart}, {Fischer}, {Furniss}, {Glencross},
  {Greenhouse}, {Griffin}, {Gry}, {Harwood}, {Hazell}, {Joubert}, {King},
  {Lim}, {Liseau}, {Long}, {Lorenzetti}, {Molinari}, {Murray}, {Naylor},
  {Nisini}, {Norman}, {Omont}, {Orfei}, {Patrick}, {Pequignot}, {Pouliquen},
  {Price}, {Nguyen-Q-Rieu}, {Rogers}, {Robinson}, {Saisse}, {Saraceno},
  {Serra}, {Sidher}, {Smith}, {Smith}, {Spinoglio}, {Swinyard}, {Texier},
  {Towlson}, {Trams}, {Unger}, \& {White}}]{Clegg_1996}
{Clegg}, P.~E., {Ade}, P.~A.~R., {Armand}, C., {et~al.} 1996, \aap, 315, L38

\bibitem[{{Comaschi} \& {Ferrara}(2016)}]{CF_2016}
{Comaschi}, P., \& {Ferrara}, A. 2016, \mnras, 455, 725

\bibitem[{{Cooray} \& {Sheth}(2002)}]{CS_2002}
{Cooray}, A., \& {Sheth}, R. 2002, \physrep, 372, 1

\bibitem[{{Cormier} {et~al.}(2019){Cormier}, {Abel}, {Hony}, {Lebouteiller},
  {Madden}, {Polles}, {Galliano}, {De Looze}, {Galametz}, \&
  {Lambert-Huyghe}}]{Cormier_2019}
{Cormier}, D., {Abel}, N.~P., {Hony}, S., {et~al.} 2019, \aap, 626, A23

\bibitem[{{Crawford} {et~al.}(1985){Crawford}, {Genzel}, {Townes}, \&
  {Watson}}]{Crawford1985}
{Crawford}, M.~K., {Genzel}, R., {Townes}, C.~H., \& {Watson}, D.~M. 1985,
  \apj, 291, 755

\bibitem[{{Crites} {et~al.}(2014){Crites}, {Bock}, {Bradford}, {Chang},
  {Cooray}, {Duband}, {Gong}, {Hailey-Dunsheath}, {Hunacek}, \&
  {Koch}}]{Crites_2014SPIE}
{Crites}, A.~T., {Bock}, J.~J., {Bradford}, C.~M., {et~al.} 2014, in Society of
  Photo-Optical Instrumentation Engineers (SPIE) Conference Series, Vol. 9153,
  Millimeter, Submillimeter, and Far-Infrared Detectors and Instrumentation for
  Astronomy VII, 91531W

\bibitem[{{Croxall} {et~al.}(2017){Croxall}, {Smith}, {Pellegrini}, {Groves},
  {Bolatto}, {Herrera-Camus}, {Sand strom}, {Draine}, {Wolfire}, {Armus},
  {Boquien}, {Brandl}, {Dale}, {Galametz}, {Hunt}, {Kennicutt}, {Kreckel},
  {Rigopoulou}, {van der Werf}, \& {Wilson}}]{Croxall_2017}
{Croxall}, K.~V., {Smith}, J.~D., {Pellegrini}, E., {et~al.} 2017, \apj, 845,
  96

\bibitem[{{Cucciati} {et~al.}(2012){Cucciati}, {Tresse}, {Ilbert}, {Le
  F{\`e}vre}, {Garilli}, {Le Brun}, {Cassata}, {Franzetti}, {Maccagni},
  {Scodeggio}, {Zucca}, {Zamorani}, {Bardelli}, {Bolzonella}, {Bielby},
  {McCracken}, {Zanichelli}, \& {Vergani}}]{Cucciati_2012}
{Cucciati}, O., {Tresse}, L., {Ilbert}, O., {et~al.} 2012, \aap, 539, A31

\bibitem[{{Dame} {et~al.}(2001){Dame}, {Hartmann}, \& {Thaddeus}}]{Dame2001}
{Dame}, T.~M., {Hartmann}, D., \& {Thaddeus}, P. 2001, \apj, 547, 792

\bibitem[{{De Looze} {et~al.}(2014){De Looze}, {Cormier}, {Lebouteiller},
  {Madden}, {Baes}, {Bendo}, {Boquien}, {Boselli}, {Clements}, {Cortese},
  {Cooray}, {Galametz}, {Galliano}, {Graci{\'a}-Carpio}, {Isaak}, {Karczewski},
  {Parkin}, {Pellegrini}, {R{\'e}my-Ruyer}, {Spinoglio}, {Smith}, \&
  {Sturm}}]{DL_2014}
{De Looze}, I., {Cormier}, D., {Lebouteiller}, V., {et~al.} 2014, \aap, 568,
  A62

\bibitem[{{Decarli} {et~al.}(2016){Decarli}, {Walter}, {Aravena}, {Carilli},
  {Bouwens}, {da Cunha}, {Daddi}, {Ivison}, {Popping}, {Riechers}, {Smail},
  {Swinbank}, {Weiss}, {Anguita}, {Assef}, {Bauer}, {Bell}, {Bertoldi},
  {Chapman}, {Colina}, {Cortes}, {Cox}, {Dickinson}, {Elbaz},
  {G{\'o}nzalez-L{\'o}pez}, {Ibar}, {Infante}, {Hodge}, {Karim}, {Le Fevre},
  {Magnelli}, {Neri}, {Oesch}, {Ota}, {Rix}, {Sargent}, {Sheth}, {van der Wel},
  {van der Werf}, \& {Wagg}}]{Decarli_2016}
{Decarli}, R., {Walter}, F., {Aravena}, M., {et~al.} 2016, \apj, 833, 69

\bibitem[{{Decarli} {et~al.}(2019){Decarli}, {Walter},
  {G{\'o}nzalez-L{\'o}pez}, {Aravena}, {Boogaard}, {Carilli}, {Cox}, {Daddi},
  {Popping}, {Riechers}, {Uzgil}, {Weiss}, {Assef}, {Bacon}, {Bauer},
  {Bertoldi}, {Bouwens}, {Contini}, {Cortes}, {da Cunha}, {D{\'\i}az-Santos},
  {Elbaz}, {Inami}, {Hodge}, {Ivison}, {Le F{\`e}vre}, {Magnelli}, {Novak},
  {Oesch}, {Rix}, {Sargent}, {Smail}, {Swinbank}, {Somerville}, {van der Werf},
  {Wagg}, \& {Wisotzki}}]{Decarli_2019}
{Decarli}, R., {Walter}, F., {G{\'o}nzalez-L{\'o}pez}, J., {et~al.} 2019, \apj,
  882, 138

\bibitem[{{Dewdney} {et~al.}(2013){Dewdney}, {Turner}, R., R., {Lazio}, \&
  {Cornwell}}]{SKA_2013}
{Dewdney}, P., {Turner}, W., R., M., {et~al.} 2013, {SKA1 System Baseline
  Design}

\bibitem[{{D{\'{\i}}az-Santos} {et~al.}(2017){D{\'{\i}}az-Santos}, {Armus},
  {Charmandaris}, {Lu}, {Stierwalt}, {Stacey}, {Malhotra}, {van der Werf},
  {Howell}, {Privon}, {Mazzarella}, {Goldsmith}, {Murphy}, {Barcos-Mu{\~n}oz},
  {Linden}, {Inami}, {Larson}, {Evans}, {Appleton}, {Iwasawa}, {Lord},
  {Sanders}, \& {Surace}}]{DiazSantos_2017}
{D{\'{\i}}az-Santos}, T., {Armus}, L., {Charmandaris}, V., {et~al.} 2017, \apj,
  846, 32

\bibitem[{{Dickman} {et~al.}(1986){Dickman}, {Snell}, \&
  {Schloerb}}]{Dickman1986}
{Dickman}, R.~L., {Snell}, R.~L., \& {Schloerb}, F.~P. 1986, \apj, 309, 326

\bibitem[{{Dor{\'e}} {et~al.}(2014){Dor{\'e}}, {Bock}, {Ashby}, {Capak},
  {Cooray}, {de Putter}, {Eifler}, {Flagey}, {Gong}, \& {Habib}}]{Dore_2014}
{Dor{\'e}}, O., {Bock}, J., {Ashby}, M., {et~al.} 2014, arXiv e-prints,
  arXiv:1412.4872

\bibitem[{{Draine}(2011)}]{Draine_2011}
{Draine}, B.~T. 2011, {Physics of the Interstellar and Intergalactic Medium}
  (Princeton, NJ: Princeton University Press)

\bibitem[{{Draine} {et~al.}(2007){Draine}, {Dale}, {Bendo}, {Gordon}, {Smith},
  {Armus}, {Engelbracht}, {Helou}, {Kennicutt}, {Li}, {Roussel}, {Walter},
  {Calzetti}, {Moustakas}, {Murphy}, {Rieke}, {Bot}, {Hollenbach}, {Sheth}, \&
  {Teplitz}}]{Draine_2007}
{Draine}, B.~T., {Dale}, D.~A., {Bendo}, G., {et~al.} 2007, \apj, 663, 866

\bibitem[{{Driver} {et~al.}(2007){Driver}, {Popescu}, {Tuffs}, {Liske},
  {Graham}, {Allen}, \& {de Propris}}]{Driver_2007}
{Driver}, S.~P., {Popescu}, C.~C., {Tuffs}, R.~J., {et~al.} 2007, \mnras, 379,
  1022

\bibitem[{{Dunne} {et~al.}(2011){Dunne}, {Gomez}, {da Cunha}, {Charlot}, {Dye},
  {Eales}, {Maddox}, {Rowlands}, {Smith}, {Auld}, {Baes}, {Bonfield}, {Bourne},
  {Buttiglione}, {Cava}, {Clements}, {Coppin}, {Cooray}, {Dariush}, {de Zotti},
  {Driver}, {Fritz}, {Geach}, {Hopwood}, {Ibar}, {Ivison}, {Jarvis}, {Kelvin},
  {Pascale}, {Pohlen}, {Popescu}, {Rigby}, {Robotham}, {Rodighiero}, {Sansom},
  {Serjeant}, {Temi}, {Thompson}, {Tuffs}, {van der Werf}, \&
  {Vlahakis}}]{Dunne_2011}
{Dunne}, L., {Gomez}, H.~L., {da Cunha}, E., {et~al.} 2011, \mnras, 417, 1510

\bibitem[{{Endo} {et~al.}(2019){Endo}, {Karatsu}, {Tamura}, {Oshima},
  {Taniguchi}, {Takekoshi}, {Asayama}, {Bakx}, {Bosma}, {Bueno}, {Chin},
  {Fujii}, {Fujita}, {Huiting}, {Ikarashi}, {Ishida}, {Ishii}, {Kawabe},
  {Klapwijk}, {Kohno}, {Kouchi}, {Llombart}, {Maekawa}, {Murugesan},
  {Nakatsubo}, {Naruse}, {Ohtawara}, {Pascual Laguna}, {Suzuki}, {Suzuki},
  {Thoen}, {Tsukagoshi}, {Ueda}, {de Visser}, {van der Werf}, {Yates},
  {Yoshimura}, {Yurduseven}, \& {Baselmans}}]{Endo_2019NatAs}
{Endo}, A., {Karatsu}, K., {Tamura}, Y., {et~al.} 2019, Nature Astronomy, 3,
  989

\bibitem[{{Foreman-Mackey}(2016)}]{FM_2016}
{Foreman-Mackey}, D. 2016, The Journal of Open Source Software, 1, 24

\bibitem[{{Foreman-Mackey} {et~al.}(2013){Foreman-Mackey}, {Hogg}, {Lang}, \&
  {Goodman}}]{FM_2013}
{Foreman-Mackey}, D., {Hogg}, D.~W., {Lang}, D., \& {Goodman}, J. 2013,
  Publications of the Astronomical Society of the Pacific, 125, 306

\bibitem[{{Fu} {et~al.}(2013){Fu}, {Kauffmann}, {Huang}, {Yates}, {Moran},
  {Heckman}, {Dav{\'e}}, {Guo}, \& {Henriques}}]{Fu_2013}
{Fu}, J., {Kauffmann}, G., {Huang}, M.-l., {et~al.} 2013, \mnras, 434, 1531

\bibitem[{{Furlanetto} \& {Lidz}(2007)}]{FL_2007}
{Furlanetto}, S.~R., \& {Lidz}, A. 2007, \apj, 660, 1030

\bibitem[{{Furlanetto} {et~al.}(2017){Furlanetto}, {Mirocha}, {Mebane}, \&
  {Sun}}]{Furlanetto_2017}
{Furlanetto}, S.~R., {Mirocha}, J., {Mebane}, R.~H., \& {Sun}, G. 2017, \mnras,
  472, 1576

\bibitem[{{Furlanetto} {et~al.}(2006){Furlanetto}, {Oh}, \&
  {Briggs}}]{FOB_2006}
{Furlanetto}, S.~R., {Oh}, S.~P., \& {Briggs}, F.~H. 2006, \physrep, 433, 181

\bibitem[{{Furlanetto} {et~al.}(2004){Furlanetto}, {Zaldarriaga}, \&
  {Hernquist}}]{FZH_2004}
{Furlanetto}, S.~R., {Zaldarriaga}, M., \& {Hernquist}, L. 2004, \apj, 613, 1

\bibitem[{{Gao} {et~al.}(2011){Gao}, {Frenk}, {Boylan-Kolchin}, {Jenkins},
  {Springel}, \& {White}}]{Gao_2011}
{Gao}, L., {Frenk}, C.~S., {Boylan-Kolchin}, M., {et~al.} 2011, \mnras, 410,
  2309

\bibitem[{{Genzel} {et~al.}(2012){Genzel}, {Tacconi}, {Combes}, {Bolatto},
  {Neri}, {Sternberg}, {Cooper}, {Bouch{\'e}}, {Bournaud}, {Burkert},
  {Comerford}, {Cox}, {Davis}, {F{\"o}rster Schreiber}, {Garcia-Burillo},
  {Gracia-Carpio}, {Lutz}, {Naab}, {Newman}, {Saintonge}, {Shapiro}, {Shapley},
  \& {Weiner}}]{Genzel2012}
{Genzel}, R., {Tacconi}, L.~J., {Combes}, F., {et~al.} 2012, \apj, 746, 69

\bibitem[{{Goldsmith} {et~al.}(2015){Goldsmith}, {Y{\i}ld{\i}z}, {Langer}, \&
  {Pineda}}]{Goldsmith_2015}
{Goldsmith}, P.~F., {Y{\i}ld{\i}z}, U.~A., {Langer}, W.~D., \& {Pineda}, J.~L.
  2015, \apj, 814, 133

\bibitem[{{Gong} {et~al.}(2012){Gong}, {Cooray}, {Silva}, {Santos}, {Bock},
  {Bradford}, \& {Zemcov}}]{Gong_2012}
{Gong}, Y., {Cooray}, A., {Silva}, M., {et~al.} 2012, \apj, 745, 49

\bibitem[{{Gong} {et~al.}(2017){Gong}, {Cooray}, {Silva}, {Zemcov}, {Feng},
  {Santos}, {Dore}, \& {Chen}}]{Gong_2017}
{Gong}, Y., {Cooray}, A., {Silva}, M.~B., {et~al.} 2017, \apj, 835, 273

\bibitem[{{Gruppioni} {et~al.}(2013){Gruppioni}, {Pozzi}, {Rodighiero},
  {Delvecchio}, {Berta}, {Pozzetti}, {Zamorani}, {Andreani}, {Cimatti},
  {Ilbert}, {Le Floc'h}, {Lutz}, {Magnelli}, {Marchetti}, {Monaco}, {Nordon},
  {Oliver}, {Popesso}, {Riguccini}, {Roseboom}, {Rosario}, {Sargent},
  {Vaccari}, {Altieri}, {Aussel}, {Bongiovanni}, {Cepa}, {Daddi},
  {Dom{\'\i}nguez-S{\'a}nchez}, {Elbaz}, {F{\"o}rster Schreiber}, {Genzel},
  {Iribarrem}, {Magliocchetti}, {Maiolino}, {Poglitsch}, {P{\'e}rez
  Garc{\'\i}a}, {Sanchez-Portal}, {Sturm}, {Tacconi}, {Valtchanov}, {Amblard},
  {Arumugam}, {Bethermin}, {Bock}, {Boselli}, {Buat}, {Burgarella},
  {Castro-Rodr{\'\i}guez}, {Cava}, {Chanial}, {Clements}, {Conley}, {Cooray},
  {Dowell}, {Dwek}, {Eales}, {Franceschini}, {Glenn}, {Griffin},
  {Hatziminaoglou}, {Ibar}, {Isaak}, {Ivison}, {Lagache}, {Levenson}, {Lu},
  {Madden}, {Maffei}, {Mainetti}, {Nguyen}, {O'Halloran}, {Page}, {Panuzzo},
  {Papageorgiou}, {Pearson}, {P{\'e}rez-Fournon}, {Pohlen}, {Rigopoulou},
  {Rowan-Robinson}, {Schulz}, {Scott}, {Seymour}, {Shupe}, {Smith}, {Stevens},
  {Symeonidis}, {Trichas}, {Tugwell}, {Vigroux}, {Wang}, {Wright}, {Xu},
  {Zemcov}, {Bardelli}, {Carollo}, {Contini}, {Le F{\'e}vre}, {Lilly},
  {Mainieri}, {Renzini}, {Scodeggio}, \& {Zucca}}]{Gruppioni_2013}
{Gruppioni}, C., {Pozzi}, F., {Rodighiero}, G., {et~al.} 2013, \mnras, 432, 23

\bibitem[{{Hailey-Dunsheath} {et~al.}(2014){Hailey-Dunsheath}, {Shirokoff},
  {Barry}, {Bradford}, {Chattopadhyay}, {Day}, {Doyle}, {Hollister}, {Kovacs},
  {LeDuc}, {Mauskopf}, {McKenney}, {Monroe}, {O'Brient}, {Padin}, {Reck},
  {Swenson}, {Tucker}, \& {Zmuidzinas}}]{HD_2014SPIE}
{Hailey-Dunsheath}, S., {Shirokoff}, E., {Barry}, P.~S., {et~al.} 2014, in
  Society of Photo-Optical Instrumentation Engineers (SPIE) Conference Series,
  Vol. 9153, \procspie, 91530M

\bibitem[{{Herrera-Camus} {et~al.}(2016){Herrera-Camus}, {Bolatto}, {Smith},
  {Draine}, {Pellegrini}, {Wolfire}, {Croxall}, {de Looze}, {Calzetti},
  {Kennicutt}, {Crocker}, {Armus}, {van der Werf}, {Sandstrom}, {Galametz},
  {Brandl}, {Groves}, {Rigopoulou}, {Walter}, {Leroy}, {Boquien}, {Tabatabaei},
  \& {Beirao}}]{HC_2016}
{Herrera-Camus}, R., {Bolatto}, A., {Smith}, J.~D., {et~al.} 2016, \apj, 826,
  175

\bibitem[{{Ho} {et~al.}(2009){Ho}, {Altamirano}, {Chang}, {Chang}, {Chang},
  {Chen}, {Chen}, {Chen}, {Han}, \& {Ho}}]{Ho_2009}
{Ho}, P. T.~P., {Altamirano}, P., {Chang}, C.-H., {et~al.} 2009, \apj, 694,
  1610

\bibitem[{{Hughes} {et~al.}(2015){Hughes}, {Foyle}, {Schirm}, {Parkin}, {De
  Looze}, {Wilson}, {Bendo}, {Baes}, {Fritz}, {Boselli}, {Cooray}, {Cormier},
  {Karczewski}, {Lebouteiller}, {Lu}, {Madden}, {Spinoglio}, \&
  {Viaene}}]{Hughes_2015}
{Hughes}, T.~M., {Foyle}, K., {Schirm}, M.~R.~P., {et~al.} 2015, \aap, 575, A17

\bibitem[{Hunter(2007)}]{Hunter_2007}
Hunter, J.~D. 2007, Computing In Science \& Engineering, 9, 90

\bibitem[{{Ivison} {et~al.}(2011){Ivison}, {Papadopoulos}, {Smail}, {Greve},
  {Thomson}, {Xilouris}, \& {Chapman}}]{Ivison2011}
{Ivison}, R.~J., {Papadopoulos}, P.~P., {Smail}, I., {et~al.} 2011, \mnras,
  412, 1913

\bibitem[{{Jones} {et~al.}(2001){Jones}, {Oliphant}, {Peterson},
  {et~al.}}]{Jones_2001}
{Jones}, E., {Oliphant}, T., {Peterson}, P., {et~al.} 2001, {SciPy}: Open
  source scientific tools for {Python}

\bibitem[{{Keating} {et~al.}(2016){Keating}, {Marrone}, {Bower}, {Leitch},
  {Carlstrom}, \& {DeBoer}}]{Keating_2016}
{Keating}, G.~K., {Marrone}, D.~P., {Bower}, G.~C., {et~al.} 2016, \apj, 830,
  34

\bibitem[{{Kennicutt}(1998)}]{Kennicutt_1998}
{Kennicutt}, Jr., R.~C. 1998, \araa, 36, 189

\bibitem[{{Kovetz} {et~al.}(2017){Kovetz}, {Viero}, {Lidz}, {Newburgh},
  {Rahman}, {Switzer}, {Kamionkowski}, {Aguirre}, {Alvarez}, \&
  {Bock}}]{Kovetz_2017}
{Kovetz}, E.~D., {Viero}, M.~P., {Lidz}, A., {et~al.} 2017, arXiv e-prints,
  arXiv:1709.09066

\bibitem[{{Kravtsov} {et~al.}(2004){Kravtsov}, {Berlind}, {Wechsler}, {Klypin},
  {Gottl{\"o}ber}, {Allgood}, \& {Primack}}]{Kravtsov_2004}
{Kravtsov}, A.~V., {Berlind}, A.~A., {Wechsler}, R.~H., {et~al.} 2004, \apj,
  609, 35

\bibitem[{{Kravtsov} {et~al.}(2018){Kravtsov}, {Vikhlinin}, \&
  {Meshcheryakov}}]{KVM_2018}
{Kravtsov}, A.~V., {Vikhlinin}, A.~A., \& {Meshcheryakov}, A.~V. 2018,
  Astronomy Letters, 44, 8

\bibitem[{{Krumholz} {et~al.}(2009){Krumholz}, {McKee}, \&
  {Tumlinson}}]{KMT_2009}
{Krumholz}, M.~R., {McKee}, C.~F., \& {Tumlinson}, J. 2009, \apj, 693, 216

\bibitem[{{Lagache}(2018)}]{Lagache_2018IAUS}
{Lagache}, G. 2018, in IAU Symposium, Vol. 333, Peering towards Cosmic Dawn,
  ed. V.~{Jeli{\'c}} \& T.~{van der Hulst}, 228--233

\bibitem[{{Li} {et~al.}(2019){Li}, {Narayanan}, \& {Dav{\'e}}}]{Li_2019}
{Li}, Q., {Narayanan}, D., \& {Dav{\'e}}, R. 2019, \mnras, 490, 1425

\bibitem[{{Li} {et~al.}(2016){Li}, {Wechsler}, {Devaraj}, \&
  {Church}}]{Li_2016}
{Li}, T.~Y., {Wechsler}, R.~H., {Devaraj}, K., \& {Church}, S.~E. 2016, \apj,
  817, 169

\bibitem[{{Lidz} {et~al.}(2009){Lidz}, {Zahn}, {Furlanetto}, {McQuinn},
  {Hernquist}, \& {Zaldarriaga}}]{Lidz_2009}
{Lidz}, A., {Zahn}, O., {Furlanetto}, S.~R., {et~al.} 2009, \apj, 690, 252

\bibitem[{{Loeb} \& {Furlanetto}(2013)}]{LF_2013}
{Loeb}, A., \& {Furlanetto}, S.~R. 2013, {The First Galaxies in the Universe}
  (Princeton University Press)

\bibitem[{{Madau} \& {Dickinson}(2014)}]{MD_2014}
{Madau}, P., \& {Dickinson}, M. 2014, \araa, 52, 415

\bibitem[{{Madau} {et~al.}(1997){Madau}, {Meiksin}, \& {Rees}}]{MMR_1997}
{Madau}, P., {Meiksin}, A., \& {Rees}, M.~J. 1997, \apj, 475, 429

\bibitem[{{Malhotra} {et~al.}(1997){Malhotra}, {Helou}, {Stacey}, {Hollenbach},
  {Lord}, {Beichman}, {Dinerstein}, {Hunter}, {Lo}, {Lu}, {Rubin},
  {Silbermann}, {Thronson}, \& {Werner}}]{Malhotra1997}
{Malhotra}, S., {Helou}, G., {Stacey}, G., {et~al.} 1997, \apjl, 491, L27

\bibitem[{{Maniyar} {et~al.}(2018){Maniyar}, {B{\'e}thermin}, \&
  {Lagache}}]{Maniyar_2018}
{Maniyar}, A.~S., {B{\'e}thermin}, M., \& {Lagache}, G. 2018, \aap, 614, A39

\bibitem[{{Mashian} {et~al.}(2015){Mashian}, {Sternberg}, \&
  {Loeb}}]{Mashian_2015}
{Mashian}, N., {Sternberg}, A., \& {Loeb}, A. 2015, Journal of Cosmology and
  Astro-Particle Physics, 2015, 028

\bibitem[{{McBride} {et~al.}(2009){McBride}, {Fakhouri}, \&
  {Ma}}]{McBride_2009}
{McBride}, J., {Fakhouri}, O., \& {Ma}, C.-P. 2009, \mnras, 398, 1858

\bibitem[{{McCracken} {et~al.}(2015){McCracken}, {Wolk}, {Colombi},
  {Kilbinger}, {Ilbert}, {Peirani}, {Coupon}, {Dunlop}, {Milvang-Jensen},
  {Caputi}, {Aussel}, {B{\'e}thermin}, \& {Le F{\`e}vre}}]{McCracken_2015}
{McCracken}, H.~J., {Wolk}, M., {Colombi}, S., {et~al.} 2015, \mnras, 449, 901

\bibitem[{{McKee} \& {Krumholz}(2010)}]{MK_2010}
{McKee}, C.~F., \& {Krumholz}, M.~R. 2010, \apj, 709, 308

\bibitem[{{M{\'e}nard} \& {Fukugita}(2012)}]{Menard_2012}
{M{\'e}nard}, B., \& {Fukugita}, M. 2012, \apj, 754, 116

\bibitem[{{M{\'e}nard} {et~al.}(2010){M{\'e}nard}, {Scranton}, {Fukugita}, \&
  {Richards}}]{Menard_2010}
{M{\'e}nard}, B., {Scranton}, R., {Fukugita}, M., \& {Richards}, G. 2010,
  \mnras, 405, 1025

\bibitem[{{Mo} {et~al.}(2010){Mo}, {van den Bosch}, \& {White}}]{Mo_book}
{Mo}, H., {van den Bosch}, F.~C., \& {White}, S. 2010, {Galaxy Formation and
  Evolution} (Cambridge University Press)

\bibitem[{{Mu{\~n}oz} \& {Oh}(2016)}]{Munoz2016}
{Mu{\~n}oz}, J.~A., \& {Oh}, S.~P. 2016, \mnras, 463, 2085

\bibitem[{{Murray} {et~al.}(2013){Murray}, {Power}, \& {Robotham}}]{MPR_2013}
{Murray}, S.~G., {Power}, C., \& {Robotham}, A.~S.~G. 2013, Astronomy and
  Computing, 3, 23

\bibitem[{{Navarro} {et~al.}(1997){Navarro}, {Frenk}, \& {White}}]{NFW_1997}
{Navarro}, J.~F., {Frenk}, C.~S., \& {White}, S. D.~M. 1997, \apj, 490, 493

\bibitem[{{Newburgh} {et~al.}(2016){Newburgh}, {Bandura}, {Bucher}, {Chang},
  {Chiang}, {Cliche}, {Dav{\'e}}, {Dobbs}, {Clarkson}, \&
  {Ganga}}]{Newburgh_2016SPIE}
{Newburgh}, L.~B., {Bandura}, K., {Bucher}, M.~A., {et~al.} 2016, in Society of
  Photo-Optical Instrumentation Engineers (SPIE) Conference Series, Vol. 9906,
  Ground-based and Airborne Telescopes VI, 99065X

\bibitem[{{Padmanabhan}(2018)}]{Padmanabhan_2018}
{Padmanabhan}, H. 2018, \mnras, 475, 1477

\bibitem[{{Padmanabhan}(2019)}]{Padmanabhan_2019}
---. 2019, \mnras, 488, 3014

\bibitem[{{Padmanabhan} {et~al.}(2017){Padmanabhan}, {Refregier}, \&
  {Amara}}]{Padmanabhan_2017}
{Padmanabhan}, H., {Refregier}, A., \& {Amara}, A. 2017, \mnras, 469, 2323

\bibitem[{{Pallottini} {et~al.}(2019){Pallottini}, {Ferrara}, {Decataldo},
  {Gallerani}, {Vallini}, {Carniani}, {Behrens}, {Kohandel}, \&
  {Salvadori}}]{Pallottini_2019}
{Pallottini}, A., {Ferrara}, A., {Decataldo}, D., {et~al.} 2019, \mnras, 487,
  1689

\bibitem[{{Pereira-Santaella} {et~al.}(2014){Pereira-Santaella}, {Spinoglio},
  {van der Werf}, \& {Piqueras L{\'o}pez}}]{PS_2014}
{Pereira-Santaella}, M., {Spinoglio}, L., {van der Werf}, P.~P., \& {Piqueras
  L{\'o}pez}, J. 2014, \aap, 566, A49

\bibitem[{{Planck Collaboration Int.~XVII}(2014)}]{Planck_Int_XVII}
{Planck Collaboration Int.~XVII}. 2014, \aap, 566, A55

\bibitem[{{Planck Collaboration XIII}(2016)}]{Planck2016XIII}
{Planck Collaboration XIII}. 2016, \aap, 594, A13

\bibitem[{{Planck Collaboration XXX}(2014)}]{planckXXX}
{Planck Collaboration XXX}. 2014, \aap, 571, A30

\bibitem[{{Popping} {et~al.}(2015){Popping}, {Behroozi}, \&
  {Peeples}}]{Popping_2015}
{Popping}, G., {Behroozi}, P.~S., \& {Peeples}, M.~S. 2015, \mnras, 449, 477

\bibitem[{{Popping} {et~al.}(2012){Popping}, {Caputi}, {Somerville}, \&
  {Trager}}]{Popping_2012}
{Popping}, G., {Caputi}, K.~I., {Somerville}, R.~S., \& {Trager}, S.~C. 2012,
  \mnras, 425, 2386

\bibitem[{{Popping} {et~al.}(2019){Popping}, {Narayanan}, {Somerville},
  {Faisst}, \& {Krumholz}}]{Popping_2019}
{Popping}, G., {Narayanan}, D., {Somerville}, R.~S., {Faisst}, A.~L., \&
  {Krumholz}, M.~R. 2019, \mnras, 482, 4906

\bibitem[{{Pritchard} \& {Loeb}(2012)}]{PL_2012}
{Pritchard}, J.~R., \& {Loeb}, A. 2012, Reports on Progress in Physics, 75,
  086901

\bibitem[{{Pullen} {et~al.}(2013){Pullen}, {Chang}, {Dor{\'e}}, \&
  {Lidz}}]{pullen2013}
{Pullen}, A.~R., {Chang}, T.-C., {Dor{\'e}}, O., \& {Lidz}, A. 2013, \apj, 768,
  15

\bibitem[{{Pullen} {et~al.}(2014){Pullen}, {Dor{\'e}}, \& {Bock}}]{Pullen_2014}
{Pullen}, A.~R., {Dor{\'e}}, O., \& {Bock}, J. 2014, \apj, 786, 111

\bibitem[{{Pullen} {et~al.}(2018){Pullen}, {Serra}, {Chang}, {Dor{\'e}}, \&
  {Ho}}]{Pullen_2018}
{Pullen}, A.~R., {Serra}, P., {Chang}, T.-C., {Dor{\'e}}, O., \& {Ho}, S. 2018,
  \mnras, 478, 1911

\bibitem[{{Riechers} {et~al.}(2019){Riechers}, {Pavesi}, {Sharon}, {Hodge},
  {Decarli}, {Walter}, {Carilli}, {Aravena}, {da Cunha}, {Daddi}, {Dickinson},
  {Smail}, {Capak}, {Ivison}, {Sargent}, {Scoville}, \& {Wagg}}]{Riechers_2019}
{Riechers}, D.~A., {Pavesi}, R., {Sharon}, C.~E., {et~al.} 2019, \apj, 872, 7

\bibitem[{{Robertson} {et~al.}(2015){Robertson}, {Ellis}, {Furlanetto}, \&
  {Dunlop}}]{Robertson_2015}
{Robertson}, B.~E., {Ellis}, R.~S., {Furlanetto}, S.~R., \& {Dunlop}, J.~S.
  2015, \apjl, 802, L19

\bibitem[{{Rowan-Robinson} {et~al.}(2016){Rowan-Robinson}, {Oliver}, {Wang},
  {Farrah}, {Clements}, {Gruppioni}, {Marchetti}, {Rigopoulou}, \&
  {Vaccari}}]{RR_2016}
{Rowan-Robinson}, M., {Oliver}, S., {Wang}, L., {et~al.} 2016, \mnras, 461,
  1100

\bibitem[{{Rubin} {et~al.}(2009){Rubin}, {Hony}, {Madden}, {Tielens},
  {Meixner}, {Indebetouw}, {Reach}, {Ginsburg}, {Kim}, {Mochizuki}, {Babler},
  {Block}, {Bracker}, {Engelbracht}, {For}, {Gordon}, {Hora}, {Leitherer},
  {Meade}, {Misselt}, {Sewilo}, {Vijh}, \& {Whitney}}]{Rubin_2009}
{Rubin}, D., {Hony}, S., {Madden}, S.~C., {et~al.} 2009, \aap, 494, 647

\bibitem[{{Rybak} {et~al.}(2019){Rybak}, {Calistro Rivera}, {Hodge}, {Smail},
  {Walter}, {van der Werf}, {da Cunha}, {Chen}, {Dannerbauer}, {Ivison},
  {Karim}, {Simpson}, {Swinbank}, \& {Wardlow}}]{Rybak_2019}
{Rybak}, M., {Calistro Rivera}, G., {Hodge}, J.~A., {et~al.} 2019, \apj, 876,
  112

\bibitem[{{Saintonge} {et~al.}(2011){Saintonge}, {Kauffmann}, {Kramer},
  {Tacconi}, {Buchbender}, {Catinella}, {Fabello}, {Graci{\'a}-Carpio}, {Wang},
  {Cortese}, {Fu}, {Genzel}, {Giovanelli}, {Guo}, {Haynes}, {Heckman},
  {Krumholz}, {Lemonias}, {Li}, {Moran}, {Rodriguez-Fernandez}, {Schiminovich},
  {Schuster}, \& {Sievers}}]{Saintonge2011}
{Saintonge}, A., {Kauffmann}, G., {Kramer}, C., {et~al.} 2011, \mnras, 415, 32

\bibitem[{{Sandstrom} {et~al.}(2013){Sandstrom}, {Leroy}, {Walter}, {Bolatto},
  {Croxall}, {Draine}, {Wilson}, {Wolfire}, {Calzetti}, {Kennicutt}, {Aniano},
  {Donovan Meyer}, {Usero}, {Bigiel}, {Brinks}, {de Blok}, {Crocker}, {Dale},
  {Engelbracht}, {Galametz}, {Groves}, {Hunt}, {Koda}, {Kreckel}, {Linz},
  {Meidt}, {Pellegrini}, {Rix}, {Roussel}, {Schinnerer}, {Schruba}, {Schuster},
  {Skibba}, {van der Laan}, {Appleton}, {Armus}, {Brandl}, {Gordon}, {Hinz},
  {Krause}, {Montiel}, {Sauvage}, {Schmiedeke}, {Smith}, \&
  {Vigroux}}]{Sandstrom2013}
{Sandstrom}, K.~M., {Leroy}, A.~K., {Walter}, F., {et~al.} 2013, \apj, 777, 5

\bibitem[{{Santos} {et~al.}(2015){Santos}, {Bull}, {Alonso}, {Camera},
  {Ferreira}, {Bernardi}, {Maartens}, {Viel}, {Villaescusa-Navarro}, \&
  {Abdalla}}]{Santos_2015}
{Santos}, M., {Bull}, P., {Alonso}, D., {et~al.} 2015, in Advancing
  Astrophysics with the Square Kilometre Array (AASKA14), 19

\bibitem[{{Savaglio}(2006)}]{Savaglio_2006NJPh}
{Savaglio}, S. 2006, New Journal of Physics, 8, 195

\bibitem[{{Serra} {et~al.}(2016){Serra}, {Dor{\'e}}, \& {Lagache}}]{Serra_2016}
{Serra}, P., {Dor{\'e}}, O., \& {Lagache}, G. 2016, \apj, 833, 153

\bibitem[{{Shang} {et~al.}(2012){Shang}, {Haiman}, {Knox}, \&
  {Oh}}]{Shang_2012}
{Shang}, C., {Haiman}, Z., {Knox}, L., \& {Oh}, S.~P. 2012, \mnras, 421, 2832

\bibitem[{{Silva} {et~al.}(2018){Silva}, {Zaroubi}, {Kooistra}, \&
  {Cooray}}]{Silva_2018}
{Silva}, B.~M., {Zaroubi}, S., {Kooistra}, R., \& {Cooray}, A. 2018, \mnras,
  475, 1587

\bibitem[{{Silva} {et~al.}(2015){Silva}, {Santos}, {Cooray}, \&
  {Gong}}]{Silva_2015}
{Silva}, M., {Santos}, M.~G., {Cooray}, A., \& {Gong}, Y. 2015, \apj, 806, 209

\bibitem[{{Silva} {et~al.}(2013){Silva}, {Santos}, {Gong}, {Cooray}, \&
  {Bock}}]{Silva_2013}
{Silva}, M.~B., {Santos}, M.~G., {Gong}, Y., {Cooray}, A., \& {Bock}, J. 2013,
  \apj, 763, 132

\bibitem[{{Solomon} {et~al.}(1987){Solomon}, {Rivolo}, {Barrett}, \&
  {Yahil}}]{Solomon1987}
{Solomon}, P.~M., {Rivolo}, A.~R., {Barrett}, J., \& {Yahil}, A. 1987, \apj,
  319, 730

\bibitem[{{Spinoglio} {et~al.}(2012){Spinoglio}, {Dasyra}, {Franceschini},
  {Gruppioni}, {Valiante}, \& {Isaak}}]{Spinoglio_2012}
{Spinoglio}, L., {Dasyra}, K.~M., {Franceschini}, A., {et~al.} 2012, \apj, 745,
  171

\bibitem[{{Stacey} {et~al.}(1991){Stacey}, {Geis}, {Genzel}, {Lugten},
  {Poglitsch}, {Sternberg}, \& {Townes}}]{Stacey1991}
{Stacey}, G.~J., {Geis}, N., {Genzel}, R., {et~al.} 1991, \apj, 373, 423

\bibitem[{{Stacey} {et~al.}(2018){Stacey}, {Aravena}, {Basu}, {Battaglia},
  {Beringue}, {Bertoldi}, {Bond}, {Breysse}, {Bustos}, \&
  {Chapman}}]{Stacey_2018SPIE}
{Stacey}, G.~J., {Aravena}, M., {Basu}, K., {et~al.} 2018, in Society of
  Photo-Optical Instrumentation Engineers (SPIE) Conference Series, Vol. 10700,
  Ground-based and Airborne Telescopes VII, 107001M

\bibitem[{{Sun} \& {Furlanetto}(2016)}]{SF_2016}
{Sun}, G., \& {Furlanetto}, S.~R. 2016, \mnras, 460, 417

\bibitem[{{Sun} {et~al.}(2018){Sun}, {Moncelsi}, {Viero}, {Silva}, {Bock},
  {Bradford}, {Chang}, {Cheng}, {Cooray}, {Crites}, {Hailey-Dunsheath},
  {Uzgil}, {Hunacek}, \& {Zemcov}}]{Sun_2018}
{Sun}, G., {Moncelsi}, L., {Viero}, M.~P., {et~al.} 2018, \apj, 856, 107

\bibitem[{{Switzer}(2017)}]{EXCLAIM_2017}
{Switzer}, E. 2017, {Measuring the Cosmological Evolution of Gas and Galaxies
  with the EXperiment for Cryogenic Large-aperture Intensity Mapping
  (EXCLAIM)}, NASA APRA Proposal

\bibitem[{{Switzer} {et~al.}(2019){Switzer}, {Anderson}, {Pullen}, \&
  {Yang}}]{Switzer_2019}
{Switzer}, E.~R., {Anderson}, C.~J., {Pullen}, A.~R., \& {Yang}, S. 2019, \apj,
  872, 82

\bibitem[{{Switzer} {et~al.}(2013){Switzer}, {Masui}, {Bandura}, {Calin},
  {Chang}, {Chen}, {Li}, {Liao}, {Natarajan}, {Pen}, {Peterson}, {Shaw}, \&
  {Voytek}}]{Switzer_2013}
{Switzer}, E.~R., {Masui}, K.~W., {Bandura}, K., {et~al.} 2013, \mnras, 434,
  L46

\bibitem[{{Tayal}(2008)}]{Tayal_2008}
{Tayal}, S.~S. 2008, \aap, 486, 629

\bibitem[{{Tayal}(2011)}]{Tayal_2011}
---. 2011, \apjs, 195, 12

\bibitem[{{Thacker} {et~al.}(2013){Thacker}, {Cooray}, {Smidt}, {De Bernardis},
  {Mitchell-Wynne}, {Amblard}, {Auld}, {Baes}, {Clements}, {Dariush}, {De
  Zotti}, {Dunne}, {Eales}, {Hopwood}, {Hoyos}, {Ibar}, {Jarvis}, {Maddox},
  {Micha{\l}owski}, {Pascale}, {Scott}, {Serjeant}, {Smith}, {Valiante}, \&
  {van der Werf}}]{Thacker2013}
{Thacker}, C., {Cooray}, A., {Smidt}, J., {et~al.} 2013, \apj, 768, 58

\bibitem[{{Tielens}(2005)}]{Tielens_2005}
{Tielens}, A.~G.~G.~M. 2005, {The Physics and Chemistry of the Interstellar
  Medium} (Cambridge University Press)

\bibitem[{{Tielens} \& {Hollenbach}(1985)}]{Tielens1985}
{Tielens}, A.~G.~G.~M., \& {Hollenbach}, D. 1985, \apj, 291, 722

\bibitem[{{Tinker} {et~al.}(2008){Tinker}, {Kravtsov}, {Klypin}, {Abazajian},
  {Warren}, {Yepes}, {Gottl{\"o}ber}, \& {Holz}}]{Tinker_2008}
{Tinker}, J., {Kravtsov}, A.~V., {Klypin}, A., {et~al.} 2008, \apj, 688, 709

\bibitem[{{Tinker} \& {Wetzel}(2010)}]{TW_2010}
{Tinker}, J.~L., \& {Wetzel}, A.~R. 2010, \apj, 719, 88

\bibitem[{{Togi} \& {Smith}(2016)}]{Togi2016}
{Togi}, A., \& {Smith}, J.~D.~T. 2016, \apj, 830, 18

\bibitem[{{Uzgil} {et~al.}(2019){Uzgil}, {Carilli}, {Lidz}, {Walter},
  {Thyagarajan}, {Decarli}, {Aravena}, {Bertoldi}, {Cortes},
  {Gonz{\'a}lez-L{\'o}pez}, {Inami}, {Popping}, {Van der Werf}, {Wagg}, \&
  {Weiss}}]{Uzgil_2019}
{Uzgil}, B., {Carilli}, C., {Lidz}, A., {et~al.} 2019, arXiv e-prints,
  arXiv:1911.00028

\bibitem[{{Uzgil} {et~al.}(2014){Uzgil}, {Aguirre}, {Bradford}, \&
  {Lidz}}]{Uzgil_2014}
{Uzgil}, B.~D., {Aguirre}, J.~E., {Bradford}, C.~M., \& {Lidz}, A. 2014, \apj,
  793, 116

\bibitem[{{Viero} {et~al.}(2013{\natexlab{a}}){Viero}, {Wang}, {Zemcov},
  {Addison}, {Amblard}, {Arumugam}, {Aussel}, {B{\'e}thermin}, {Bock}, \&
  {Boselli}}]{Viero_2013CIB}
{Viero}, M.~P., {Wang}, L., {Zemcov}, M., {et~al.} 2013{\natexlab{a}}, \apj,
  772, 77

\bibitem[{{Viero} {et~al.}(2013{\natexlab{b}}){Viero}, {Moncelsi}, {Quadri},
  {Arumugam}, {Assef}, {B{\'e}thermin}, {Bock}, {Bridge}, {Casey}, {Conley},
  {Cooray}, {Farrah}, {Glenn}, {Heinis}, {Ibar}, {Ikarashi}, {Ivison}, {Kohno},
  {Marsden}, {Oliver}, {Roseboom}, {Schulz}, {Scott}, {Serra}, {Vaccari},
  {Vieira}, {Wang}, {Wardlow}, {Wilson}, {Yun}, \& {Zemcov}}]{Viero_2013}
{Viero}, M.~P., {Moncelsi}, L., {Quadri}, R.~F., {et~al.} 2013{\natexlab{b}},
  \apj, 779, 32

\bibitem[{{Visbal} \& {Loeb}(2010)}]{VL_2010}
{Visbal}, E., \& {Loeb}, A. 2010, \jcap, 11, 016

\bibitem[{Walt {et~al.}(2011)Walt, Colbert, \& Varoquaux}]{Walt_2011}
Walt, S. v.~d., Colbert, S.~C., \& Varoquaux, G. 2011, Computing in Science and
  Engg., 13, 22

\bibitem[{{Walter} {et~al.}(2019){Walter}, {Carilli}, {Decarli}, {Riechers},
  {Aravena}, {Bauer}, {Bertoldi}, {Bolatto}, {Boogaard}, {Bouwens},
  {Burgarella}, {Casey}, {Cooray}, {Cortes}, {Cox}, {Daddi}, {Darling},
  {Emonts}, {Gonzalez Lopez}, {Hodge}, {Inami}, {Ivison}, {Kovetz}, {Le
  F{\`e}vre}, {Magnelli}, {Marrone}, {Murphy}, {Narayanan}, {Novak}, {Oesch},
  {Pavesi}, {Diaz Santos}, {Sargent}, {Scott}, {Scoville}, {Stacey}, {Wagg},
  {van der Werf}, {Uzgil}, {Weiss}, \& {Yun}}]{Walter_2019BAAS}
{Walter}, F., {Carilli}, C., {Decarli}, R., {et~al.} 2019, \baas, 51, 442

\bibitem[{{Wetzel} {et~al.}(2009){Wetzel}, {Cohn}, \& {White}}]{Wetzel_2009}
{Wetzel}, A.~R., {Cohn}, J.~D., \& {White}, M. 2009, \mnras, 395, 1376

\bibitem[{{Wolz} {et~al.}(2017){Wolz}, {Blake}, \& {Wyithe}}]{Wolz_2017}
{Wolz}, L., {Blake}, C., \& {Wyithe}, J.~S.~B. 2017, \mnras, 470, 3220

\bibitem[{{Wright} {et~al.}(1991){Wright}, {Mather}, {Bennett}, {Cheng},
  {Shafer}, {Fixsen}, {Eplee}, {Isaacman}, {Read}, {Boggess}, {Gulkis},
  {Hauser}, {Janssen}, {Kelsall}, {Lubin}, {Meyer}, {Moseley}, {Murdock},
  {Silverberg}, {Smoot}, {Weiss}, \& {Wilkinson}}]{Wright1991}
{Wright}, E.~L., {Mather}, J.~C., {Bennett}, C.~L., {et~al.} 1991, \apj, 381,
  200

\bibitem[{{Wu} \& {Dor{\'e}}(2017)}]{WD_2017}
{Wu}, H.-Y., \& {Dor{\'e}}, O. 2017, \mnras, 467, 4150

\bibitem[{{Xu} {et~al.}(2015){Xu}, {Wang}, \& {Chen}}]{Xu_2015}
{Xu}, Y., {Wang}, X., \& {Chen}, X. 2015, \apj, 798, 40

\bibitem[{{Yang} {et~al.}(2019){Yang}, {Pullen}, \& {Switzer}}]{YPS_2019}
{Yang}, S., {Pullen}, A.~R., \& {Switzer}, E.~R. 2019, \mnras, 489, L53

\bibitem[{{Young Owl} {et~al.}(2002){Young Owl}, {Meixner}, {Fong}, {Haas},
  {Rudolph}, \& {Tielens}}]{YoungOwl2002}
{Young Owl}, R.~C., {Meixner}, M.~M., {Fong}, D., {et~al.} 2002, \apj, 578, 885

\bibitem[{{Yue} \& {Ferrara}(2019)}]{YF_2019}
{Yue}, B., \& {Ferrara}, A. 2019, \mnras, 490, 1928

\bibitem[{{Yue} {et~al.}(2015){Yue}, {Ferrara}, {Pallottini}, {Gallerani}, \&
  {Vallini}}]{Yue_2015}
{Yue}, B., {Ferrara}, A., {Pallottini}, A., {Gallerani}, S., \& {Vallini}, L.
  2015, \mnras, 450, 3829

\end{thebibliography}

\end{document}